\documentclass[12pt,a4paper]{article}
\usepackage[latin1]{inputenc}
\usepackage{times}
\usepackage{a4wide}
\usepackage{amsfonts}
\usepackage{amssymb}
\usepackage{amsmath}
\usepackage{ifpdf}
\ifpdf
\usepackage[pdftex]{graphicx}
\usepackage[pdftex,unicode,implicit]{hyperref}
\hypersetup{%
  pdftitle    = {The general gaugings of maximal d=9 supergravity},
  pdfkeywords = {Supersymmetry, supergravity, symmetry, gauge },
  pdfauthor   = {J.J. Fernandez-Melgarejo, T. Ortin and E. Torrente-Lujan},
  pdfcreator  = {pdf\LaTeXe\ with package \flqq hyperref\frqq},
  pdfproducer = {pdf\LaTeXe\ with package \flqq hyperref\frqq},
  pdfpagemode = None,  %change this to FullScreen if you want...
  pdffitwindow= true,  % Open the pdf file without the sidebars open.
  unicode     = true,
  plainpages  = true,
  colorlinks  = true,  % links have no different colors..
  citecolor   = blue,
  urlcolor    = red,
  linkcolor   = black
}
\newcommand{\hepth}[1]{arXiv:{\tt
\href{http://www.arXiv.org/abs/hep-th/#1}{hep-th/#1}}}

\newcommand{\arxiv}[1]{{\tt
\href{http://www.arXiv.org/abs/#1}{arXiv:#1}}}
\else
  \usepackage[dvips]{graphicx}
  \usepackage[unicode,implicit]{hyperref}
  \newcommand{\hepth}[1]{arXiv:{\tt hep-th/#1}}

  \newcommand{\arxiv}[1]{{\tt arXiv:#1}}
\fi
\makeatletter
\@addtoreset{equation}{section}
\makeatother

\makeindex
\begin{document}

\begin{flushright}
\small
IFT-UAM/CSIC-11-18\\
UM-TH/11-09\\
%\texttt{arXiv:YYMM.NNNN}\\
June  $9^{\rm th}$, 2011\\
\normalsize
\end{flushright}

\begin{center}

\vspace{1.5cm}

{\LARGE {\bf The General Gaugings of Maximal\\[1cm] $d=9$ Supergravity}}

\vspace{1cm}

\begin{center}

{\sl\large J.J.~Fern\'andez-Melgarejo$^{\dagger}$}
\footnote{E-mail: {\tt jj.fernandezmelgarejo@um.es}},
{\sl\large T.~Ort\'{\i}n$^{\diamond}$}
\footnote{E-mail: {\tt Tomas.Ortin@csic.es}}
{\sl\large and E.~Torrente-Luj\'an$^{\dagger}$}
\footnote{E-mail: {\tt torrente@cern.ch}}

\vspace{1cm}

${}^{\dagger}${\it Grupo de F\'{\i}sica Te\'orica y Cosmolog\'{\i}a, Dept. de F\'{\i}sica,\\
  U.~Murcia, Campus de Espinardo, E-30100-Murcia,  Spain}\\

\vspace{.6cm}

${}^{\diamond}${\it Instituto de F\'{\i}sica Te\'orica UAM/CSIC\\
C/ Nicol\'as Cabrera, 13-15,  C.U.~Cantoblanco,  E-28049-Madrid, Spain}\\

\end{center}

\vspace{1cm}

{\bf Abstract}

\begin{quotation}

  {\small 
    We use the embedding tensor method to construct the most general maximal
    gauged/massive supergravity in $d=9$ dimensions and to determine its
    extended field content. Only the 8 independent deformation parameters
    (embedding tensor components, mass parameters etc.)  identified by
    Bergshoeff \textit{et al.} (an $SL(2,\mathbb{R})$ triplet, two doublets
    and a singlet) can be consistently introduced in the theory, but their
    simultaneous use is subject to a number of quadratic constraints. These
    constraints have to be kept and enforced because they cannot be used to
    solve some deformation parameters in terms of the rest. The deformation
    parameters are associated to the possible 8-forms of the theory, and the
    constraints are associated to the 9-forms, all of them transforming in the
    conjugate representations. We also give the field strengths and the gauge
    and supersymmetry transformations for the electric fields in the most
    general case.  We compare these results with the predictions of the
    $E_{11}$ approach, finding that the latter predicts one additional doublet
    of 9-forms, analogously to what happens in $N=2$ $d=4,5,6$ theories. 
}

\end{quotation}

\end{center}

\newpage
%%%%%%%%%%%%%%%%%%%%%%%%%%%%%%%%%%%%%%%%%%%%%%%%%%%%%%%%%%%%%%%%%%%%%%
%%%%%%%%%%%%%%%%%%%%%%%%%%%%%%%%%%%%%%%%%%%%%%%%%%%%%%%%%%%%%%%%%%%%%%
%%%%%%%%%%%%%%%%%%%%%%%%%%%%%%%%%%%%%%%%%%%%%%%%%%%%%%%%%%%%%%%%%%%%%%
%%%%%%%%%%%%%%%%%%%%%%%%%%%%%%%%%%%%%%%%%%%%%%%%%%%%%%%%%%%%%%%%%%%%%%
\pagestyle{plain}
%%%%%%%%%%%%%%%%%%%%%%%%%%%%%%%%%%%%%%%%%%%%%%%%%%%%%%%%%%%%%%%%%%%%%%
%%%%%%%%%%%%%%%%%%%%%%%%%%%%%%%%%%%%%%%%%%%%%%%%%%%%%%%%%%%%%%%%%%%%%%
%%%%%%%%%%%%%%%%%%%%%%%%%%%%%%%%%%%%%%%%%%%%%%%%%%%%%%%%%%%%%%%%%%%%%%
%%%%%%%%%%%%%%%%%%%%%%%%%%%%%%%%%%%%%%%%%%%%%%%%%%%%%%%%%%%%%%%%%%%%%%
%%%%%%%%%%%%%%%%%%%%%%%%%%%%%%%%%%%%%%%%%%%%%%%%%%%%%%%%%%%%%%%%%%%%%%

\tableofcontents

\newpage

%%%%%%%%%%%%%%%%%%%%%%%%%%%%%%%%%%%%%%%%%%%%%%%%%%%%%%%%%%%%%%%%%%%%%%
%%%%%%%%%%%%%%%%%%%%%%%%%%%%%%%%%%%%%%%%%%%%%%%%%%%%%%%%%%%%%%%%%%%%%%
%%%%%%%%%%%%%%%%%%%%%%%%%%%%%%%%%%%%%%%%%%%%%%%%%%%%%%%%%%%%%%%%%%%%%%
%%%%%%%%%%%%%%%%%%%%%%%%%%%%%%%%%%%%%%%%%%%%%%%%%%%%%%%%%%%%%%%%%%%%%%

\section{Introduction}

The discovery of the relation between RR $(p+1)$-form potentials in
10-dimensional type~II supergravity theories and D-branes
\cite{Polchinski:1995mt} made it possible to associate most of the fields of
the string low-energy effective field theories (supergravity theories in
general) to extended objects (\textit{branes}) of diverse kinds: fundamental,
Dirichlet, solitonic, Kaluza-Klein etc. This association has been fruitfully
used in two directions: to infer the existence of new supergravity fields from
the known existence in the String Theory of a given brane or string state and
\textit{vice versa}. Thus, the knowledge of the existence of D$p$-branes with
large values of $p$ made it necessary to learn how to deal consistently with
the magnetic duals of the RR fields that were present in the standard
formulations of the supergravity theories constructed decades before, because
in general it is impossible to dualize and rewrite the theory in terms of the
dual magnetic fields. The existence of NS-NS $(p+1)$-forms in the supergravity
theories that could also be dualized made it necessary to include solitonic
branes dual to the fundamental ones (strings, basically). It was necessary to
include all the objects and fields that could be reached from those already
known by U-duality transformations and this effort led to the discovery of new
branes and the introduction of the \textit{democratic} formulations of the
type~II supergravities \cite{Bergshoeff:2001pv} dealing simultaneously with
all the relevant electric and magnetic supergravity fields in a consistent
way.

The search for all the extended states of String Theory has motivated the
search for all the fields that can be consistently introduced in the
corresponding Supergravity Theories, a problem that has no simple answer for
the $d$-, $(d-1)$ and $(d-2)$-form fields, which are not the duals of electric
fields already present in the standard formulation, at least in any obvious
way. The branes that would couple to them can play important r\^oles in String
Theory models, which makes this search more interesting.

As mentioned before, U-duality arguments have been used to find new
supergravity fields but U-duality can only reach new fields belonging to the
same orbits as the known fields. To find other possible fields, a systematic
study of the possible consistent supersymmetry transformation rules for
$p$-forms as been carried out in the 10-dimensional maximal supergravities in
Refs.~\cite{Bergshoeff:1999bx,Bergshoeff:2001pv,Bergshoeff:2005ac,Bergshoeff:2006qw,Bergshoeff:2010mv,Greitz:2011da}
but this procedure is long and not systematic. The conjectured $E_{11}$
symmetry \cite{Julia:1997cy,West:2001as,Riccioni:2009xr} can be used to
determine the bosonic extended field content of maximal supergravity in
different dimensions\footnote{Smaller Ka\v{c}-Moody algebras can be used in
  supergravities with smaller number of supercharges such as $N=2$ theories in
  $d=4,5,6$ dimensions \cite{Kleinschmidt:2008jj}.}. Thee results have been
recently used to construct the U-duality-covariant Wess-Zumino terms of all
possible branes in all dimensions \cite{Bergshoeff:2010xc,Bergshoeff:2011zk}.
In this approach supersymmetry is not explicitly taken into account, only
through the U-duality group. 

Another possible systematic approach to this problem (that does not take
supersymmetry into account explicitly either) is provided by the
embedding-tensor formalism \footnote{For recent reviews see
  Refs.~\cite{Trigiante:2007ki,Weidner:2006rp,Samtleben:2008pe}.}.  This
formalism, introduced in
Refs.~\cite{Cordaro:1998tx,deWit:2002vt,deWit:2003hq,deWit:2005hv,deWit:2005ub}
allows the study of the most general deformations of field theories and, in
particular, of supergravity theories
\cite{deWit:2004nw,Samtleben:2005bp,Schon:2006kz,de
  Wit:2007mt,Bergshoeff:2007vb,deWit:2008ta,Bergshoeff:2008bh,Hartong:2009az,Huebscher:2010ib}. One
of the main features of this formalism is that it requires the systematic
introduction of new higher-rank potentials which are related by St\"uckelberg
gauge transformations. This structure is known as the \textit{tensor
  hierarchy} of the theory
\cite{deWit:2005hv,deWit:2005ub,deWit:2008ta,Bergshoeff:2009ph,deWit:2009zv,Hartong:2009vc}
and can be taken as the (bosonic) extended field content of the theory. In
Supergravity Theories one may need to take into account additional constraints
on the possible gaugings, but, if the gauging is allowed by supersymmetry,
then gauge invariance will require the introduction of all the fields in the
associated tensor hierarchy and, since gauge invariance is a \textit{sine qua
  non} condition for supersymmetry, the tensor hierarchy will be automatically
compatible with supersymmetry. Furthermore, if we set to zero all the
deformation parameters (gauge coupling constants, Romans-like mass parameters
\cite{Romans:1985tz} etc.) the fields that we have introduced will remain in
the undeformed theory.

This formalism, therefore, provides another systematic way of finding the
extended field content of Supergravity Theories. However, it cannot be used in
the most interesting cases, $N=1,d=11$ and $N=2A,B,d=10$ Supergravity, because
these theories cannot be gauged because they do not have 1-forms ($N=1,d=11$
and $N=2B,d=10$) or the 1-form transforms under the only (Abelian) global
symmetry ($N=2A,d=10$). Only $N=2A,d=10$ can be deformed through the
introduction of Romans' mass parameter, but the consistency of this
deformation does not seem to require the introduction of any higher-rank
potentials. The dimensional reduction to $d=9$ of these theories, though, has
3 vector fields, and their embedding-tensor formalism can be used to study all
its possible gaugings and find its extended field content. 

Some gaugings of the maximal $d=9$ supergravity have been obtained in the past
by generalized dimensional reduction \cite{Scherk:1979zr} of the
10-dimensional theories with respect to the $SL(2,\mathbb{R})$ global symmetry
of the $N=2B$ theory \cite{Lavrinenko:1997qa,Meessen:1998qm,Gheerardyn:2001jj}
or other rescaling symmetries \cite{Howe:1997qt}\footnote{An $SO(2)$-gauged
  version of the theory was directly constructed in
  Ref.~\cite{Nishino:2002zi}.}. All these possibilities were systematically
and separately studied in Ref.~\cite{Bergshoeff:2002nv}, taking into account
the dualities that relate the possible deformation parameters introduced with
the generalized dimensional reductions. However, the possible combinations of
deformations were not studied, and, as we will explain, some of the
higher-rank fields are associated to the constraints on the combinations of
deformations. Furthermore, we do not know if other deformations, with no
higher-dimensional origin (such as Romans' massive deformation of the
$N=2A,d=10$ supergravity) are possible.

Our goal in this paper will be to make a systematic study of all these
possibilities using the embedding-tensor formalism plus supersymmetry to
identify the extended-field content of the theory, finding the r\^ole played
by the possible 7-, 8- and 9-form potentials, and compare the results with the
prediction of the $E_{11}$ approach. We expect to get at least compatible
results, as in the $N=2,d=4,5,6$ cases studied in \cite{Huebscher:2010ib} and
\cite{Kleinschmidt:2008jj}.

This paper is organized as follows: in Section~\ref{sec-undeformed} we review
the undeformed maximal 9-dimensional supergravity and its global
symmetries. In Section~\ref{sec-deformation} we study the possible
deformations of the theory using the embedding-tensor formalism and checking
the closure of the local supersymmetry algebra for each electric $p$-form of
the theory. In Section~\ref{sec-summary} we summarize the results of the
previous section describing the possible deformations and the constraints they
must satisfy. We discuss the relations between those results and the possible
7- 8- and 9-form potentials of the theory and how these results compare with
those obtained in the literature using the $E_{11}$ approach.
Section~\ref{sec-conclusions} contains our conclusions.  Our conventions are
briefly discussed in Appendix~\ref{app-conventions}. The Noether currents of
the undeformed theory are given in Appendix~\ref{sec-noether}. A summary of
our results for the deformed theory (deformed field strengths, gauge
transformations and covariant derivatives, supersymmetry transformations etc.)
is contained in Appendix\ref{sec-final}.

%%%%%%%%%%%%%%%%%%%%%%%%%%%%%%%%%%%%%%%%%%%%%%%%%%%%%%%%%%%%%%%%%%%%%%
%%%%%%%%%%%%%%%%%%%%%%%%%%%%%%%%%%%%%%%%%%%%%%%%%%%%%%%%%%%%%%%%%%%%%%
%%%%%%%%%%%%%%%%%%%%%%%%%%%%%%%%%%%%%%%%%%%%%%%%%%%%%%%%%%%%%%%%%%%%%%
%%%%%%%%%%%%%%%%%%%%%%%%%%%%%%%%%%%%%%%%%%%%%%%%%%%%%%%%%%%%%%%%%%%%%%

\section{Maximal $d=9$ supergravity: the undeformed theory}
\label{sec-undeformed}

There is only one undeformed (\textit{i.e.}~ungauged, massless) maximal
(\textit{i.e.}~$N=2$, containing no dimensionful parameters in their action,
apart from the overall Newton constant) 9-dimensional supergravity
\cite{Gates:1984kr}. Both the dimensional reduction of the massless
$N=2A,d=10$ theory and that of the $N=2B,d=10$ theory on a circle give the
same undeformed $N=2,d=9$ theory, a property related to the T~duality between
type~IIA and~IIB string theories compactified on circles
\cite{Dai:1989ua,Dine:1989vu} and from which the type~II Buscher rules can be
derived \cite{Bergshoeff:1995as}.

The fundamental (\textit{electric}) fields of this theory are, 

\begin{equation}
  \left\{ e_{\mu}{}^{a},  \varphi, \tau\equiv  \chi+ie^{-\phi},
    A^{I}{}_{\mu}, B^{i}{}_{\mu\nu}, C_{\mu\nu\rho}, 
\psi_{\mu}, \tilde{\lambda}, \lambda,  \right\}\, .
\end{equation}

\noindent
where $I=0,\mathbf{i}$, with $\mathbf{i,j,k}=1,2$ and
$i,j,k=1,2$\footnote{Sometimes we need to distinguish the indices $1,2$ of the
  1-forms (and their dual 6-forms) from those of the 2-forms (and their dual
  5-forms). We will use boldface indices for the former and their associated
  gauge parameters.}.  The complex scalar $\tau$ parametrizes an
$SL(2,\mathbb{R})/U(1)$ coset that can also be described through the symmetric
$SL(2,\mathbb{R})$ matrix

\begin{equation}
\mathcal{M} 
\equiv 
e^{\phi}  
\left(
\begin{array}{cc}
|\tau|^{2}  & \chi \\
& \\
\chi & 1 \\
\end{array}
\right)\, ,
\hspace{1cm}
\mathcal{M}^{-1} 
\equiv 
e^{\phi}  
\left(
\begin{array}{cc}
1 & - \chi  \\
& \\
- \chi & |\tau|^{2}   \\
\end{array}
\right)\, .
\end{equation}

The undeformed field strengths of the electric $p$-forms are, in our
conventions\footnote{We use the shorthand notation $A^{IJ}\equiv A^{I}\wedge
  A^{J}$, $B^{ijk}\equiv B^{i}\wedge B^{j}\wedge B^{k}$
  etc.}${}^{,}$\footnote{The relation between these fields and those of
  Refs.~\cite{Meessen:1998qm} and \cite{Bergshoeff:2002nv} are given in
  Appendix~\ref{sec-relationwithotherconventions}.}

\begin{eqnarray}
F^{I}
& = & 
dA^{I}\, ,
\\
& & \nonumber \\  
\label{eq:Hiundeformed}
H^{i}
& = & 
dB^{i}
+\tfrac{1}{2}\delta^{i}{}_{\mathbf{i} }(A^{0}\wedge F^{\mathbf{i}}
+A^{\mathbf{i}}\wedge F^{0})\, ,
\\
& & \nonumber \\
G
& = & 
d[C -\tfrac{1}{6} \varepsilon_{\mathbf{ij}}A^{0\mathbf{ij}}] 
-\varepsilon_{\mathbf{i}j}
F^{\mathbf{i}}  \wedge\left( B^{j} 
  +\tfrac{1}{2}\delta^{j}{}_{\mathbf{j}}A^{0\mathbf{j}} \right)\, ,
\end{eqnarray}

\noindent
and are invariant under the undeformed gauge transformations

\begin{eqnarray}
\delta_{\Lambda} A^{I} 
& = & 
-d\Lambda^{I}\, ,
\\
& & \nonumber \\
\delta_{\Lambda} B^{i} 
& = & 
-d\Lambda^{i}
+\delta^{i}{}_{\mathbf{i}} \left[
\Lambda^{\mathbf{i}} F^{0}
+\Lambda^{0} F^{\mathbf{i}}
+\tfrac{1}{2} 
\left(A^{0} \wedge \delta_{\Lambda}A^{\mathbf{i}} 
+A^{\mathbf{i}} \wedge \delta_{\Lambda}A^{0}\right)
\right]
\, ,
\\
& & \nonumber \\
\delta_{\Lambda} [C -\tfrac{1}{6} \varepsilon_{\mathbf{ij}}A^{0\mathbf{ij}}]
& = &
-d\Lambda
-\varepsilon_{\mathbf{i}j} 
\left( 
F^{\mathbf{i}}\wedge \Lambda^{j}
+\Lambda^{\mathbf{i}}\wedge H^{j}
- \delta_{\Lambda}A^{\mathbf{i}}\wedge B^{j}
\right.
\nonumber \\
& & \nonumber \\
& & 
\left.
+\tfrac{1}{2} \delta^{j}{}_{\mathbf{j}} A^{0\mathbf{i}}\wedge \delta_{\Lambda}A^{\mathbf{j}}
\right)\, .
\end{eqnarray}

The bosonic action is, in these conventions, given by

\begin{equation}
\label{eq:undeformedaction}
\begin{array}{rcl}
S & = & 
{\displaystyle \int}
\biggl \{
-\star R +\tfrac{1}{2}d\varphi \wedge \star d\varphi   
+\tfrac{1}{2}\left[ d\phi \wedge \star d\phi +e^{2\phi} d\chi \wedge \star d\chi\right]
+\tfrac{1}{2} e^{\frac{4}{\sqrt{7}}\varphi} F^{0}\wedge \star F^{0}
\\
& & \\
& & 
+\tfrac{1}{2} e^{\frac{3}{\sqrt{7}}\varphi} (\mathcal{M}^{-1})_{\mathbf{ij}}
F^{\mathbf{i}}\wedge \star F^{\mathbf{j}}
+\tfrac{1}{2} e^{-\frac{1}{\sqrt{7}}\varphi} (\mathcal{M}^{-1})_{ij}
H^{i}\wedge \star H^{j}
+\tfrac{1}{2} e^{\frac{2}{\sqrt{7}}\varphi} G \wedge \star G
\\
& & \\
& & 
-\tfrac{1}{2}
\left[
G+\varepsilon_{\mathbf{i}j}A^{\mathbf{i}}\wedge 
\left(H^{j} -\tfrac{1}{2}\delta^{j}{}_{\mathbf{j}}A^{\mathbf{j}}\wedge F^{0}\right)
\right] \wedge
\left\{
\left[
G+\varepsilon_{\mathbf{i}j}A^{\mathbf{i}}\wedge 
\left(H^{j} -\tfrac{1}{2}\delta^{j}{}_{\mathbf{j}}A^{\mathbf{j}}\wedge F^{0}\right)
\right]\wedge A^{0}
\right.
\\
& & \\
& & 
\left.
-\varepsilon_{ij}
\left(H^{i} -\delta^{i}{}_{\mathbf{i}}A^{\mathbf{i}}\wedge F^{0}\right)
\wedge
\left(B^{j} -\tfrac{1}{2}\delta^{j}{}_{\mathbf{j}}A^{0\mathbf{j}}\right)
\right\} 
\biggr \}\, .
\end{array}
\end{equation}

The kinetic term for the $SL(2,\mathbb{R})$ scalars $\phi$ and $\chi$ can be
written in the alternative forms

\begin{equation}
\tfrac{1}{2}\left[ d\phi \wedge \star d\phi +e^{2\phi} d\chi \wedge \star d\chi\right]
=
\frac{d\tau \wedge \star d\bar{\tau}}{2(\Im {\rm m}\tau)^{2}}
= 
\tfrac{1}{4} \mathrm{Tr}\left[ d\mathcal{M}\mathcal{M}^{-1} \wedge \star 
d\mathcal{M}\mathcal{M}^{-1}\right]\, ,
\end{equation}

\noindent
the last of which is manifestly $SL(2,\mathbb{R})$-invariant. The Chern-Simons
term of the action (the last two lines of Eq.~(\ref{eq:undeformedaction})) can
also be written in the alternative form

\begin{equation}
\begin{array}{l}
-\tfrac{1}{2}
d\left[ C -\tfrac{1}{6}\varepsilon_{\mathbf{ij}}A^{0\mathbf{ij}}
-\varepsilon_{\mathbf{i}j}A^{\mathbf{i}}\wedge B^{j}
\right] \wedge
\left\{
d\left[ C -\tfrac{1}{6}\varepsilon_{\mathbf{ij}}A^{0\mathbf{ij}}
-\varepsilon_{\mathbf{i}j}A^{\mathbf{i}}\wedge B^{j}
\right]\wedge A^{0}
\right.
\\
\\
\left.
-\varepsilon_{ij}
d\left(B^{i} -\tfrac{1}{2}\delta^{i}{}_{\mathbf{i}}A^{0\mathbf{i}}\right)
\wedge
\left(B^{j} -\tfrac{1}{2}\delta^{j}{}_{\mathbf{j}}A^{0\mathbf{j}}\right)
\right\} 
\, ,
\end{array}  
\end{equation}

\noindent
that has an evident 11-dimensional origin.

The equations of motion of the scalars, derived from the action above, are

\begin{eqnarray}
d\star d \varphi
-\tfrac{2}{\sqrt{7}} e^{\frac{4}{\sqrt{7}}\varphi} F^{0}\wedge \star F^{0}
-\tfrac{3}{2\sqrt{7}} e^{\frac{3}{\sqrt{7}}\varphi} 
(\mathcal{M}^{-1})_{\mathbf{ij}}F^{\mathbf{i}}\wedge \star F^{\mathbf{j}}
& & \nonumber \\
& & \nonumber \\
+\tfrac{1}{2\sqrt{7}} e^{-\frac{1}{\sqrt{7}}\varphi} (\mathcal{M}^{-1})_{ij}
H^{i}\wedge \star H^{j}
-\tfrac{1}{\sqrt{7}} e^{\frac{2}{\sqrt{7}}\varphi} G \wedge \star G
& = & 0\, ,\\
& & \nonumber \\
d \left[\star\frac{d\bar{\tau}}{(\Im {\rm m} \tau)^{2}} \right]
-i \frac{d\tau \wedge \star d\bar{\tau}}{(\Im {\rm m}\tau)^{3}}
-\partial_{\tau}(\mathcal{M}^{-1})_{\mathbf{ij}}
\left[F^{\mathbf{i}}\wedge \star F^{\mathbf{j}}+H^{i}\wedge \star H^{j} \right]
& = & 0\, ,
\end{eqnarray}

\noindent
and those of the fundamental $p$-forms ($p\geq 1$), after some algebraic
manipulations, take the form

\begin{align}
\label{eq:eomA0}
d\left(e^{\frac{4}{\sqrt{7}}\varphi}\star F^{0} \right)
& = 
-e^{-\frac{1}{\sqrt{7}}\varphi} \mathcal{M}^{-1}_{\mathbf{i}j} 
F^{\mathbf{i}} \wedge \star H^{j} +\tfrac{1}{2}G \wedge G\, ,
\\  
& \nonumber \\
\label{eq:eomAi}
d\left(e^{\frac{3}{\sqrt{7}}\varphi}\mathcal{M}^{-1}_{\mathbf{ij}}
\star F^{\mathbf{j}} \right)
& = 
-e^{\frac{3}{\sqrt{7}}\varphi} \mathcal{M}^{-1}_{\mathbf{i}j} 
F^{0} \wedge \star H^{j}
+\varepsilon_{\mathbf{i}j} e^{\frac{2}{\sqrt{7}}\varphi}H^{j}\wedge \star G\, ,
\\  
& \nonumber \\
\label{eq:eomBi}
d\left(e^{-\frac{1}{\sqrt{7}}\varphi}\mathcal{M}^{-1}_{ij}\star H^{j} \right)
& = 
\varepsilon_{\mathbf{i}j} e^{\frac{2}{\sqrt{7}}\varphi}F^{\mathbf{j}} \wedge \star G
-\varepsilon_{ij}H^{j}\wedge G\, ,
\\
& \nonumber \\ 
\label{eq:eomC} 
d\left(e^{\frac{2}{\sqrt{7}}\varphi}\star G \right)
& = 
F^{0} \wedge G+ \tfrac{1}{2}\varepsilon_{ij}H^{i}\wedge H^{j}\, .
\end{align}

%%%%%%%%%%%%%%%%%%%%%%%%%%%%%%%%%%%%%%%%%%%%%%%%%%%%%%%%%%%%%%%%%%%%%%
%%%%%%%%%%%%%%%%%%%%%%%%%%%%%%%%%%%%%%%%%%%%%%%%%%%%%%%%%%%%%%%%%%%%%%
%%%%%%%%%%%%%%%%%%%%%%%%%%%%%%%%%%%%%%%%%%%%%%%%%%%%%%%%%%%%%%%%%%%%%%
%%%%%%%%%%%%%%%%%%%%%%%%%%%%%%%%%%%%%%%%%%%%%%%%%%%%%%%%%%%%%%%%%%%%%%

\subsection{Global symmetries}
\label{sec-global}

The undeformed theory has as (classical) global symmetry group
$SL(2,\mathbb{R})\times (\mathbb{R}^{+})^{2}$. The $(\mathbb{R}^{+})^{2}$
symmetries correspond to scalings of the fields, the first of which, that we
will denote by $\alpha$\footnote{This discussion follows closely that of
  Ref.~\cite{Bergshoeff:2002nv} in which the higher-dimensional origin of each
  symmetry is also studied. In particular, we use the same names and
  definitions for the scaling symmetries and we reproduce the table of scaling
  weights for the electric fields.}, acts on the metric and only leaves the
equations of motion invariant while the second of them, which we will denote
by $\beta$, leaves invariant both the metric and the action. The $\beta$
rescaling corresponds to the so-called \textit{trombone symmetry} which may
not survive to higher-derivative string corrections. 

One can also discuss two more scaling symmetries $\gamma$ and $\delta$, but
$\gamma$ is just a subgroup of $SL(2,\mathbb{R})$ and $\delta$ is related to
the other scaling symmetries by

\begin{equation}
\tfrac{4}{9} \alpha - \tfrac{8}{3} \beta -\gamma -\tfrac{1}{2} \delta=0\, .
\label{eq:rel}
\end{equation}

We will take $\alpha$ and $\beta$ as the independent symmetries.  The weights
of the electric fields under all the scaling symmetries are given in
Table~\ref{fundamental9d_weights}. We can see that each of the three gauge
fields $A^{I}{}_{\mu}$ has zero weight under {\it two} (linear combinations)
of these three symmetries: one is a symmetry of the action, the other is a
symmetry of the equations of motion only. The 1-form that has zero weight
under a given rescaling is precisely the one that can be used to gauge that
rescaling, but this kind of conditions are automatically taken into account by
the embedding-tensor formalism and we will not have to discuss them in
detail.

\begin{table}[ht]
\centering
\hspace{-1cm}
{\footnotesize
\begin{tabular}{||c||c|c|c|c|c|c|c|c|c|c|c|c|c|c|c||}
\hline 
\hline 
  $\mathbb{R}^+$ & $e_\mu{}^a$  & $e^\varphi$ & $e^\phi$ & $\chi$ & 
$A^0$ &  $A^{1}$ & $A^{2}$ & $B^{1}$ & $B^{2}$ & $C$ & $\psi_\mu$ & 
$\lambda$ & $\tilde \lambda$ & $\epsilon$ & $\mathcal{L}$  \\
\hline 
\hline 
$\alpha$ & $9/7$ & $6/\sqrt{7}$ & $0$ & $0$ & $3$ & $0$ & $0$ &
 $3$ & $3$ & $3$ & $9/14$ & $-9/14$ & $-9/14$ & $9/14$ & $9$ \\
\hline 
$\beta$ & $0$ & $\sqrt{7}/4$ & $3/4$ & $-3/4$ & $1/2$ & $-3/4$ & $0$ & $-1/4$ & 
$1/2$ & $-1/4$ & $0$ & $0$ & $0$ & $0$ &  $0$  \\
\hline 
$\gamma$ & $0$ & $0$ & $-2$ & $2$ & $0$ & $1$ & $-1$ & $1$ & $-1$ & $0$ & 
$0$ & $0$ & $0$ & $0$ &  $0$   \\
\hline 
$\delta$ & $8/7$ & $-4/\sqrt{7}$ & $0$ & $0$ & $0$ & $2$ & 
$2$ & $2$ & $2$ & $4$ & $4/7$ & $-4/7$ & $-4/7$ & $4/7$ &  $8$   \\
\hline
\hline 
\end{tabular}
}
\caption{\it The scaling weights of the electric fields of
  maximal $d=9$ supergravity.}
\label{fundamental9d_weights}
\end{table}

The action of the element of $SL(2,\mathbb{R})$ given by the matrix

\begin{equation}
\left(\Omega_{j}^{i}\right)  
=
\left( \begin{array}{cc} a&b\\ c&d \end{array} \right)\, ,
\hspace{1cm}
ad-bc=1\, ,
\end{equation}

\noindent
on the fields of the theory is

\begin{equation}
\label{SL2R9D}
\begin{array}{rclrcl}
\tau^{\prime} 
& = & 
{\displaystyle\frac{a {\tau} +b}{c{\tau} +d}}\, , 
\hspace{1.5cm} &
\mathcal{M}^{\prime}_{ij}
& = & 
\Omega_{i}{}^{k} \mathcal{M}_{kl}\Omega_{j}{}^{l}\, ,\\
& & & & &  \\
A^{\mathbf{i}\, \prime} 
& =  &  
\Omega_{\mathbf{j}}{}^{\mathbf{i}} A^{\mathbf{j}}\, , &
B^{i\, \prime} 
& =  &  
\Omega_{j}{}^{i} B^{j}\, ,\\
& & & & &  \\
\psi^{\prime}_{\mu} 
& = &
e^{\frac{i}{2}l}\psi_{{\mu}}\, ,&
{\lambda} 
& = & 
e^{\frac{3i}{2}l} \lambda\, , \\
& & & & &  \\
\tilde{\lambda}^{\prime} 
& = & 
e^{-\frac{i}{2}l} \tilde{\lambda}\, ,&
\epsilon^{\prime} 
& = & 
e^{\frac{i}{2}l} \epsilon\, .\\
\end{array}
\end{equation}

\noindent
where 

\begin{equation}
e^{2il} \equiv \frac{c \, {\tau}^*+d}{c\, {\tau}+d}\, .
\end{equation}

\noindent
The rest of the fields ($e^{a}{}_{\mu},\varphi,A^{0}{}_{\mu},C_{\mu\nu\rho}$),
are invariant under $SL(2,\mathbb{R})$.

We are going to label the 5 generators of these global symmetries by
$T_{A}$, $A=1,\cdots,5$. $\{T_{1},T_{2},T_{3}\}$ will be the 3
generators of $SL(2,\mathbb{R})$ (collectively denoted by $\{T_{m}\}$,
$m=1,2,3$), and $T_{4}$ and $T_{5}$ will be, respectively, the
generators of the rescalings $\alpha$ and $\beta$.  Our choice for the
generators of $SL(2,\mathbb{R})$ acting on the doublets of 1-forms
$A^{\mathbf{i}}$ and 2-forms $B^{i}$ is

\begin{equation}
T_{1} = \tfrac{1}{2}\sigma^{3}\, ,
\hspace{1cm}
T_{2} = \tfrac{1}{2}\sigma^{1}\, ,
\hspace{1cm}
T_{3} = \tfrac{i}{2}\sigma^{2}\, ,
\end{equation}

\noindent
where the $\sigma^{m}$ are the standard Pauli matrices, so 

\begin{equation}
[T_{1},T_{2}] = T_{3}\, ,
\hspace{1cm}
[T_{2},T_{3}] = -T_{1}\, ,
\hspace{1cm}
[T_{3},T_{1}] = -T_{2}\, .
\end{equation}

\noindent
Then, the $3\times 3$ matrices corresponding to generators acting
(contravariantly) on the 3 1-forms $A^{I}$ (and covariantly on their dual
6-forms $\tilde{A}_{I}$ to be introduced later) are

\begin{equation}
  \begin{array}{ccc}
\left((T_{1})_{J}{}^{I}\right)
=
\tfrac{1}{2}
\left(
  \begin{array}{c|c}
 0  & 0 \\  \hline 0  & \sigma^{3}  \\ 
  \end{array}
\right)\, ,
\hspace{.5cm}
&
\left((T_{2})_{J}{}^{I}\right)
=
\tfrac{1}{2}
\left(
  \begin{array}{c|c}
 0  & 0 \\  \hline 0  & \sigma^{1}  \\ 
  \end{array}
\right)\, ,
\hspace{.5cm}
&
\left((T_{3})_{J}{}^{I}\right)
=
\tfrac{1}{2}
\left(
  \begin{array}{c|c}
 0  & 0 \\  \hline 0  & i\sigma^{2}  \\ 
  \end{array}
\right)\, ,
\\
& & \\
\left((T_{4})_{J}{}^{I}\right)
=
\mathrm{diag}(3,0,0)\, ,
\hspace{.5cm}
&
\left((T_{5})_{J}{}^{I}\right)
=
\mathrm{diag}(1/2,-3/4,0)\, .
&
\\
\end{array}
\end{equation}

\noindent
We will sometimes denote this representation by $T^{(3)}_{A}$.  The $2\times
2$ matrices corresponding to generators acting (contravariantly) on the
doublet of 2-forms $B^{i}$ (and covariantly on their dual 5-forms
$\tilde{B}_{i}$ to be introduced later) are

\begin{equation}
  \begin{array}{ccc}
\left((T_{1})_{j}{}^{i}\right)
=
\tfrac{1}{2}\sigma^{3}\, ,
\hspace{.5cm}
&
\left((T_{2})_{j}{}^{i}\right)
=
\tfrac{1}{2}\sigma^{1}\, ,
\hspace{.5cm}
&
\left((T_{3})_{j}{}^{i}\right)
=
\tfrac{i}{2}\sigma^{2}\, ,
\\
& & \\
\left((T_{4})_{j}{}^{i}\right)
=
\mathrm{diag}(3,3)\, ,
\hspace{.5cm}
&
\left((T_{5})_{j}{}^{i}\right)
=
\mathrm{diag}(-1/4,1/2)\, .
&
\\
\end{array}
\end{equation}

\noindent
We will denote this representation by $T^{(2)}_{A}$.  The generators
that act on the 3-form $C$ (sometimes denoted by $T^{(1)}_{A}$) are

\begin{equation}
T_{1}= T_{2}= T_{3} = 0\, ,  
\hspace{.5cm}
T_{4}= 3\, ,  
\hspace{.5cm}
T_{5}= -1/4\, .  
\end{equation}

\noindent
We will also need the generators that act on the magnetic 4-form $\tilde{C}$
(see next section), also denoted by $T^{(\tilde{1})}_{A}$

\begin{equation}
\tilde{T}_{1}= \tilde{T}_{2}= \tilde{T}_{3} = 0\, ,  
\hspace{.5cm}
\tilde{T}_{4}= 6\, ,  
\hspace{.5cm}
\tilde{T}_{5}= 1/4\, .  
\end{equation}

We define the structure constants $f_{AB}{}^{C}$ by

\begin{equation}
[T_{A},T_{B}] = f_{AB}{}^{C} T_{C}\, .  
\end{equation}

The symmetries of the theory are isometries of the scalar manifold
($\mathbb{R}\times SL(2,\mathbb{R}/U(1)$). The Killing vector associated to
the generator $T_{A}$ will be denoted by $k_{A}$ and will be normalized so
that their Lie brackets are given by

\begin{equation}
[k_{A},k_{B}] = -f_{AB}{}^{C} k_{C}\, .  
\end{equation}

The $SL(2,\mathbb{R})/U(1)$ factor of the scalar manifold is a K\"ahler space
with K\"ahler potential, K\"ahler metric and K\"ahler 1-form, respectively
given by

\begin{equation}
\mathcal{K}= -\log \Im{\rm m}\tau =\phi\, ,  
\hspace{.5cm}
\mathcal{G}_{\tau\tau^{*}}
=
\partial_{\tau}  
\partial_{\tau^{*}}\mathcal{K} = \tfrac{1}{4}e^{2\phi}\, ,  
\hspace{.5cm}
\mathcal{Q} 
= \tfrac{1}{2i}\left(\partial_{\tau}\mathcal{K}d\tau - \mathrm{c.c.}\right)
= \tfrac{1}{2}e^{\phi}d\chi\, .  
\end{equation}

In general, the isometries of the K\"ahler metric only leave invariant the
K\"ahler potential up to K\"ahler transformations :

\begin{equation}
\pounds_{k_{m}}\mathcal{K} 
= 
k_{m}{}^{\tau}\partial_{\tau}\mathcal{K}  +\mathrm{c.c.}
=
\lambda_{m}(\tau) +\mathrm{c.c.}\, ,
\hspace{1cm}
\pounds_{k_{m}}\mathcal{Q} = -\tfrac{i}{2}d\lambda_{m}\, , 
\end{equation}

\noindent
where the $\lambda_{m}$ are holomorphic functions of the coordinates that
satisfy the equivariance property

\begin{equation}
\pounds_{k_{m}}\lambda_{n} 
-  
\pounds_{k_{n}}\lambda_{m}
=
-f_{mn}{}^{p}\lambda_{p}\, . 
\end{equation}

Then, for each of the $SL(2,\mathbb{R})$ Killing vectors $k_{m}$, $m=1,2,3$,
it is possible to find a real \textit{Killing prepotential} or \textit{momentum
  map} $\mathcal{P}_{m}$ such that 

\begin{equation}
\label{eq:momentummapsdef}
  \begin{array}{rcl}
k_{m\, \tau^{*}} 
& = & 
\mathcal{G}_{\tau^{*}\tau}k_{m}{}^{\tau} 
=i\partial_{\tau^{*}}\mathcal{P}_{m}\, ,
\\
& & \\
k_{m}{}^{\tau} \partial_{\tau}\mathcal{K} 
& = & 
i\mathcal{P}_{m}+\lambda_{m}\, , 
\\
& & \\
\pounds_{k_{m}}\mathcal{P}_{n}
& = & 
-f_{mn}{}^{p}\mathcal{P}_{p}\, .
\end{array}
\end{equation}

The non-vanishing components of all the Killing vectors are\footnote{The
  holomorphic and anti-holomorphic components are defined by
  $k=k^{\tau}\partial_{\tau} +\mathrm{c.c.}= k^{\chi}\partial_{\chi}+
  k^{\phi}\partial_{\phi}$.}

\begin{equation}
\label{eq:taukillingvectors}
k_{1}{}^{\tau}=\tau\, ,
\hspace{.7cm}  
k_{2}{}^{\tau}=\tfrac{1}{2}(1-\tau^{2})\, ,
\hspace{.7cm}  
k_{3}{}^{\tau}=\tfrac{1}{2}(1+\tau^{2})\, ,
\hspace{.7cm}  
k_{4}{}^{\tau}=0\, ,
\hspace{.7cm}  
k_{5}{}^{\tau}=-\tfrac{3}{4}\tau\, .
\end{equation}

\noindent
and

\begin{equation}
\label{eq:varphikillingvectors}
k_{4}{}^{\varphi}=6/\sqrt{7}\, ,
\hspace{1.5cm}  
k_{5}{}^{\varphi}=\sqrt{7}/4\, .
\end{equation}

\noindent
The holomorphic functions $\lambda_{m}(\tau)$ take the values

\begin{equation}
\label{eq:lambdam}
\lambda_{1}=-\tfrac{1}{2}\, ,
\hspace{1cm}
\lambda_{2}=\tfrac{1}{2}\tau\, ,
\hspace{1cm}
\lambda_{3}=-\tfrac{1}{2}\tau\, ,
\end{equation}

\noindent
and the momentum maps are given by:

\begin{equation}
\label{eq:momentummaps}
\mathcal{P}_{1}= \tfrac{1}{2}e^{\phi}\chi\, ,
\hspace{1cm}
\mathcal{P}_{2}= \tfrac{1}{4}e^{\phi}(1-|\tau|^{2})\, ,
\hspace{1cm}
\mathcal{P}_{3}= \tfrac{1}{4}e^{\phi}(1+|\tau|^{2})\, .
\end{equation}

These objects will be used in the construction of $SL(2,\mathbb{R})$-covariant
derivatives for the fermions.

%%%%%%%%%%%%%%%%%%%%%%%%%%%%%%%%%%%%%%%%%%%%%%%%%%%%%%%%%%%%%%%%%%%%%%
%%%%%%%%%%%%%%%%%%%%%%%%%%%%%%%%%%%%%%%%%%%%%%%%%%%%%%%%%%%%%%%%%%%%%%
%%%%%%%%%%%%%%%%%%%%%%%%%%%%%%%%%%%%%%%%%%%%%%%%%%%%%%%%%%%%%%%%%%%%%%
%%%%%%%%%%%%%%%%%%%%%%%%%%%%%%%%%%%%%%%%%%%%%%%%%%%%%%%%%%%%%%%%%%%%%%

\subsection{Magnetic fields}
\label{sec-magnetic}

As it is well known, for each $p$-form potential with $p>0$ one can define a
\textit{magnetic} dual which in $d-9$ dimensions will be a $(7-p)$-form
potential. Then, we will have magnetic 4-, 5- and 6-form potentials in the
theory. 

A possible way to define those potentials and identify their $(8-p)$-form
field strengths consists in writing the equations of motion of the $p$-forms
as total derivatives. Let us take, for instance,  the equation of motion of
the 3-form $C$ Eq.~(\ref{eq:eomC}). It can be written as 

\begin{equation}
\label{eq:eomC2}
  \begin{array}{rcl}
d{\displaystyle\frac{\partial \mathcal{L}}{\partial  G}}   
= 
d \biggl \{
e^{\frac{2}{\sqrt{7}}\varphi}  \star G
-
\left[
G+\varepsilon_{\mathbf{i}j}A^{\mathbf{i}}\wedge 
\left(H^{j} -\tfrac{1}{2}\delta^{j}{}_{\mathbf{j}}A^{\mathbf{j}}\wedge F^{0}\right)
\right]\wedge A^{0}
& & \\
& & \\
+\tfrac{1}{2}\varepsilon_{ij}
\left(H^{i} -\delta^{i}{}_{\mathbf{i}}A^{\mathbf{i}}\wedge F^{0}\right)
\wedge
\left(B^{j} -\tfrac{1}{2}\delta^{j}{}_{\mathbf{j}}A^{0\mathbf{j}}\right)
\biggr \}
& = & 0\, .  
\end{array}
\end{equation}

\noindent 
We can transform this equation of motion into a Bianchi identity by replacing
the combination of fields on which the total derivative acts by the total
derivative of a 4-form which we choose for the sake of
convenience\footnote{With this definition $\tilde{G}$ will have exactly the
  same form that we will obtain from the embedding tensor formalism.}

\begin{equation}
  \begin{array}{rcl}
d\left[\tilde{C} -C\wedge A^{0} -\tfrac{3}{4}
\varepsilon_{\mathbf{i}j}A^{0\mathbf{i}}\wedge B^{j} \right]
& \equiv &
e^{\frac{2}{\sqrt{7}}\varphi}  \star G
-
\left[
G+\varepsilon_{\mathbf{i}j}A^{\mathbf{i}}\wedge 
\left(H^{j} -\tfrac{1}{2}\delta^{j}{}_{\mathbf{j}}A^{\mathbf{j}}\wedge F^{0}\right)
\right]\wedge A^{0}
\\
& & \\
& & 
+\tfrac{1}{2}\varepsilon_{ij}
\left(H^{i} -\delta^{i}{}_{\mathbf{i}}A^{\mathbf{i}}\wedge F^{0}\right)
\wedge
\left(B^{j} -\tfrac{1}{2}\delta^{j}{}_{\mathbf{j}}A^{0\mathbf{j}}\right)\, ,
\end{array}
\end{equation}

\noindent
where $\tilde{C}$ will be the magnetic 4-form. This relation can be put in the
form of a duality relation

\begin{equation}
e^{\frac{2}{\sqrt{7}}\varphi}  \star G
=
\tilde{G}\, ,
\end{equation}

\noindent
where we have defined the magnetic 5-form field strength

\begin{equation}
\label{eq:tildeGundeformed}
\tilde{G}
\equiv
d\tilde{C}
+C \wedge F^{0}   
-\tfrac{1}{24}\varepsilon_{\mathbf{ij}} A^{0\mathbf{ij}} \wedge F^{0} 
- \varepsilon_{ij} \left(H^{i} -\tfrac{1}{2} dB^{i}\right) \wedge B^{j}\, .
\end{equation}

\noindent
The equation of motion for $\tilde{C}$ is just the Bianchi identity of $G$
rewritten in terms of $\tilde{G}$.

% The equations of motion of the 2-form doublet $B^{i}$ can also be rewritten as
% a total derivative if we use the magnetic 4-form $\tilde{C}$: varying the
% action with respect to $B^{i}$ we get

% \begin{equation}
% \frac{\delta S}{\delta B^{i}} 
% = 
% -d \frac{\partial S}{\partial H^{i}}   
% +\varepsilon_{i\mathbf{j}}dA^{\mathbf{j}} \wedge 
% \frac{\partial \mathcal{L}}{\partial  G}
% +\tfrac{1}{2}\varepsilon_{ij}
% d\left[ C -\tfrac{1}{6}\varepsilon_{\mathbf{kl}}A^{0\mathbf{kl}}
% -\varepsilon_{\mathbf{k}l}A^{\mathbf{k}}\wedge B^{l}
% \right]
% \wedge 
% d\left(B^{j} -\tfrac{1}{2}\delta^{j}{}_{\mathbf{j}}A^{0\mathbf{j}}\right)\, ,
% \end{equation}

% \noindent
% and using our previous definition

% \begin{equation}
% \frac{\partial \mathcal{L}}{\partial  G} =   
% d\left[\tilde{C} -C\wedge A^{0} -\tfrac{3}{4}
% \varepsilon_{\mathbf{k}l}A^{0\mathbf{k}}\wedge B^{} \right]\, ,
% \end{equation}

% \noindent
% we see that $\delta S/ \delta B^{i}$ is the total derivative of doublet of
% 6-forms that we can rewrite as Bianchi identities if we define a doublet of
% magnetic 5-form potentials $\tilde{B}_{i}$. The definition is not unique and
% we will not attempt to make a definite choice here.

% A similar story is valid for the 1-forms' equations of motion, if we make use
% of some definition of the magnetic 5-form doublet $\tilde{B}_{i}$. Thus, we
% can define three magnetic 6-forms $\tilde{A}_{I}$ in this theory.

In a similar fashion we can define a doublet of 5-forms $\tilde{B}_{i}$ with
field strengths denoted by $\tilde{H}_{i}$, and a singlet and a doublet of
6-forms $\tilde{A}_{0},\tilde{A}_{\mathbf{i}}$ with field strengths denoted,
respectively, by $\tilde{F}_{0}$ and $\tilde{F}_{\mathbf{i}}$. The field
strengths can be chosen to have the form

\begin{align}
\tilde{H}_{i} 
& = 
d\tilde{B}_{i} -\delta_{ij} B^{j}\wedge G
+\delta_{i\mathbf{j}}\tilde{C} \wedge F^{\mathbf{j}}
+\tfrac{1}{2}\delta_{i\mathbf{j}} \left(A^{0} \wedge F^{\mathbf{j}}
+A^{\mathbf{j}} \wedge F^{0}\right)\wedge C
\nonumber \\
& \nonumber \\
& 
+\tfrac{1}{2}\delta_{ij}\varepsilon_{k\mathbf{l}}B^{jk} \wedge F^{\mathbf{l}}
\, ,
\\
& \nonumber \\
\tilde{F}_{0} 
& =  
d\tilde{A}_{0} +\tfrac{1}{2} C\wedge G
-\varepsilon_{\mathbf{i}j}F^{\mathbf{i}} \wedge 
\left(\delta^{jk}\tilde{B}_{k} -\tfrac{2}{3}B^{j}\wedge C \right)
\nonumber \\
& \nonumber \\
& 
-\tfrac{1}{18} \varepsilon_{\mathbf{ij}} A^{\mathbf{ij}} \wedge 
\left(
\tilde{G} -F^{0} \wedge C -\tfrac{1}{2} \varepsilon_{kl}B^{k} \wedge H^{l}
\right)
\nonumber \\
& \nonumber \\
& 
-\tfrac{1}{6}\varepsilon_{\mathbf{i}j}
A^{\mathbf{i}}\wedge \left(
B^{j} \wedge G - C \wedge H^{j} -\tfrac{2}{3} \delta^{j}{}_{\mathbf{j}}
\tilde{C} \wedge F^{\mathbf{j}}
-\varepsilon_{k\mathbf{l}}B^{jk} \wedge F^{\mathbf{l}}
\right)\, ,
\\
& \nonumber \\
\tilde{F}_{\mathbf{i}}  
& = 
d\tilde{A}_{\mathbf{i}} 
+\delta_{\mathbf{i}j} \left(
B^{j} +\tfrac{7}{18} \delta^{j}{}_{\mathbf{k}}A^{0\mathbf{k}} 
\right) 
\wedge \tilde{G}
-\delta_{\mathbf{i}}{}^{j}F^{0} \wedge \tilde{B}_{j}
-\tfrac{1}{9}
\delta_{\mathbf{ij}}\left(
8A^{0} \wedge F^{\mathbf{j}} +  A^{\mathbf{j}}\wedge F^{0}
\right) 
\wedge \tilde{C}
\nonumber \\
& \nonumber \\
& 
-\tfrac{1}{3}\delta_{\mathbf{i}j} \varepsilon_{lm}
 \left(
B^{j} +\tfrac{1}{3} \delta^{j}{}_{\mathbf{k}}A^{0\mathbf{k}} 
\right) 
\wedge B^{l} \wedge H^{m}
-\tfrac{1}{6}\delta_{\mathbf{i}j} \varepsilon_{\mathbf{k}l}
\left(
A^{0} \wedge H^{j} -B^{j} \wedge F^{0}
\right)
\wedge A^{\mathbf{k}} \wedge B^{l}
\nonumber \\
& \nonumber \\
& 
-\tfrac{1}{9} A^{0} \wedge F^{0} \wedge
\delta_{\mathbf{ij}}\left(
\tfrac{7}{2} A^{\mathbf{j}} \wedge C
+\delta^{\mathbf{j}}{}_{k}\varepsilon_{\mathbf{lm}}
A^{\mathbf{lm}} \wedge B^{k}
\right)\, ,
\end{align}

\noindent
and the duality relations are

\begin{align}
 \tilde{H}_{i}
& =
e^{-\frac{1}{\sqrt{7}}\varphi}\mathcal{M}^{-1}_{ij}\star H^{j}\, ,
\\
& \nonumber \\
\tilde{F}_{0}
& = 
e^{\frac{4}{\sqrt{7}}\varphi}\star F^{0}\, ,
\\
& \nonumber \\
\tilde{F}_{\mathbf{i}}
& = 
e^{\frac{3}{\sqrt{7}}\varphi}\mathcal{M}^{-1}_{\mathbf{ij}}
\star F^{\mathbf{j}}\, .
\end{align}

The situation is summarized in Table~\ref{xxx1}. The scaling weights of the
magnetic fields are given in Table~\ref{dual9d_weights}.

\begin{table}[h]
\begin{center}
\begin{tabular}{||c|c|c|c|c|c|c|c|c|c||}
\hline
0 & 1 & 2 & 3 & 4 & 5 & 6 & 7 & 8 & 9\\ \hline \hline
$j_{A}$ &  $ A^{I}$ & $B^{i}$ & $C$ & $\tilde{C}$ & $\tilde{B}_{i}$ & $\tilde{A}_{I}$ 
& $\tilde{A}_{(7)}^{A}$ & $\tilde{A}_{(8)}$ & $\tilde{A}_{(9)}$ \\ 
\hline
 & $ F^{I}$ & $H^{i}$ & $G$ & $\tilde{G}$ & $\tilde{H}_{i}$ & $\tilde{F}_{I}$ 
& $\tilde{F}_{(8)}^{A}$ & $\tilde{F}_{(9)}$ & \\ 
\hline \hline
\end{tabular}
\caption{\it Electric and magnetic forms and their field strengths.}
\label{xxx1}
\end{center}
\end{table}

\begin{table}[ht]
\centering
\begin{tabular}{||c||c|c|c|c|c|c||}
\hline 
\hline 
  $\mathbb{R}^{+}$ &  $\tilde{C}$ &  
$\tilde{B}_{2}$ & $\tilde{B}_{1}$ & $\tilde{A}_{2}$ & $\tilde{A}_{1}$ & 
$\tilde{A}_{0}$  \\
 \hline
\hline 
$\alpha$ & 6 &6 &6 &9 &9 &6  \\ 
\hline 
$\beta$ & $1/4$ & $-1/2$&$+1/4$&0&$+3/4$ &$-1/2$  \\ 
\hline 
$\gamma$ & 0&1 &-1 &1 &-1 &0      \\  
\hline 
$\delta$ & 4&6 &6 &6 &6 &8        \\ 
\hline
\hline 
\end{tabular}
\caption{\it The scaling weights of the  magnetic fields of
  maximal $d=9$ supergravity can be determined by requiring that
  the sum of the weights of the electric and magnetic potentials 
  equals that of the Lagrangian. The scaling weights of the 7-, 8-
  and 9-forms can be determined in the same way after we find the entities
  they are dual to (Noether currents, embedding-tensor components and
  constraints, see Section~\ref{sec-summary}).}
\label{dual9d_weights}
\end{table}

This dualization procedure is made possible by the gauge symmetries associated
to all the $p$-form potentials for $p>0$ (actually, by the existence of gauge
transformations with constant parameters) and, therefore, it always works for
massless $p$-forms with $p>0$ and generically fails for 0-form
fields. However, in maximal supergravity theories at least, there is a global
symmetry group that acts on the scalar manifold and whose dimension is larger
than that of the scalar manifold. Therefore, there is one Noether 1-form
current $j_{A}$ associated to each of the generators of the global symmetries
of the theory $T_{A}$. These currents are conserved on-shell,
\textit{i.~e.}~they satisfy

\begin{displaymath}
d\star j_{A}=0\, ,  
\end{displaymath}

\noindent
on-shell, and we can define a $(d-2)$-form potential $\tilde{A}_{(d-2)}^{A}$ by

\begin{displaymath}
d\tilde{A}_{(d-2)}^{A}= G^{AB}\star j_{B}\, ,  
\end{displaymath}

\noindent
where $G^{AB}$ is the inverse Killing metric of the global symmetry group, so
that the conservation law (dynamical) becomes a Bianchi identity.

Thus, while the dualization procedure indicates that for each electric
$p$-form with $p>0$ there is a dual magnetic $(7-p)$-form transforming in the
conjugate representation, it tells us that there are as many magnetic
$(d-2)$-form duals of the scalars as the dimension of the global group (and
not of as the dimension of the scalar manifold) and that they transform in the
co-adjoint representation. Actually, since there is no need to have scalar
fields in order to have global symmetries, it is possible to define magnetic
$(d-2)$-form potentials even in the total absence of scalars\footnote{See
  Refs.~\cite{Hartong:2009az,Huebscher:2010ib} for examples.}.

According to these general arguments, which are in agreement with the general
results of the embedding-tensor formalism
\cite{Bergshoeff:2009ph,Hartong:2009vc,Hartong:2009az,Huebscher:2010ib}, we
expect a triplet of 7-form potentials $\tilde{A}_{(7)}^{m}$ associated to the
$SL(2,\mathbb{R})$ factor of the global symmetry group \cite{Meessen:1998qm}
and two singlets $\tilde{A}_{(7)}^{4},\tilde{A}_{(7)}^{5}$ associated to the
rescalings $\alpha,\beta$ (see Table~\ref{xxx1}).

Finding or just determining the possible magnetic $(d-1)$- and $d$-form
potentials in a given theory is more complicated. In the embedding-tensor
formalism it is natural to expect as many $(d-1)$-form potentials as
deformation parameters (embedding-tensor components, mass parameters etc.) can
be introduced in the theory since the r\^ole of the $(d-1)$-forms in the
action is that of being Lagrange multipliers enforcing their
constancy\footnote{The embedding-tensor formalism gives us a reason to
  introduce the $(d-1)$-form potentials based on the deformation parameters
  but the $(d-1)$-form potentials do not disappear when the deformation
  parameters are set equal to zero.}. The number of deformation parameters
that can be introduced in this theory is, as we are going to see, very large,
but there are many constraints that they have to satisfy to preserve gauge and
supersymmetry invariance. Furthermore, there are many St\"uckelberg shift
symmetries acting on the possible $(d-1)$-form potentials.  Solving the
constraints leaves us with the independent deformation parameters that we can
denote by $m_{\sharp}$ and, correspondingly, with a reduced number of
$(d-1)$-form potentials $\tilde{A}_{(d-1)}^{\sharp}$ on which only a few St\"uckelberg
symmetries (or none at all) act\footnote{The $(d-1)$-form potentials that
  ``disappear'' when we solve the constraints are evidently associated to the
  gauge-fixing of the missing St\"uckelberg symmetries.}.

The $d$-form field strengths $\tilde{F}_{(d)}^{\sharp}$ are related to the
scalar potential of the theory through the expression
\cite{Bergshoeff:2009ph,Hartong:2009vc,Hartong:2009az,Huebscher:2010ib}

\begin{equation}
  \tilde{F}_{(d)}^{\sharp}
  =
  \tfrac{1}{2}\star\frac{\partial V}{\partial m_{\sharp}}\, . 
\end{equation}

Thus, in order to find the possible 8-form potentials of this theory we need
to study its independent consistent deformations $m_{\sharp}$. We will
consider this problem in the next section.

In the embedding-tensor formalism, the $d$-form potentials are associated to
constraints of the deformation parameters since they would be the Lagrange
multipliers enforcing them in the action \cite{Bergshoeff:2007vb}. If we do
not solve any of the constraints there will be many $d$-form potentials but
there will be many St\"uckelberg symmetries acting on them as well. Thus, only
a small number of \textit{irreducible} constraints that cannot be
solved\footnote{In general, the quadratic constraints cannot be used to solve
  some deformation parameters in terms of the rest. For instance, in this
  sense, if $a$ and $b$ are two of them, a constraint of the form $ab=0$
  cannot be solved and we can call it \textit{irreducible}.} and of associated
$d$-forms may be expected in the end, but we have to go through the whole
procedure to identify them. This identification will be one of the main
results of the following section.

However, this is not the end of the story for the possible 9-forms. As it was
shown in Ref.~\cite{Huebscher:2010ib} in 4- 5- and 6-dimensional cases, in the
ungauged case one can find more $d$-forms with consistent supersymmetric
transformation rules than predicted by the embedding-tensor formalism. Those
additional fields are predicted by the Ka\v{c}-Moody approach
\cite{Kleinschmidt:2008jj}. However, after gauging, the new fields do not have
consistent, independent, supersymmetry transformation rules to all orders in
fermions\footnote{The insufficience of first-order in fermions checks was
  first noticed in Ref.~\cite{Bergshoeff:2010mv}.}, and have to be combined
with other $d$-forms, so that, in the end, only the number of $d$-forms
predicted by the embedding-tensor formalism survive. 

This means that the results obtained via the embedding-tensor formalism for
the 9-forms have to be interpreted with special care and have to be compared
with the results obtained with other approaches.

The closure of the local supersymmetry algebra needs to be checked on all the
fields in the tensor hierarchy predicted by the embedding-tensor formalism
and, in particular, on the 9-forms to all orders in fermions.  However, given
that gauge invariance is requirement for local supersymmetry invariance, we
expect consistency in essentially all cases with the possible exception of the
9-forms, according to the above discussion. In the next section we will do
this for the electric fields of the theory.

%%%%%%%%%%%%%%%%%%%%%%%%%%%%%%%%%%%%%%%%%%%%%%%%%%%%%%%%%%%%%%%%%%%%%%
%%%%%%%%%%%%%%%%%%%%%%%%%%%%%%%%%%%%%%%%%%%%%%%%%%%%%%%%%%%%%%%%%%%%%%
%%%%%%%%%%%%%%%%%%%%%%%%%%%%%%%%%%%%%%%%%%%%%%%%%%%%%%%%%%%%%%%%%%%%%%
%%%%%%%%%%%%%%%%%%%%%%%%%%%%%%%%%%%%%%%%%%%%%%%%%%%%%%%%%%%%%%%%%%%%%%

\section{Deforming the maximal $d=9$ supergravity}
\label{sec-deformation}

In this section we are going to study the possible deformations of $d=9$
supergravity, starting from its possible gaugings using the embedding-tensor
formalism and constructing the corresponding tensor hierarchy
\cite{Cordaro:1998tx,deWit:2002vt,deWit:2003hq,deWit:2005hv,deWit:2005ub,Bergshoeff:2009ph,Hartong:2009vc}
up to the 4-form potentials.

If we denote by $\Lambda^{I}(x)$ the scalar parameters of the gauge
transformations of the 1-forms $A^{I}$ and by $\alpha^{A}$ the constant
parameters of the global symmetries, we want to promote 

\begin{equation}
\label{eq:gauging}
\alpha^{A} \longrightarrow \Lambda^{I}(x)\vartheta_{I}{}^{A}\, ,
\end{equation}

\noindent
where $\vartheta_{I}{}^{A}$ is the \textit{embedding tensor}, in the
transformation rules of all the fields, and we are going to require the theory
to be covariant under the new local transformations using the 1-forms as gauge
fields. 

To achieve this goal, starting with the transformations of the
scalars, the successive introduction of higher-rank $p$-form potentials is
required, which results in the construction of a tensor hierarchy. Most of
these fields are already present in the supergravity theory or can be
identified with their magnetic duals but this procedure allows us to introduce
consistently the highest-rank fields (the $d$-, $(d-1)$- and $(d-2)$-form
potentials), which are not dual to any of the original electric
fields. Actually, as explained in Section~\ref{sec-magnetic}, the highest-rank
potentials are related to the symmetries (Noether currents), the independent
deformation parameters and the constraints that they satisfy, but we need to
determine these, which requires going through this procedure checking the
consistency with gauge and supersymmetry invariance at each step.

Thus, we are going to require invariance under the new gauge transformations
for the scalar fields and we are going to find that we need new couplings to
the gauge 1-form fields (as usual). Then we will study the modifications of
the supersymmetry transformation rules of the scalars and fermion fields which
are needed to ensure the closure of the local supersymmetry algebra on the
scalars. Usually we do not expect modifications in the bosons' supersymmetry
transformations, but the fermions' transformations need to be modified by
replacing derivatives and field strengths by covariant derivatives and
covariant field strengths and, furthermore, by adding \textit{fermion
  shifts}. The local supersymmetry algebra will close provided that we impose
certain constraints on the embedding tensor components and on the fermion
shifts. 

Repeating this procedure on the 1-forms (which requires the coupling to the
2-forms) etc.~we will find a set of constraints that we can solve, determining
the independent components of the deformation tensors\footnote{As we are going
  to see, besides the embedding tensor, one can introduce many other
  deformation tensors.} and the fermions shifts. Some constraints (typically
quadratic in deformation parameters) have to be left unsolved and we will have
to take them into account towards the end of this procedure.

As a result we will identify the independent deformations of the theory and
the constraints that they satisfy. From this we will be able to extract
information about the highest-rank potentials in the tensor hierarchy.

%%%%%%%%%%%%%%%%%%%%%%%%%%%%%%%%%%%%%%%%%%%%%%%%%%%%%%%%%%%%%%%%%%%%%%
%%%%%%%%%%%%%%%%%%%%%%%%%%%%%%%%%%%%%%%%%%%%%%%%%%%%%%%%%%%%%%%%%%%%%%
%%%%%%%%%%%%%%%%%%%%%%%%%%%%%%%%%%%%%%%%%%%%%%%%%%%%%%%%%%%%%%%%%%%%%%
%%%%%%%%%%%%%%%%%%%%%%%%%%%%%%%%%%%%%%%%%%%%%%%%%%%%%%%%%%%%%%%%%%%%%%

\subsection{The 0-forms $\varphi, \tau$}

Under the global symmetry group, the scalars transform according to

\begin{equation}
\delta_{\alpha}\varphi =\alpha^{A}k_{A}{}^{\varphi}\, ,
\hspace{1cm}
\delta_{\alpha}\tau =\alpha^{A}k_{A}{}^{\tau}\, , 
\end{equation}

\noindent
where the $\alpha^{A}$ are the constant parameters of the transformations,
labeled by $A=1,\cdots,5$, and where $k_{A}{}^{\varphi}$ and $k_{A}{}^{\tau}$
are the corresponding components of the Killing vectors of the scalar
manifold, given in Eq.~(\ref{eq:varphikillingvectors})
(Eq.~(\ref{eq:taukillingvectors})).

According to the general prescription Eq.~(\ref{eq:gauging}), we want to gauge
these symmetries making the theory invariant under the local transformations

\begin{equation}
\label{eq:scalargaugetransformations}
\delta_{\Lambda}\varphi =\Lambda^{I}\vartheta_{I}{}^{A}k_{A}{}^{\varphi}\, ,
\hspace{1cm}
\delta_{\Lambda}\tau =\Lambda^{I}\vartheta_{I}{}^{A}k_{A}{}^{\tau}\, , 
\end{equation}

\noindent
where $\Lambda^{I}(x)$, $I=0,\mathbf{1},\mathbf{2}$, are the 0-form gauge
parameters of the 1-form gauge fields $A^{I}$ and $\vartheta_{I}{}^{A}$ is the
embedding tensor. 

To construct gauge-covariant field strengths for the scalars it is enough to
replace their derivatives by covariant derivatives.

%%%%%%%%%%%%%%%%%%%%%%%%%%%%%%%%%%%%%%%%%%%%%%%%%%%%%%%%%%%%%%%%%%%%%%
%%%%%%%%%%%%%%%%%%%%%%%%%%%%%%%%%%%%%%%%%%%%%%%%%%%%%%%%%%%%%%%%%%%%%%
%%%%%%%%%%%%%%%%%%%%%%%%%%%%%%%%%%%%%%%%%%%%%%%%%%%%%%%%%%%%%%%%%%%%%%
%%%%%%%%%%%%%%%%%%%%%%%%%%%%%%%%%%%%%%%%%%%%%%%%%%%%%%%%%%%%%%%%%%%%%%

\subsubsection{Covariant derivatives}

The covariant derivatives of the scalars have the standard form

\begin{equation}
\mathfrak{D}\varphi = d\varphi +A^{I}\vartheta_{I}^{A}k_{A}{}^{\varphi}\, ,  
\hspace{1cm}
\mathfrak{D}\tau = d\tau +A^{I}\vartheta_{I}^{A}k_{A}{}^{\tau}\, ,  
\end{equation}

\noindent
and they transform covariantly provided that the 1-form gauge fields transform
as

\begin{equation}
\label{eq:deltaAI}
\delta_{\Lambda}A^{I}
= 
-\mathfrak{D}\Lambda^{I} +Z^{I}{}_{i}\Lambda^{i}\, ,  
\end{equation}

\noindent
where the $\Lambda^{i}$, $i=1,2$, are two possible 1-form gauge parameters and
$Z^{I}{}_{i}$ is a possible new deformation parameter that must satisfy
the orthogonality constraint

\begin{equation}
\label{eq:constraint2}
\vartheta_{I}{}^{A}Z^{I}{}_{i} =0\, .  
\end{equation}

\noindent
Furthermore, it is necessary that the embedding tensor satisfies the standard
quadratic constraint

\begin{equation}
\label{eq:standardquadraticconstraint}
\vartheta_{I}{}^{A}T_{A\,  J}{}^{K}\vartheta_{K}{}^{C}
-\vartheta_{I}{}^{A}\vartheta_{J}{}^{B}f_{AB}{}^{C}=0\, ,  
\end{equation}

\noindent
that expresses the gauge-invariance of the embedding tensor. 

As a general rule, all the deformation tensors have to be gauge-invariant and
we can anticipate that we will have to impose the constraint that expresses
the gauge-invariance of $Z^{I}{}_{i}$, namely 

\begin{equation}
\label{eq:constraint3}
X_{J\, K}{}^{I}Z^{K}{}_{i}
-X_{J\, i}{}^{j}Z^{I}{}_{j} =  0\, ,
\end{equation}

\noindent
where 

\begin{equation}
\label{eq:XIJKXIjk}
X_{I\, J}{}^{K} \equiv \vartheta_{I}{}^{A}T_{A\, J}{}^{K}\, ,
\hspace{1cm}
X_{J\, i}{}^{j} \equiv \vartheta_{J}{}^{A}T_{A\, i}{}^{j}\, .
\end{equation}

%%%%%%%%%%%%%%%%%%%%%%%%%%%%%%%%%%%%%%%%%%%%%%%%%%%%%%%%%%%%%%%%%%%%%%
%%%%%%%%%%%%%%%%%%%%%%%%%%%%%%%%%%%%%%%%%%%%%%%%%%%%%%%%%%%%%%%%%%%%%%
%%%%%%%%%%%%%%%%%%%%%%%%%%%%%%%%%%%%%%%%%%%%%%%%%%%%%%%%%%%%%%%%%%%%%%
%%%%%%%%%%%%%%%%%%%%%%%%%%%%%%%%%%%%%%%%%%%%%%%%%%%%%%%%%%%%%%%%%%%%%%

\subsubsection{Supersymmetry transformations of the fermion fields}

We will assume for simplicity that the supersymmetry transformations of the
fermion fields in the deformed theory have essentially the same form as in the
undeformed theory but covariantized (derivatives and field strengths) and,
possibly, with the addition of fermion shifts which we add in the most general
form:

\begin{eqnarray}
\delta_{\epsilon}\psi_{\mu}
& = & 
\mathfrak{D}_{\mu}\epsilon
+f\gamma_{\mu} \epsilon
+k\gamma_{\mu} \epsilon^{*}
+\tfrac{i}{8\cdot 2!}e^{-\frac{2}{\sqrt{7}}\varphi}
\left(\tfrac{5}{7}\gamma_{\mu}\gamma^{(2)} 
-\gamma^{(2)}\gamma_{\mu} \right)F^{0}\epsilon  
\nonumber \\
& & \nonumber \\
& & 
-\tfrac{1}{8\cdot 2!}e^{\frac{3}{2\sqrt{7}}\varphi+\frac{1}{2}\phi}
\left(\tfrac{5}{7}\gamma_{\mu}\gamma^{(2)} 
-\gamma^{(2)}\gamma_{\mu} \right)(F^{1}-\tau F^{2})\epsilon^{*}  
\nonumber \\
& & \nonumber \\
& & 
-\tfrac{i}{8\cdot 3!}e^{-\frac{1}{2\sqrt{7}}\varphi}
\left(\tfrac{3}{7}\gamma_{\mu}\gamma^{(3)} 
+\gamma^{(3)}\gamma_{\mu} \right)(H^{1}-\tau H^{2})\epsilon^{*}  
\nonumber \\
& & \nonumber \\
& & 
-\tfrac{1}{8\cdot 4!}e^{\frac{1}{\sqrt{7}}\varphi}
\left(\tfrac{1}{7}\gamma_{\mu}\gamma^{(4)} 
-\gamma^{(4)}\gamma_{\mu} \right)G \epsilon\, ,
\end{eqnarray}

\begin{eqnarray}
\delta_{\epsilon}\tilde{\lambda}
& = & 
i\not\!\!\mathfrak{D} \varphi\epsilon^{*} +\tilde{g}\epsilon  +\tilde{h}\epsilon^{*}
-\tfrac{1}{\sqrt{7}}e^{-\frac{2}{\sqrt{7}}\varphi}\not\! F^{0}\epsilon^{*}
-\tfrac{3i}{2\cdot 2!\sqrt{7}}e^{\frac{3}{2\sqrt{7}}\varphi +\frac{1}{2}\phi}
(\not\! F^{1}-\tau^{*}\not\! F^{2})\epsilon
\nonumber \\
& & \nonumber \\
& & 
-\tfrac{1}{2\cdot 3!\sqrt{7}}e^{-\frac{1}{2\sqrt{7}}\varphi +\frac{1}{2}\phi}
(\not\!\! H^{1}-\tau^{*}\not\!\! H^{2})\epsilon
-\tfrac{i}{4!\sqrt{7}}e^{\frac{1}{\sqrt{7}}\varphi}\not\! G\epsilon^{*}\, ,
\end{eqnarray}

\begin{eqnarray}
\delta_{\epsilon}\lambda
& = & 
-e^{\phi}\not\!\!\mathfrak{D} \tau\epsilon^{*} +g\epsilon  +h\epsilon^{*}
-\tfrac{i}{2\cdot 2!}e^{\frac{3}{2\sqrt{7}}\varphi +\frac{1}{2}\phi}
(\not\! F^{1}-\tau\not\! F^{2})\epsilon
\nonumber \\
& & \nonumber \\
& & 
+\tfrac{1}{2\cdot 3!}e^{-\frac{1}{2\sqrt{7}}\varphi +\frac{1}{2}\phi}
(\not\!\! H^{1}-\tau\not\!\! H^{2})\epsilon\, .
\end{eqnarray}

\noindent
In these expressions, $f,k,g,h,\tilde{g},\tilde{h}$ are six functions of the
scalars and deformation parameters to be determined, the covariant field
strengths have the general form predicted by the tensor hierarchy (to be
determined) and the covariant derivatives of the scalars have the forms given
above. Furthermore, in $\delta_{\epsilon}\psi_{\mu}$, 
$\mathfrak{D}_{\mu}\epsilon$ stands for the Lorentz- and gauge-covariant
derivative of the supersymmetry parameter, which turns out to be given by

\begin{equation}
\mathfrak{D}_{\mu}\epsilon  
\equiv
\left\{ 
\nabla_{\mu}
+\tfrac{i}{2}
\left[
\tfrac{1}{2}e^{\phi}
\mathfrak{D}^{5}_{\mu}\chi
+A^{I}{}_{\mu}\vartheta_{I}{}^{m}\mathcal{P}_{m}
\right]
+\tfrac{9}{14}\gamma_{\mu}\not\!\!A^{I}\vartheta_{I}{}^{4}
\right\}\epsilon
\end{equation}

\noindent
where $\mathcal{P}_{m}$ $1,2,3$ are the momentum maps of the holomorphic
Killing vectors of $SL(2,\mathbb{R})$, defined in
Eq.~(\ref{eq:momentummapsdef}) and given in Eq.~(\ref{eq:momentummaps}),
$\nabla_{\mu}$ is the Lorentz-covariant derivative and

\begin{equation}
\mathfrak{D}^{5}_{\mu}\chi
\equiv
\partial_{\mu}\chi
-\tfrac{3}{4}A^{I}{}_{\mu}\vartheta_{I}{}^{5} \chi
\end{equation}

\noindent
is the derivative of $\chi$ covariant only with respect to the $\beta$
rescalings. it can be checked that $\mathfrak{D}_{\mu}\epsilon$ transforms
covariantly under gauge transformations if and only if the embedding tensor
satisfies the standard quadratic constraint
Eq.~(\ref{eq:standardquadraticconstraint}).

An equivalent expression for it is

\begin{equation}
\mathfrak{D}_{\mu}\epsilon  
=
\left\{ 
\nabla_{\mu}
+\tfrac{i}{2}
\left[
\tfrac{1}{2}e^{\phi}
\mathfrak{D}_{\mu}\chi
-A^{I}{}_{\mu}\vartheta_{I}{}^{m}\Im{\rm m}\lambda_{m}
\right]
+\tfrac{9}{14}\gamma_{\mu}\not\!\!A^{I}\vartheta_{I}{}^{4}
\right\}\epsilon\, ,
\end{equation}

\noindent
where the $\lambda_{m}$, $m=1,2,3$, of $SL(2,\mathbb{R})$ and defined in
Eq.~(\ref{eq:momentummapsdef}) and given in Eq.~(\ref{eq:lambdam}) and where
now

\begin{equation}
\mathfrak{D}_{\mu}\chi
\equiv
\partial_{\mu}\chi +A^{I}{}_{\mu}\vartheta_{I}{}^{A} k_{A}{}^{\chi}\, ,
\end{equation}

\noindent
is the total covariant derivative of $\chi$ (which is invariant under both the
$\alpha$ and $\beta$ scaling symmetries as well as under $SL(2,\mathbb{R})$).

The actual form of the $(p+1)$-form field strengths will not be needed until
the moment in which study the closure of the supersymmetry algebra on the
corresponding $p$-form potential. 

%%%%%%%%%%%%%%%%%%%%%%%%%%%%%%%%%%%%%%%%%%%%%%%%%%%%%%%%%%%%%%%%%%%%%%
%%%%%%%%%%%%%%%%%%%%%%%%%%%%%%%%%%%%%%%%%%%%%%%%%%%%%%%%%%%%%%%%%%%%%%
%%%%%%%%%%%%%%%%%%%%%%%%%%%%%%%%%%%%%%%%%%%%%%%%%%%%%%%%%%%%%%%%%%%%%%
%%%%%%%%%%%%%%%%%%%%%%%%%%%%%%%%%%%%%%%%%%%%%%%%%%%%%%%%%%%%%%%%%%%%%%

\subsubsection{Closure of the supersymmetry algebra on the
 0-forms $\varphi,\tau$}

We assume that the supersymmetry transformations of the scalars are the same as
in the undeformed theory

\begin{eqnarray}
\delta_{\epsilon}\varphi 
& = & 
-\tfrac{i}{4}\bar{\epsilon}\tilde{\lambda}^{*}+\mathrm{h.c.}\, ,
\\
& & \nonumber \\  
\delta_{\epsilon}\tau 
& = & 
-\tfrac{1}{2}e^{-\phi}\bar{\epsilon}^{*}\lambda\, .
\end{eqnarray}

To lowest order in fermions, the commutator of two supersymmetry
transformations gives

%(\textbf{for} $\mathbf{\beta_{3}=-1,\beta_{1}=\beta_{2}}$)

\begin{eqnarray}
\label{eq:commutatoronvarphi1}
\left[\delta_{\epsilon_{1}},\delta_{\epsilon_{2}}\right]\varphi
& = &   
\xi^{\mu}\mathfrak{D}_{\mu}\varphi 
+\Re{\rm e}(\tilde{h}) b
-\Im{\rm m}(\tilde{g})c +\Re{\rm e}(\tilde{g}) d\, ,
\\
& & \nonumber \\
\label{eq:commutatorontau1}
\left[\delta_{\epsilon_{1}},\delta_{\epsilon_{2}}\right]\tau
& = &   
\xi^{\mu}\mathfrak{D}_{\mu}\tau +e^{-\phi}\left[g(c-id) -ihb\right]\, ,
\end{eqnarray}

\noindent
where $\xi^{\mu}$ is one of the spinor bilinears defined in
Appendix~\ref{sec-bilinears} that clearly plays the r\^ole of parameter of the
general coordinate transformations and $a,b,c,d$ are the scalar bilinears
defined in the same appendix.

In the right hand side of these commutators, to lowest order in fermions, we
expect a general coordinate transformation (the Lie derivative $\pounds_{\xi}$
of the scalars with respect to $\xi^{\mu}$) and a gauge transformation which
has the form of Eq.~(\ref{eq:scalargaugetransformations}) for the
scalars. Therefore, the above expressions should be compared with

\begin{eqnarray}
\label{eq:commutatoronvarphi2}
\left[\delta_{\epsilon_{1}},\delta_{\epsilon_{2}}\right]\varphi
& = &   
\pounds_{\xi}\varphi +\Lambda^{I}\vartheta_{I}{}^{A}k_{A}{}^{\varphi}\, ,
\\
& & \nonumber \\
\label{eq:commutatorontau2}
\left[\delta_{\epsilon_{1}},\delta_{\epsilon_{2}}\right]\tau
& = &   
\pounds_{\xi}\tau +\Lambda^{I}\vartheta_{I}{}^{A}k_{A}{}^{\tau}\, ,
\end{eqnarray}

\noindent
from which we get the relations

\begin{eqnarray}
\label{eq:relation1}
\Re{\rm e}(\tilde{h}) b
-\Im{\rm m}(\tilde{g})c +\Re{\rm e}(\tilde{g}) d  
& = & 
(\Lambda^{I}-a^{I})\vartheta_{I}{}^{A}k_{A}{}^{\varphi}
\, ,
\\
& & \nonumber \\
\label{eq:relation2}
g(c-id) -ihb  
& = & 
e^{\phi}(\Lambda^{I}-a^{I})\vartheta_{I}{}^{A}k_{A}{}^{\tau}
\, ,
\end{eqnarray}

\noindent
which would allow us to determine the fermion shift functions if we knew the
gauge parameters $\Lambda^{I}$. In order to determine the $\Lambda^{I}$s we
have to close the supersymmetry algebra on the 1-forms. In these expressions
and in those that will follow, we use the shorthand notation

\begin{equation}
a^{I} \equiv \xi^{\mu}A^{I}{}_{\mu}\, ,\hspace{.5cm}
b^{i}{}_{\mu} \equiv \xi^{\nu}B^{i}{}_{\nu\mu}\, ,\hspace{.5cm}
c_{\mu\nu} \equiv \xi^{\rho}C_{\rho\mu\nu}\, ,\hspace{.5cm}
\mathrm{etc.} 
\end{equation}

%%%%%%%%%%%%%%%%%%%%%%%%%%%%%%%%%%%%%%%%%%%%%%%%%%%%%%%%%%%%%%%%%%%%%%
%%%%%%%%%%%%%%%%%%%%%%%%%%%%%%%%%%%%%%%%%%%%%%%%%%%%%%%%%%%%%%%%%%%%%%
%%%%%%%%%%%%%%%%%%%%%%%%%%%%%%%%%%%%%%%%%%%%%%%%%%%%%%%%%%%%%%%%%%%%%%
%%%%%%%%%%%%%%%%%%%%%%%%%%%%%%%%%%%%%%%%%%%%%%%%%%%%%%%%%%%%%%%%%%%%%%

\subsection{The 1-forms $A^{I}$}

The next step in this procedure is to consider the 1-forms that we just
introduced to construct covariant derivatives for the scalars.

%%%%%%%%%%%%%%%%%%%%%%%%%%%%%%%%%%%%%%%%%%%%%%%%%%%%%%%%%%%%%%%%%%%%%%
%%%%%%%%%%%%%%%%%%%%%%%%%%%%%%%%%%%%%%%%%%%%%%%%%%%%%%%%%%%%%%%%%%%%%%
%%%%%%%%%%%%%%%%%%%%%%%%%%%%%%%%%%%%%%%%%%%%%%%%%%%%%%%%%%%%%%%%%%%%%%
%%%%%%%%%%%%%%%%%%%%%%%%%%%%%%%%%%%%%%%%%%%%%%%%%%%%%%%%%%%%%%%%%%%%%%

\subsubsection{The 2-form field strengths $F^{I}$}

The gauge transformations of the 1-forms are given in Eq.~(\ref{eq:deltaAI})
and we first need to determine their covariant field strengths. A general
result of the embedding-tensor formalism tells us that we need to introduce
2-form potentials in the covariant field strengths. In this case only have the
$SL(2,\mathbb{R})$ doublet $B^{i}$ at our disposal and, therefore, the 2-form
field strengths have the form

\begin{equation}
\label{eq:FI}
F^{I} = dA^{I} +\tfrac{1}{2}X_{JK}{}^{I}A^{J}\wedge A^{K}
+Z^{I}{}_{i}B^{i}\, ,  
\end{equation}

\noindent
where $X_{JK}{}^{I}$ has been defined in Eq.~(\ref{eq:XIJKXIjk}) and
$Z^{I}{}_{i}$ is precisely the deformation tensor we introduced in
Eq.~(\ref{eq:deltaAI}). $F^{I}$ will transform covariantly under
Eq.~(\ref{eq:deltaAI}) if simultaneously the 2-forms $B^{i}$ transform
according to

\begin{equation}
\label{eq:deltaBi}
\delta_{\Lambda}B^{i}
= 
-\mathfrak{D}\Lambda^{i} 
-2h_{IJ}{}^{i}\left[\Lambda^{I}F^{J} 
+\tfrac{1}{2}A^{I}\wedge \delta_{\Lambda}A^{J}\right] 
+Z^{i}\Lambda\, ,
\end{equation}

\noindent
where $h_{IJ}{}^{i}$ and $Z^{i}$ are two possible new deformation tensors the
first of which must satisfy the constraint

\begin{equation}
\label{eq:constraint4}
X_{(JK)}{}^{I} +Z^{I}{}_{i}h_{JK}{}^{i} =  0\, ,
\end{equation}

\noindent
while $Z^{i}$ must satisfy the orthogonality constraint

\begin{equation}
\label{eq:constraint6}
Z^{I}{}_{i}Z^{i} = 0\, .
\end{equation}

\noindent
Both of them must satisfy the constraints that express their gauge invariance:

\begin{eqnarray}
\label{eq:constraint7}
X_{I\, j}{}^{i}h_{JK}{}^{j} -2X_{I(J}{}^{L}h_{K)L}{}^{i} & = & 0\, ,
\\
& & \nonumber \\
\label{eq:constraint9}
X_{I}Z^{i}- X_{I\, j}{}^{i}Z^{j} & = & 0\, ,
\end{eqnarray}

\noindent
where 

\begin{equation}
X_{I} \equiv \vartheta_{I}{}^{A}T^{(1)}_{A}\, .
\end{equation}

%%%%%%%%%%%%%%%%%%%%%%%%%%%%%%%%%%%%%%%%%%%%%%%%%%%%%%%%%%%%%%%%%%%%%%
%%%%%%%%%%%%%%%%%%%%%%%%%%%%%%%%%%%%%%%%%%%%%%%%%%%%%%%%%%%%%%%%%%%%%%
%%%%%%%%%%%%%%%%%%%%%%%%%%%%%%%%%%%%%%%%%%%%%%%%%%%%%%%%%%%%%%%%%%%%%%
%%%%%%%%%%%%%%%%%%%%%%%%%%%%%%%%%%%%%%%%%%%%%%%%%%%%%%%%%%%%%%%%%%%%%%

\subsubsection{Closure of the supersymmetry algebra on the 
1-forms $A^{I}$}

We assume, as we are doing with all the bosons, that the supersymmetry
transformations of the 1-forms of the theory are not deformed by the gauging,
so they take the form

\begin{eqnarray}
\delta_{\epsilon}A^{0}{}_{\mu}
& = & 
\tfrac{i}{2}e^{\frac{2}{\sqrt{7}}\varphi}\bar{\epsilon}
\left(\psi_{\mu} -\tfrac{i}{\sqrt{7}}\gamma_{\mu}\tilde{\lambda}^{*}\right) 
+\mathrm{h.c.}
\, ,
\\ 
& & \nonumber \\ 
\delta_{\epsilon}A^{\mathbf{1}}{}_{\mu}
& = & 
\tfrac{i}{2}\tau^{*}e^{-\frac{3}{2\sqrt{7}}\varphi+\frac{1}{2}\phi}
\left(
\bar{\epsilon}^{*}\psi_{\mu} 
-\tfrac{i}{4}\bar{\epsilon}\gamma_{\mu}\lambda
+\tfrac{3i}{4\sqrt{7}}\bar{\epsilon}^{*}\gamma_{\mu}\tilde{\lambda}^{*}
\right)
+\mathrm{h.c.}
\, ,
\\ 
& & \nonumber \\ 
\delta_{\epsilon}A^{\mathbf{2}}{}_{\mu}
& = & 
\tfrac{i}{2}e^{-\frac{3}{2\sqrt{7}}\varphi+\frac{1}{2}\phi}
\left(
\bar{\epsilon}^{*}\psi_{\mu} 
-\tfrac{i}{4}\bar{\epsilon}\gamma_{\mu}\lambda
+\tfrac{3i}{4\sqrt{7}}\bar{\epsilon}^{*}\gamma_{\mu}\tilde{\lambda}^{*}
\right)
+\mathrm{h.c.}
\end{eqnarray}

\noindent
The commutator of two of them gives, to lowest order in fermions,

%(\textbf{for} $\mathbf{\beta_{1}=\beta_{2}=-\beta_{4}\beta_{5}\, ,\,\,\, \beta_{5}=-1\, ,\,\,\, \alpha_{1}\beta_{1}=-1}$)

\begin{equation}
  \begin{array}{rcl}
\left[\delta_{\epsilon_{1}},\delta_{\epsilon_{2}}\right]A^{0}{}_{\mu}
& = &   
\xi^{\nu}F^{0}{}_{\nu\mu} 
-\mathfrak{D}_{\mu}\left(e^{\frac{2}{\sqrt{7}} \varphi}b\right)
+\tfrac{2}{\sqrt{7}}
e^{\frac{2}{\sqrt{7}}\varphi}
\left\{
\left[\Re{\rm e}(\tilde{h})-\sqrt{7}\, \Im{\rm m}(f)\right]\xi_{\mu}
\right.
 \\
& &  \\
& & 
\left.
+\left[\Re{\rm e}(\tilde{g})-\sqrt{7}\, \Im{\rm m}(k)\right]\sigma_{\mu}
+\left[\Im{\rm m}(\tilde{g})-\sqrt{7}\, \Re{\rm e}(k)\right]\rho_{\mu}
\right\}
\, ,
\end{array}
\end{equation}

\begin{equation}
  \begin{array}{rcl}
\left[\delta_{\epsilon_{1}},\delta_{\epsilon_{2}}\right]A^{\mathbf{1}}{}_{\mu}
& = &   
\xi^{\nu}F^{\mathbf{1}}{}_{\nu\mu} 
-\partial_{\mu}
\left[
e^{-\frac{3}{2\sqrt{7}}\varphi+\frac{1}{2}\phi}(\chi d +e^{-\phi}c)
\right]
 \\
& &  \\
& & 
\hspace{-3cm}
-A^{I}{}_{\mu}
\left[
(\tfrac{1}{2}\vartheta_{I}{}^{1}-\tfrac{3}{4}\vartheta_{I}{}^{5})
e^{-\frac{3}{2\sqrt{7}}\varphi+\frac{1}{2}\phi}(\chi d +e^{-\phi}c)
+
\tfrac{1}{2}(\vartheta_{I}{}^{2}+\vartheta_{I}{}^{3})
e^{-\frac{3}{2\sqrt{7}}\varphi\frac{1}{2}\phi}d
\right]
 \\
& & \\
& & 
\hspace{-3cm}
-2e^{-\frac{3}{2\sqrt{7}}\varphi\frac{1}{2}\phi}\left\{
\chi
\left[
\Im{\rm m}(k)+\tfrac{3}{4\sqrt{7}}\Re{\rm e}(\tilde{g}) -\tfrac{1}{4}\Re{\rm e}(g)
\right]
+e^{-\phi}
\left[
-\Re{\rm e}(k)-\tfrac{3}{4\sqrt{7}}\Im{\rm m}(\tilde{g}) -\tfrac{1}{4}\Im{\rm m}(g)
\right]
\right\}\xi_{\mu}
 \\
& &  \\
& & 
\hspace{-3cm}
-2e^{-\frac{3}{2\sqrt{7}}\varphi\frac{1}{2}\phi}
\left\{
\chi
\left[
-\Re{\rm e}(f) -\tfrac{3}{4\sqrt{7}}\Im{\rm m}(\tilde{h}) 
+\tfrac{1}{4}\Im{\rm m}(h)
\right]
+
e^{-\phi}
\left[
-\Im{\rm m}(f) -\tfrac{3}{4\sqrt{7}}\Re{\rm e}(\tilde{h}) 
-\tfrac{1}{4}\Re{\rm e}(h)
\right]
\right\}\rho_{\mu}
 \\
& &  \\
& & 
\hspace{-3cm}
-2e^{-\frac{3}{2\sqrt{7}}\varphi\frac{1}{2}\phi}
\left\{
\chi
\left[
\Im{\rm m}(f)+\tfrac{3}{4\sqrt{7}}\Re{\rm e}(\tilde{h}) 
-\tfrac{1}{4}\Re{\rm e}(h)
\right]
+e^{-\phi}
\left[
-\Re{\rm e}(f)-\tfrac{3}{4\sqrt{7}}\Im{\rm m}(\tilde{h}) 
-\tfrac{1}{4}\Im{\rm m}(h)
\right]
\right\}
\sigma_{\mu}\, ,
\end{array}
\end{equation}

\noindent
and

\begin{equation}
  \begin{array}{rcl}
\left[\delta_{\epsilon_{1}},\delta_{\epsilon_{2}}\right]A^{\mathbf{2}}{}_{\mu}
& = &   
\xi^{\nu}F^{\mathbf{2}}{}_{\nu\mu} 
-\partial_{\mu}
\left(
e^{-\frac{3}{2\sqrt{7}}\varphi+\frac{1}{2}\phi}d
\right)
 \\
& &  \\
& & 
-A^{I}{}_{\mu}
\left[
\tfrac{1}{2}(\vartheta_{I}{}^{2}-\vartheta_{I}{}^{3})
e^{-\frac{3}{2\sqrt{7}}\varphi+\frac{1}{2}\phi}(\chi d +e^{-\phi}c)
-\tfrac{1}{2}\vartheta_{I}{}^{1}
e^{-\frac{3}{2\sqrt{7}}\varphi+\frac{1}{2}\phi}d
\right]
 \\
& &  \\
& & 
-2e^{-\frac{3}{2\sqrt{7}}\varphi+\frac{1}{2}\phi}
\left[
\Im{\rm m}(k)+\tfrac{3}{4\sqrt{7}}\Re{\rm e}(\tilde{g}) -\tfrac{1}{4}\Re{\rm e}(g)
\right]\xi_{\mu}
 \\
& &  \\
& & 
-2e^{-\frac{3}{2\sqrt{7}}\varphi+\frac{1}{2}\phi}
\left[
-\Re{\rm e}(f) -\tfrac{3}{4\sqrt{7}}\Im{\rm m}(\tilde{h}) 
+\tfrac{1}{4}\Im{\rm m}(h)
\right]
\rho_{\mu}
 \\
& &  \\
& & 
-2e^{-\frac{3}{2\sqrt{7}}\varphi+\frac{1}{2}\phi}
\left[
\Im{\rm m}(f)+\tfrac{3}{4\sqrt{7}}\Re{\rm e}(\tilde{h}) 
-\tfrac{1}{4}\Re{\rm e}(h)
\right]
\sigma_{\mu}\, ,
\end{array}
\end{equation}

\noindent
where $\sigma_{\mu}$ and $\rho_{\mu}$ are spinor bilinears defined in
Appendix~\ref{sec-bilinears}.

The closure of the local supersymmetry algebra requires the commutators to
take the form

\begin{equation}
\left[\delta_{\epsilon_{1}},\delta_{\epsilon_{2}}\right]A^{I}{}_{\mu}
 =    
\pounds_{\xi}A^{I}{}_{\mu}
-\mathfrak{D}_{\mu}\Lambda^{I} +Z^{I}{}_{i}\Lambda^{i}{}_{\mu}\, ,
\end{equation}

\noindent
which will only happen if gauge parameters $\Lambda^{I}$ are given by 

\begin{equation}
\begin{array}{rcl}
\label{eq:LambdaI}
\Lambda^{0}
& = & 
a^{0}  +e^{\frac{2}{\sqrt{7}}\varphi}b\, ,
\\
& &  \\
\Lambda^{\mathbf{1}}
& = & 
a^{\mathbf{1}}
+e^{-\frac{3}{2\sqrt{7}}\varphi+\frac{1}{2}\phi}(\chi d +e^{-\phi}c)\, ,
\\
& &  \\
\Lambda^{\mathbf{2}}
& = & 
a^{\mathbf{2}} +e^{-\frac{3}{2\sqrt{7}}\varphi+\frac{1}{2}\phi}d\, ,\\
\end{array}
\end{equation}

\noindent
and the 1-form gauge parameters $\Lambda^{i}{}_{\mu}$ satisfy the relations

\begin{eqnarray}
\left[\Re{\rm e}(\tilde{h})-\sqrt{7}\, \Im{\rm m}(f)\right]\xi_{\mu}
+\left[\Re{\rm e}(\tilde{g})-\sqrt{7}\, \Im{\rm m}(k)\right]\sigma_{\mu}
+\left[\Im{\rm m}(\tilde{g})-\sqrt{7}\, \Re{\rm e}(k)\right]\rho_{\mu}
& &   
\nonumber \\
& & \nonumber \\
=
\tfrac{\sqrt{7}}{2}e^{-\frac{2}{\sqrt{7}}\varphi}Z^{0}{}_{i}
\left[
\Lambda^{i}{}_{\mu}
-(b^{i}{}_{\mu} -h_{IJ}{}^{i}a^{I}A^{J}{}_{\mu})
\right]\, ,
\label{eq:1-formshifts1}
& &  \\
& & \nonumber \\
\left\{
\chi
\left[
\Im{\rm m}(k)+\tfrac{3}{4\sqrt{7}}\Re{\rm e}(\tilde{g}) -\tfrac{1}{4}\Re{\rm e}(g)
\right]
+e^{-\phi}
\left[
-\Re{\rm e}(k)-\tfrac{3}{4\sqrt{7}}\Im{\rm m}(\tilde{g}) -\tfrac{1}{4}\Im{\rm m}(g)
\right]
\right\}\xi_{\mu}
\nonumber \\
& & \nonumber \\
\hspace{-3cm}
+
\left\{
\chi
\left[
-\Re{\rm e}(f) -\tfrac{3}{4\sqrt{7}}\Im{\rm m}(\tilde{h}) 
+\tfrac{1}{4}\Im{\rm m}(h)
\right]
+
e^{-\phi}
\left[
-\Im{\rm m}(f) -\tfrac{3}{4\sqrt{7}}\Re{\rm e}(\tilde{h}) 
-\tfrac{1}{4}\Re{\rm e}(h)
\right]
\right\}\rho_{\mu}
& & \nonumber \\
& & \nonumber \\
\hspace{-3cm}
+\left\{
\chi
\left[
\Im{\rm m}(f)+\tfrac{3}{4\sqrt{7}}\Re{\rm e}(\tilde{h}) 
-\tfrac{1}{4}\Re{\rm e}(h)
\right]
+e^{-\phi}
\left[
-\Re{\rm e}(f)-\tfrac{3}{4\sqrt{7}}\Im{\rm m}(\tilde{h}) 
-\tfrac{1}{4}\Im{\rm m}(h)
\right]
\right\}
\sigma_{\mu}\, ,
& & \nonumber \\
& & \nonumber \\
 = 
-\tfrac{1}{2}e^{+\frac{3}{2\sqrt{7}}\varphi-\frac{1}{2}\phi}Z^{\mathbf{1}}{}_{i}
\left[
\Lambda^{i}{}_{\mu}
-(b^{i}{}_{\mu} -h_{IJ}{}^{i}a^{I}A^{J}{}_{\mu})
\right]\, , 
& & \\
& & \nonumber \\
\left[
\Im{\rm m}(k)+\tfrac{3}{4\sqrt{7}}\Re{\rm e}(\tilde{g}) -\tfrac{1}{4}\Re{\rm e}(g)
\right]
\xi_{\mu}
+
\left[
-\Re{\rm e}(f) -\tfrac{3}{4\sqrt{7}}\Im{\rm m}(\tilde{h}) 
+\tfrac{1}{4}\Im{\rm m}(h)
\right]
\rho_{\mu}
& & \nonumber \\
& & \nonumber \\
\hspace{-3cm}
+\left[
\Im{\rm m}(f)+\tfrac{3}{4\sqrt{7}}\Re{\rm e}(\tilde{h}) 
-\tfrac{1}{4}\Re{\rm e}(h)
\right]
\sigma_{\mu}\, ,
& & \nonumber \\
& & \nonumber \\
 = 
-\tfrac{1}{2}e^{+\frac{3}{2\sqrt{7}}\varphi-\frac{1}{2}\phi}Z^{\mathbf{2}}{}_{i}
\left[
\Lambda^{i}{}_{\mu}
-(b^{i}{}_{\mu} -h_{IJ}{}^{i}a^{I}A^{J}{}_{\mu})
\right]\, .
\label{eq:1-formshifts2}
\end{eqnarray}

\noindent
Using the values of the parameters $\Lambda^{I}$ that we just have determined
in the relations Eqs.~(\ref{eq:relation1}) and (\ref{eq:relation2}) we can
determine some of the fermions shifts:

\begin{eqnarray}
\Re{\rm e}(\tilde{h})
& = & 
\vartheta_{0}{}^{A}k_{A}{}^{\varphi} e^{\frac{2}{\sqrt{7}}\varphi}\, ,
\\
& & \nonumber \\
\tilde{g}
& = & 
(\vartheta_{\mathbf{1}}{}^{A}\tau^{*} +\vartheta_{\mathbf{2}}{}^{A})
k_{A}{}^{\varphi} e^{-\frac{3}{2\sqrt{7}}\varphi+\tfrac{1}{2}\phi}\, ,
\\
& & \nonumber \\
h
& = &
i\vartheta_{0}{}^{A}k_{A}{}^{\tau} e^{\frac{2}{\sqrt{7}}\varphi+\phi}\, ,
\\
& & \nonumber \\
g
& = &
\vartheta_{\mathbf{1}}{}^{A}
k_{A}{}^{\tau} e^{-\frac{3}{2\sqrt{7}}\varphi+\frac{1}{2}\phi}\, .
\end{eqnarray}

As a matter of fact, $g$ is overdetermined: we get two different expression
for it that give the same value if and only if

\begin{equation}
(\vartheta_{\mathbf{1}}{}^{A}\tau +\vartheta_{\mathbf{2}}{}^{A})k_{A}{}^{\tau}  
= 
0\, ,
\end{equation}

\noindent
which, upon use of the explicit expressions of the holomorphic Killing vectors
$k_{A}{}^{\tau}$ in Section~\ref{sec-global}, leads to the following linear
constraints on the components of the embedding tensor:

\begin{equation}
\label{eq:linearconstraints1}
\begin{array}{rcl}
\vartheta_{\mathbf{2}}{}^{2}  +\vartheta_{\mathbf{2}}{}^{3}
& = & 0\, ,
\\
& & \\
\vartheta_{\mathbf{1}}{}^{2}  +\vartheta_{\mathbf{1}}{}^{3} +2\vartheta_{\mathbf{2}}{}^{1}
-\tfrac{3}{2}\vartheta_{\mathbf{2}}{}^{5}
& = & 0\, ,
\\
& &  \\
\vartheta_{\mathbf{2}}{}^{2}  -\vartheta_{\mathbf{2}}{}^{3} -2\vartheta_{\mathbf{1}}{}^{1}
+\tfrac{3}{2}\vartheta_{\mathbf{1}}{}^{5}
& = & 0\, ,
\\
& &  \\
\vartheta_{\mathbf{1}}{}^{2}  -\vartheta_{\mathbf{1}}{}^{3}
& = & 0\, .
\end{array}
\end{equation}

These constraints allow us to express 4 of the 15 components of the embedding
tensor in terms of the remaining 11, but we are only going to do this after we
take into account the constraints that we are going to find in the closure of
the local supersymmetry algebra on the doublet of 2-forms $B^{i}$.

The values of $g,h.\tilde{g},\tilde{h}$ and the above constraints are
compatible with those of the primary deformations found in
Ref.~\cite{Bergshoeff:2002nv}.

%%%%%%%%%%%%%%%%%%%%%%%%%%%%%%%%%%%%%%%%%%%%%%%%%%%%%%%%%%%%%%%%%%%%%%
%%%%%%%%%%%%%%%%%%%%%%%%%%%%%%%%%%%%%%%%%%%%%%%%%%%%%%%%%%%%%%%%%%%%%%
%%%%%%%%%%%%%%%%%%%%%%%%%%%%%%%%%%%%%%%%%%%%%%%%%%%%%%%%%%%%%%%%%%%%%%
%%%%%%%%%%%%%%%%%%%%%%%%%%%%%%%%%%%%%%%%%%%%%%%%%%%%%%%%%%%%%%%%%%%%%%

\subsection{The 2-forms $B^{i}$}

In the previous subsection we have introduced a doublet of 2-forms $B^{i}$
with given gauge transformations to construct the 2-form field strengths
$F^{I}$. We now have to construct their covariant field strengths and check
the closure of the local supersymmetry algebra on them.

%%%%%%%%%%%%%%%%%%%%%%%%%%%%%%%%%%%%%%%%%%%%%%%%%%%%%%%%%%%%%%%%%%%%%%
%%%%%%%%%%%%%%%%%%%%%%%%%%%%%%%%%%%%%%%%%%%%%%%%%%%%%%%%%%%%%%%%%%%%%%
%%%%%%%%%%%%%%%%%%%%%%%%%%%%%%%%%%%%%%%%%%%%%%%%%%%%%%%%%%%%%%%%%%%%%%
%%%%%%%%%%%%%%%%%%%%%%%%%%%%%%%%%%%%%%%%%%%%%%%%%%%%%%%%%%%%%%%%%%%%%%

\subsubsection{The 3-form field strengths $H^{i}$}

In general we need to introduce 3-form potentials to construct the covariant
3-form field strengths and, since in maximal 9-dimensional supergravity, we
only have $C$ at our disposal, the 3-form field strengths will be given by

\begin{equation}
\label{eq:Hi}
H^{i}=
\mathfrak{D}B^{i} -h_{IJ}{}^{i}A^{I}\wedge dA^{J} 
-\tfrac{1}{3} X_{[IJ}{}^{L}h_{K]L}{}^{i}A^{IJK} +Z^{i}C\, ,
\end{equation}

\noindent
and they transform covariantly under the gauge transformations of the 1- and
2-forms that we have previously determined provided if the 3-form $C$
transforms as

\begin{equation}
\label{eq:deltaC}
\delta_{\Lambda}C = -\mathfrak{D}\Lambda
+g_{Ii}
\left[
-\Lambda^{I}H^{i} -F^{I}\wedge \Lambda^{i} +\delta_{\Lambda}A^{I}\wedge B^{i}
-\tfrac{1}{3}h_{JK}{}^{i}A^{IJ}\wedge \delta_{\Lambda}A^{K}
\right] 
+Z\tilde{\Lambda}\, .
\end{equation}

\noindent
where $g_{Ii}$ and $Z$ are two possible new deformation parameters. $g_{Ii}$
must satisfy the constraint

\begin{equation}
\label{eq:constraint5}
2h_{IJ}{}^{i}Z^{J}{}_{j}  +X_{I\, j}{}^{i} +Z^{i}g_{Ij}
=  0\, ,
\end{equation}

\noindent
while $Z$ must satisfy the orthogonality constraint

\begin{equation}
\label{eq:constraint10}  
Z^{i}Z  =  0\, .  
\end{equation}

\noindent
Both must by gauge-invariant, which implies the constraints

\begin{eqnarray}
\label{eq:constraint14}  
X_{IJ}{}^{L}g_{Li}+X_{I\, i}{}^{j}g_{Jj}-X_{I}g_{Ji} & = & 0\, ,
\\
& & \nonumber \\
\label{eq:constraint18}
(X_{I}-\tilde{X}_{I})Z & = & 0\, ,
\end{eqnarray}

\noindent
where 

\begin{equation}
\tilde{X}_{I} \equiv \vartheta_{I}{}^{A}T^{(\tilde{1})}_{A}\, .  
\end{equation}

Using the constraints obeyed by the deformation parameters and the explicit
form of the 2-form field strengths $F^{I}$ we can rewrite the 3-form field
strengths in the useful form

\begin{equation}
H^{i} =
\mathfrak{D}B^{i} -h_{IJ}{}^{i}A^{I}\wedge F^{J} 
+\tfrac{1}{6} X_{[IJ}{}^{L}h_{K]L}{}^{i}A^{IJK}
-\tfrac{1}{2}X_{Ij}{}^{i}A^{I}\wedge B^{j} 
+Z^{i}(C -\tfrac{1}{2}g_{Ij}A^{I}\wedge B^{j})\, .
\end{equation}

%%%%%%%%%%%%%%%%%%%%%%%%%%%%%%%%%%%%%%%%%%%%%%%%%%%%%%%%%%%%%%%%%%%%%%
%%%%%%%%%%%%%%%%%%%%%%%%%%%%%%%%%%%%%%%%%%%%%%%%%%%%%%%%%%%%%%%%%%%%%%
%%%%%%%%%%%%%%%%%%%%%%%%%%%%%%%%%%%%%%%%%%%%%%%%%%%%%%%%%%%%%%%%%%%%%%
%%%%%%%%%%%%%%%%%%%%%%%%%%%%%%%%%%%%%%%%%%%%%%%%%%%%%%%%%%%%%%%%%%%%%%

\subsubsection{Closure of the supersymmetry algebra on the 2-forms $B^{i}$}

In the undeformed theory, the supersymmetry transformation rules for the
2-forms are

\begin{eqnarray}
\delta_{\epsilon}B^{1}
& = &
\tau^{*} e^{\frac{1}{2\sqrt{7}}\varphi +\frac{1}{2}\phi} 
\left[ 
\bar{\epsilon}^{*}\gamma_{[\mu}\psi_{\nu]} 
-\tfrac{i}{8}\bar{\epsilon}\gamma_{\mu\nu}\lambda 
-\tfrac{i}{8\sqrt{7}}\bar{\epsilon}^{*}\gamma_{\mu\nu}\tilde{\lambda}^{*}
\right]   
\nonumber \\
& & \nonumber \\
& & 
-\delta^{1}{}_{\mathbf{i}}\left(A^{0}{}_{[\mu|}\delta_{\epsilon}A^{\mathbf{i}}{}_{|\nu]}
+A^{\mathbf{i}}{}_{[\mu|}\delta_{\epsilon}A^{0}{}_{|\nu]}\right)
+\mathrm{h.c.}\, ,
\\
& & \nonumber \\
\delta_{\epsilon}B^{2}
& = &
e^{\frac{1}{2\sqrt{7}}\varphi +\frac{1}{2}\phi} 
\left[ 
\bar{\epsilon}^{*}\gamma_{[\mu}\psi_{\nu]} 
-\tfrac{i}{8}\bar{\epsilon}\gamma_{\mu\nu}\lambda 
-\tfrac{i}{8\sqrt{7}}\bar{\epsilon}^{*}\gamma_{\mu\nu}\tilde{\lambda}^{*}
\right]   
\nonumber \\
& & \nonumber \\
& & 
-\delta^{2}{}_{\mathbf{i}}
\left(A^{0}{}_{[\mu|}\delta_{\epsilon}A^{\mathbf{i}}{}_{|\nu]}
+A^{\mathbf{i}}{}_{[\mu|}\delta_{\epsilon}A^{0}{}_{|\nu]}\right)
+\mathrm{h.c.}\, .
\end{eqnarray}

\noindent
The last terms in both transformations are associated to the presence of
derivatives of $A^{\mathbf{1}}$ and $A^{\mathbf{2}}$ in the field strengths of
$B^{1}$ and $B^{2}$ in the undeformed theory (see
Eq.~(\ref{eq:Hiundeformed})). In the deformed theory, the terms $-(A^{0}\wedge
dA^{\mathbf{i}} +A^{\mathbf{i}}\wedge dA^{0})$ are replaced by more general
couplings $-h_{IJ}{}^{i}A^{I}\wedge dA^{J}$ and, therefore, it would be natural to
replace the last terms in $\delta_{\epsilon}B^{i}{}_{\mu\nu}$ by

\begin{equation}
-2h_{IJ}{}^{i}A^{I}{}_{[\mu|}\delta_{\epsilon}A^{J}{}_{|\nu]}\, .  
\end{equation}

In the commutator of two supersymmetry transformations on the 2-forms, these
terms give the right contributions to the terms
$-2h_{IJ}{}^{i}\Lambda^{I}F^{J}$ of the gauge transformations (see
Eq.~(\ref{eq:deltaBi})). However, these terms must receive other contributions
in order to be complete and it turns out that the only terms of the form
$-2h_{IJ}{}^{i}\Lambda^{I}F^{J}$ that can be completed are precisely those of
the undeformed theory, which correspond to

\begin{equation}
\label{eq:valueofhIJi}
h_{\mathbf{i}0}{}^{j} = -\tfrac{1}{2}\delta_{\mathbf{i}}{}^{j}\, .  
\end{equation}

In order to get more general $h_{IJ}{}^{i}$s it would be necessary to deform
the fermions' supersymmetry rules, something we will not do here. Furthermore,
the structure of the Chern-Simons terms of the field strengths is usually
determined by the closure of the supersymmetry algebra at higher orders in
fermions and it is highly unlikely that a more general structure of the
Chern-Simons terms will be allowed by supersymmetry. Therefore, from now on,
we will set $h_{IJ}{}^{i}$ to the above value and we will set the values of
the deformation tensors in the Chern-Simons terms of the higher-rank field
strengths, to the values of the undeformed theory. Using the above value of
$h_{IJ}{}^{i}$ in the constraints in which it occurs will help us to solve
them, sometimes completely, as we will see.  Nevertheless, we will keep using
the notation $h_{IJ}{}^{i}$ for convenience.
 
Using the identity

\begin{equation}
  \begin{array}{rcl}
\xi^{\rho}H^{i}{}_{\rho\mu\nu}
-2h_{IJ}{}^{i}A^{I}{}_{\mu}\pounds_{\xi}A^{J}{}_{\nu}
& = &
\pounds_{\xi}B^{i}{}_{\mu\nu}
-2\mathfrak{D}_{[\mu|}(b^{i}{}_{|\nu]} 
-h_{IJ}{}^{i}a^{I}A^{J}{}_{|\nu]})]
\\
& & \\
& & 
-2h_{IJ}{}^{i}a^{I}F^{J}{}_{\mu\nu}  
\\
& & \\
& & 
+Z^{i}\left( c_{\mu\nu} -g_{Ij}a^{I} B^{j}{}_{\mu\nu} 
+\tfrac{2}{3} g_{Jj}h_{IK}{}^{j}a^{I}A^{JK}{}_{\mu\nu]}\right)\, ,
\end{array}
\end{equation}

\noindent
we find that the local supersymmetry algebra closes on the $B^{i}$s in the
expected form (to lowest order in fermions) 

\begin{equation}
\left[\delta_{\epsilon_{1}},\delta_{\epsilon_{2}}\right]B^{i}{}_{\mu\nu}
 =    
\pounds_{\xi}B^{i}{}_{\mu\nu}
+\delta_{\Lambda}B^{i}{}_{\mu\nu}\, ,
\end{equation}

\noindent
where $\delta_{\Lambda}B^{i}{}_{\mu\nu}$ is the gauge transformation given in
Eq.~(\ref{eq:deltaBi}) in which the 0-form gauge parameters $\Lambda^{I}$ are
as in Eqs.~(\ref{eq:LambdaI}), the 1-form gauge parameters
$\Lambda^{i}{}_{\mu}$ are given by 

\begin{equation}
\Lambda^{i}{}_{\mu}
= 
\lambda^{i}{}_{\mu}
+b^{i}{}_{\mu}-h_{IJ}{}^{i}a^{I}A^{J}{}_{\mu}\, ,  
\end{equation}

\noindent
where

\begin{equation}
\begin{array}{rcl}
\lambda^{1}{}_{\mu}
& \equiv & 
e^{\frac{1}{2\sqrt{7}}\varphi+\frac{1}{2}\phi} (\chi\sigma_{\mu}
-e^{-\phi}\rho_{\mu})\, ,\\
& & \\
\lambda^{2}{}_{\mu}
& \equiv & 
e^{\frac{1}{2\sqrt{7}}\varphi} \sigma_{\mu}
\, ,\\    
\end{array}
\end{equation}

\noindent
and the shift term is given by

\begin{eqnarray}
Z^{1}\left[\Lambda_{\mu\nu} 
-\left( c_{\mu\nu} -g_{Ij}a^{I} B^{j}{}_{\mu\nu} 
+\tfrac{2}{3} g_{Jj}h_{IK}{}^{j}a^{I}A^{JK}{}_{\mu\nu}\right)\right]  
& &  \nonumber \\
& & \nonumber \\
&  &
\hspace{-5cm} 
= e^{\frac{1}{2\sqrt{7}}\varphi+\frac{1}{2}\phi}
\left[
\left(
\tfrac{1}{2} \Im {\rm m} (g) 
-4 \Re{\rm e} (k)
+\tfrac{1}{2\sqrt{7}} \Im {\rm m} (\tilde{g}) 
\right)\chi
\right.
\nonumber \\
& & \nonumber \\
& &
\hspace{-5cm} 
\left.
-
\left(
\tfrac{1}{2} \Re {\rm e} (g) 
+4 \Im{\rm m} (k)
-\tfrac{1}{2\sqrt{7}} \Re {\rm e} (\tilde{g}) 
\right)e^{-\phi}
\right]\xi_{\mu\nu}\, ,
\label{eq:2formshift1}
\\
& & \nonumber \\
Z^{2}\left[\Lambda_{\mu\nu} 
-\left( c_{\mu\nu} -g_{Ij}a^{I} B^{j}{}_{\mu\nu} 
-\tfrac{2}{3} g_{Jj}h_{IK}{}^{j}a^{I}A^{JK}{}_{\mu\nu}\right)\right]  
& &  \nonumber \\
& & \nonumber \\
&  &
\hspace{-5cm} 
= e^{\frac{1}{2\sqrt{7}}\varphi+\frac{1}{2}\phi}
\left(
\tfrac{1}{2} \Im {\rm m} (g) 
-4 \Re{\rm e} (k)
+\tfrac{1}{2\sqrt{7}} \Im {\rm m} (\tilde{g}) 
\right)\xi_{\mu\nu}\, .
\label{eq:2formshift2}
\end{eqnarray}

Now, let us analyze the constraints that involve $h_{IJ}{}^{i}$. From those
that only involve the embedding tensor we find seven linear constraints that
imply those in Eqs.~(\ref{eq:linearconstraints1}) and that can be used to
eliminate seven components of the embedding tensor:

\begin{equation}
  \begin{array}{rclrclrcl}
\vartheta_{\mathbf{2}}{}^{1} & = & 0\, ,\,\,\,  &
\vartheta_{\mathbf{1}}{}^{2}& = & \tfrac{3}{4}\vartheta_{\mathbf{2}}{}^{5}\, ,\,\,\,  & 
\vartheta_{\mathbf{1}}{}^{3}& = & \tfrac{3}{4}\vartheta_{\mathbf{2}}{}^{5}\, ,\,\,\,  \\
 & & & & & & & & \\
 \vartheta_{\mathbf{1}}{}^{1}& = & \tfrac{3}{2}\vartheta_{\mathbf{1}}{}^{5}\, ,\,\,\, & 
\vartheta_{\mathbf{2}}{}^{2}& = & \tfrac{3}{4}\vartheta_{\mathbf{1}}{}^{5}\, ,\,\,\,\hspace{.5cm}  &
\vartheta_{\mathbf{2}}{}^{3}& = & -\tfrac{3}{4}\vartheta_{\mathbf{1}}{}^{5}\, ,\,\,\,  \\
 & & & & & & & & \\
\vartheta_{0}{}^{4}& = & -\tfrac{1}{6}\vartheta_{0}{}^{5}\, ,\hspace{.5cm}  & 
& & & & & \\
\end{array}
\end{equation}

\noindent
leaving the eight components (a triplet of $SL(2,\mathbb{R})$ in the upper
component, a singlet and two doublets of $SL(2,\mathbb{R})$ in the lower
components)

\begin{equation}
\vartheta_{0}{}^{m}\, ,\,\,\, m=1,2,3\, ,\,\,\,
\vartheta_{0}{}^{5}\, ,\,\,\,
\vartheta_{\mathbf{i}}{}^{4}\, ,\,\,\,
\vartheta_{\mathbf{i}}{}^{5}\, ,\,\,\, \mathbf{i}=\mathbf{1},\mathbf{2}\, ,
\end{equation}

\noindent
as the only independent ones.  These components correspond to the eight
deformation parameters of the primary deformations studied in
Ref.~\cite{Bergshoeff:2002nv}. More precisely, the relation between them are

\begin{equation}
  \begin{array}{rclrclrcl}
\vartheta_{0}{}^{m} & = &  m_{m}\, ,\,\,\, (m=1,2,3)\hspace{.5cm}& 
\vartheta_{\mathbf{1}}{}^{4} & = & -m_{11}\, ,\hspace{.5cm}& 
\vartheta_{\mathbf{1}}{}^{5} & = & \tilde{m}_{4}\, ,\,\,\,
\\
& & & & & & & & \\
\vartheta_{0}{}^{5} & = & -\tfrac{16}{3}m_{\rm IIB}\, ,\,\,\,&
\vartheta_{\mathbf{2}}{}^{4} & = & m_{\rm IIA}\, ,\,\,\,& 
\vartheta_{\mathbf{2}}{}^{5} & = & m_{4}\, . 
\end{array}
\end{equation}

From the constraints that relate $h_{IJ}{}^{i}$ to $Z^{I}{}_{i},Z^{i}$ and
$g_{Ii}$ we can determine all these tensors, up to a constant $\zeta$,  in
terms of the independent components of the embedding tensor:

\begin{equation}
\label{eq:valuesofsometensors}
  \begin{array}{rclrcl}
Z^{\mathbf{i}}{}_{j} 
& = & 
\vartheta_{0}{}^{m}(T_{m})_{\mathbf{j}}{}^{\mathbf{i}} 
-\tfrac{3}{4}\vartheta_{0}{}^{5}\delta_{j}{}^{1}\delta_{1}{}^{\mathbf{i}}\,
,\,\,\, &
Z^{0}{}_{i} & = & 3\vartheta_{\mathbf{i}}{}^{4}
+\tfrac{1}{2}\vartheta_{\mathbf{i}}{}^{5}\, ,\\ 
& & & & & \\
g_{0i} & = & 0\, ,\,\,\, &
g_{\mathbf{i}j} & = & \varepsilon_{\mathbf{i}j}\, .\,\,\, \\
\end{array}
\end{equation}
 
The constant $\zeta$ is the coefficient of a Chern-Simons term in the 4-form
field strength and, therefore, will be completely determined by supersymmetry.

Finally, using all these results in
Eqs.~(\ref{eq:1-formshifts1}-\ref{eq:1-formshifts2}) we find 

\begin{eqnarray}
k
& = 
& -\tfrac{9i}{14} e^{-\frac{3}{2\sqrt{7}}\varphi +\frac{1}{2}\phi}
(\vartheta_{\mathbf{1}}{}^{4}\tau +\vartheta_{\mathbf{2}}{}^{4})\, ,  
\\
& & \nonumber \\
\Im{\rm m} (f) 
& = & 
\tfrac{3}{28} \vartheta_{0}{}^{5}e^{\frac{2}{\sqrt{7}}\varphi}\, ,
\\
& & \nonumber \\
\Re{\rm e} (f) +\tfrac{3}{4\sqrt{7}}\Im {\rm m}(\tilde{h}) 
& = & 
\tfrac{1}{4}e^{\frac{2}{\sqrt{7}}\varphi +\phi} 
\left\{
\tfrac{1}{2}(\vartheta_{0}{}^{2}+\vartheta_{0}{}^{3})
+(\vartheta_{0}{}^{1}-\tfrac{3}{4}\vartheta_{0}{}^{5}) \chi
\right.
\nonumber \\
& & \nonumber \\
& & 
\left.
-\tfrac{1}{2}(\vartheta_{0}{}^{2}-\vartheta_{0}{}^{3})|\tau|^{2}
\right\}\, ,
\end{eqnarray}

\noindent
which determines almost completely all the fermion shifts. We find that, in
order to determine completely $\Re{\rm e} (f)$ and $\Im {\rm m}(\tilde{h})$,
separately, one must study the closure of the supersymmetry algebra on the
fermions of the theory or on the bosons at higher order in fermions. The
result is  

\begin{eqnarray}
\Re{\rm e} (f)
& = & 
\tfrac{1}{14}e^{\frac{2}{\sqrt{7}}\varphi}\vartheta_{0}{}^{m}\mathcal{P}_{m}\, ,
\\
& & \nonumber \\
\Im {\rm m}(\tilde{h})  
& = & 
\tfrac{4}{\sqrt{7}}e^{\frac{2}{\sqrt{7}}\varphi}\vartheta_{0}{}^{m}\mathcal{P}_{m}\, .
\end{eqnarray}

All these results are collected in Appendix~\ref{sec-final}.

%%%%%%%%%%%%%%%%%%%%%%%%%%%%%%%%%%%%%%%%%%%%%%%%%%%%%%%%%%%%%%%%%%%%%%
%%%%%%%%%%%%%%%%%%%%%%%%%%%%%%%%%%%%%%%%%%%%%%%%%%%%%%%%%%%%%%%%%%%%%%
%%%%%%%%%%%%%%%%%%%%%%%%%%%%%%%%%%%%%%%%%%%%%%%%%%%%%%%%%%%%%%%%%%%%%%
%%%%%%%%%%%%%%%%%%%%%%%%%%%%%%%%%%%%%%%%%%%%%%%%%%%%%%%%%%%%%%%%%%%%%%

\subsection{The 3-form $C$}

In the next step we are going to consider the last of the fundamental,
electric $p$-forms of the theory, the 3-form $C$, whose gauge transformation
is given in Eq.~(\ref{eq:deltaC}).

%%%%%%%%%%%%%%%%%%%%%%%%%%%%%%%%%%%%%%%%%%%%%%%%%%%%%%%%%%%%%%%%%%%%%%
%%%%%%%%%%%%%%%%%%%%%%%%%%%%%%%%%%%%%%%%%%%%%%%%%%%%%%%%%%%%%%%%%%%%%%
%%%%%%%%%%%%%%%%%%%%%%%%%%%%%%%%%%%%%%%%%%%%%%%%%%%%%%%%%%%%%%%%%%%%%%
%%%%%%%%%%%%%%%%%%%%%%%%%%%%%%%%%%%%%%%%%%%%%%%%%%%%%%%%%%%%%%%%%%%%%%

\subsubsection{The 4-form field strength $G$}

The 4-form field strength $G$ is given by 

\begin{equation}
\label{eq:G}
G = \mathfrak{D}C -g_{Ii}\left(F^{I}-\tfrac{1}{2}Z^I{}_jB^j\right)\wedge B^{i} 
-\tfrac{1}{3} h_{IK}{}^{i}g_{Ji}A^{IJ}\wedge dA^{K}  +Z\tilde{C}\, ,
\end{equation}

\noindent
and it is covariant under general gauge transformations provided that the 4-form
$\tilde{C}$ transforms as

\begin{equation}
\label{eq:deltatildeC}
\begin{array}{rcl}
\delta_{\Lambda}\tilde{C}
& = &  
-\mathfrak{D}\tilde{\Lambda} 
-\tilde{g}_{I}\left[\Lambda^{I}G +C\wedge\delta_{\Lambda} A^{I} 
+F^{I}\wedge\Lambda +\tfrac{1}{12}g_{Ji}h_{KL}{}^{i}A^{IJK}\wedge
\delta_{\Lambda} A^{L}\right]
\\
& & \\
& & 
-\tilde{g}_{ij}[2H^{i}\wedge \Lambda^{j}-B^{i}\wedge\delta_{\Lambda} B^{j}
+2h_{IJ}{}^{i}B^{j}\wedge A^{I}\wedge\delta_{\Lambda} A^{J}]
\\
& & \\
& & 
-\tilde{g}_{IJK}\left[3\Lambda^{I} F^{JK} +2(F^{I}-Z^{I}{}_{i}B^{i})\wedge
  A^{J}\wedge\delta_{\Lambda} A^{K}
  -\tfrac{1}{4}X_{LM}{}^{J}A^{ILM}\wedge\delta_{\Lambda} A^{K}\right]
\\
& & \\
& & 
+Z^{i}\tilde{\Lambda}_{i}\, ,
\end{array}
\end{equation}

\noindent
where the new deformation tensors that we have introduced,
$\tilde{g}_{I},\tilde{g}_{ij}=-\tilde{g}_{ji}$ and
$\tilde{g}_{IJK}=\tilde{g}_{(IJK)}$, are subject to the constraints

\begin{eqnarray}
\label{eq:constraint11}  
g_{I[i} Z^{I}{}_{j]} +Z \tilde{g}_{ij} & = & 0\, ,\\
& & \nonumber \\
\label{eq:constraint12}  
X_{I}+g_{Ii}Z^{i}+Z\tilde{g}_{I} & = & 0\, ,\\
& & \nonumber \\
\label{eq:constraint13}  
h_{(IJ}{}^{i}g_{K)i}-Z \tilde{g}_{IJK} & = & 0\, ,
\end{eqnarray}

\noindent
plus the constraints that express the gauge invariance of the new deformation
parameters 

\begin{eqnarray}
\label{eq:constraint15}  
\tilde{X}_{I}\tilde{g}_{J} - X_{I\, J}{}^{K}\tilde{g}_{K}
& = & 0\, ,\\
& & \nonumber \\
\label{eq:constraint16}  
\tilde{X}_{I}\tilde{g}_{ij} -2X_{I\, [i|}{}^{k}\tilde{g}_{k|j]}
& = & 0\, ,\\
& & \nonumber \\
\label{eq:constraint17}  
\tilde{X}_{I}\tilde{g}_{JKL} -3X_{I\, (J}{}^{M}\tilde{g}_{KL)M}
& = & 0\, .  
\end{eqnarray}

%%%%%%%%%%%%%%%%%%%%%%%%%%%%%%%%%%%%%%%%%%%%%%%%%%%%%%%%%%%%%%%%%%%%%%
%%%%%%%%%%%%%%%%%%%%%%%%%%%%%%%%%%%%%%%%%%%%%%%%%%%%%%%%%%%%%%%%%%%%%%
%%%%%%%%%%%%%%%%%%%%%%%%%%%%%%%%%%%%%%%%%%%%%%%%%%%%%%%%%%%%%%%%%%%%%%
%%%%%%%%%%%%%%%%%%%%%%%%%%%%%%%%%%%%%%%%%%%%%%%%%%%%%%%%%%%%%%%%%%%%%%

\subsubsection{Closure of the supersymmetry algebra on the 3-form $C$}

Taking into account the form of $\delta_{\epsilon}C_{\mu\nu\rho}$ in the
undeformed case and the form of the field strength $G$, we arrive at the
following Ansatz for the supersymmetry transformation of the 3-form $C$:

\begin{equation}
\delta_{\epsilon}C_{\mu\nu\rho} 
= 
-\tfrac{3}{2} e^{-\frac{1}{\sqrt{7}}\varphi} \bar{\epsilon} \gamma_{[\mu\nu}
\left(\psi_{\rho]} +\tfrac{i}{6\sqrt{7}}\tilde{\lambda}^{*}
\right)   
+\mathrm{h.c.}
+ 3\delta_{\epsilon}A^{I}{}_{[\mu|}
\left(
g_{Ii}B^{i}{}_{|\nu\rho]}
+\tfrac{2}{3} h_{IJ}{}^{i} g_{Ki} A^{JK}{}_{|\nu\rho]}
\right)\, .
\end{equation}

\noindent
The last two terms are written in terms of the tensors $g_{Ii}$ and
$h_{IJ}{}^{i}$. In the undeformed theory these tensors have values which are
determined by supersymmetry (at orders in fermions higher than we are
considering here) and that cannot be changed in the deformed theory, as we
already discussed when we considered the 2-forms for $h_{IJ}{}^{i}$. Thus,
$h_{IJ}{}^{i}$ is given by Eq.~(\ref{eq:valueofhIJi}) and $g_{Ii}$ is given by
Eqs.~(\ref{eq:valuesofsometensors}) with $\zeta=+1$

Using the identity

\begin{equation}
\begin{array}{rcl}
\xi^{\sigma}G_{\sigma\mu\nu\rho} 
+3\pounds_{\xi}A^{I}{}_{[\mu|}
\left[g_{Ii} B^{i}{}_{|\nu\rho]}
+\tfrac{2}{3} h_{IJ}{}^{i}g_{Ki} A^{JK}{}_{|\nu\rho]}
\right] 
& = & 
\\
& & \\
& & \hspace{-7cm}
=\pounds_{\xi}C_{\mu\nu\rho}   
-3 \mathfrak{D}_{[\mu|}
\left[
\left( c_{|\nu\rho]} -g_{Ij}a^{I} B^{j}{}_{|\nu\rho]} 
+\tfrac{2}{3} g_{Jj}h_{IK}{}^{j}a^{I}A^{JK}{}_{|\nu\rho]}\right)\right]  
\\
& & \\
& & \hspace{-7cm}
+g_{Ii}
\left[
-a^{I} H^{i}{}_{\mu\nu\rho} 
-3F^{I}{}_{[\mu\nu|}(b^{i}{}_{|\rho]} -h_{JK}{}^{i}a^{J}A^{K}{}_{|\rho]}) 
\right]
\\
& & \\
& & \hspace{-7cm}
+Z
\left\{
\tilde{c}_{\mu\nu\rho} -\tilde{g}_{I}a^{I} C_{\mu\nu\rho} 
+3\tilde{g}_{ij}B^{i}{}_{[\mu\nu|}(b^{j}{}_{|\rho]}
-h_{JK}{}^{j} a^{J}A^{K}{}_{\rho})
-12\tilde{g}_{IJK}a^{I}A^{J}{}_{[\mu}\partial_{\nu}A^{K}{}_{\rho]} 
\right.
\\
& & \\
& & \hspace{-7cm}
\left.
+3h_{IJ}{}^{i}\tilde{g}_{ij}a^{I}A^{J}{}_{[\mu}B^{j}{}_{\nu\rho]}
-\tfrac{1}{4}
\left(
h_{IJ}{}^{i}g_{Ki}\tilde{g}_{L} +3 X_{JK}{}^{M}\tilde{g}_{ILM}
\right)a^{I}A^{JKL}{}_{\mu\nu\rho}
\right\}\, ,
\end{array}
\end{equation}

\noindent
one can see that the local supersymmetry algebra  closes into a general
coordinate transformation plus a gauge transformation of $C$ of the form  
Eq.~(\ref{eq:deltaC}) with

\begin{equation}
\label{eq:Lambda}
\Lambda_{\mu\nu} 
= e^{\frac{1}{\sqrt{7}}\varphi} \xi_{\mu\nu}
+\left( c_{\mu\nu} -g_{Ij}a^{I} B^{j}{}_{\mu\nu} 
-\tfrac{2}{3} g_{Jj}h_{IK}{}^{j}a^{I}A^{JK}{}_{\mu\nu}\right)\, ,
\end{equation}

\noindent
and with the identification

\begin{equation}
\begin{array}{rcl}
Z \left\{
\tilde{\Lambda}_{\mu\nu\rho} -\tilde{c}_{\mu\nu\rho} 
+\tilde{g}_{I}a^{I}C_{\mu\nu\rho} +3\tilde{g}_{ij} B^{i}{}_{[\mu\nu|}
\left(b^{j}{}_{|\rho]} -h_{JK}{}^{j}a^{J}A^{K}{}_{|\rho]} \right)
-12\tilde{g}_{IJK}a^{I}A^{J}{}_{[\mu}\partial_{\nu}A^{K}{}_{\rho]}
\right.
& & \\
& & \\
\left.
-3\tilde{g}_{ij} h_{IJ}{}^{i}a^{I}A^{J}{}_{[\mu}B^{j}{}_{\nu\rho]}
+\tfrac{1}{4} \left(\tilde{g}_{L}g_{Ki}h_{IJ}{}^{i}
+3\tilde{g}_{ILN}X_{JK}{}^{N} \right) a^{I}A^{JKL}{}_{\mu\nu\rho}
\right\}
& & \\
& & \\
= 
6 e^{-\frac{1}{\sqrt{7}}\varphi} 
\left[ 
\Im \mathrm{m} (f) +\tfrac{1}{6\sqrt{7}} \Re\mathrm{e}(\tilde{h})
\right] \zeta_{\mu\nu\rho}\, .   \\
\end{array}
\end{equation}

\noindent
Comparing Eq.~(\ref{eq:Lambda}) with Eqs.~(\ref{eq:2formshift1}) and
(\ref{eq:2formshift2}) we find that 

\begin{equation}
Z^{1}   =  X_{\mathbf{2}} =
3\vartheta_{\mathbf{2}}{}^{4} -\tfrac{1}{4}\vartheta_{\mathbf{2}}{}^{5}\, ,
\hspace{1cm}
Z^{2} = -X_{\mathbf{1}} =  
-3\vartheta_{\mathbf{1}}{}^{4}+\tfrac{1}{4}\vartheta_{\mathbf{1}}{}^{5} \, .
\end{equation}

To make further progress it is convenient to compute the 5-form $\tilde{G}$
since it will contain the tensors
$\tilde{g}_{I},\tilde{g}_{ij},\tilde{g}_{IJK}$ that appear in the above
expression. These tensors cannot be deformed (just as it happens with
$h_{IJ}{}^{i}$) and their values can be found by comparing the general form of
$\tilde{G}$ with the value found by duality, Eq.~(\ref{eq:tildeGundeformed}).

The generic form of the magnetic 5-form field strength $\tilde{G}$ is

\begin{equation}
\label{eq:tildeG}
\begin{array}{rcl}
\tilde{G} 
& = & 
\mathfrak{D}\tilde{C}
-\tilde{g}_{J}\left[
(F^{J}-Z^{J}{}_{j}B^{j})\wedge C
+\tfrac{1}{12}g_{Kj}h_{MN}{}^{j}A^{JKM}\wedge dA^{N}\right]
\\
& & \\
& & 
+2\tilde{g}_{ij}\left(H^{i}-\tfrac{1}{2}\mathfrak{D}B^{i}\right)\wedge B^{j}
-\tilde{g}_{JKL}\left(A^{J}\wedge dA^{KL}
+\tfrac{3}{4}X_{MN}{}^{L}A^{JMN}\wedge dA^{K}\right)
\\
& & \\
& & 
+Z^{i}\tilde{B}_{i}\, ,\\
\end{array}
\end{equation}

\noindent
and comparing this generic expression with Eq.~(\ref{eq:tildeGundeformed}) we
find that 

\begin{equation}
\tilde{g}_{I} = -\delta_{I}{}^{0}\, ,
\hspace{1cm}
\tilde{g}_{ij} = -\tfrac{1}{2}\varepsilon_{ij}\, ,
\hspace{1cm}
\tilde{g}_{IJK} = 0\, .  
\end{equation}

\noindent
Plugging these values into the constraints that involve $Z$
Eqs.~(\ref{eq:constraint10}),(\ref{eq:constraint18}), and
(\ref{eq:constraint11}-\ref{eq:constraint13}) we find that it must be related
to $\vartheta_{0}{}^{5}$ by 

\begin{equation}
Z = -\tfrac{3}{4} \vartheta_{0}{}^{5}\, ,  
\end{equation}

\noindent
and that $\vartheta_{0}{}^{5}$ must satisfy the two doublets of quadratic constraints

\begin{align}
\vartheta_{\mathbf{i}}{}^{4}\vartheta_{0}{}^{5}  
& = 
0\, ,
\\
& \nonumber \\
\vartheta_{\mathbf{i}}{}^{5}\vartheta_{0}{}^{5}  
& = 
0\, .
\end{align}

\noindent
Plugging our results into all the other constraints between deformation
tensors, we find that all of them are satisfied provided that the quadratic
constraints

\begin{align}
\varepsilon^{\mathbf{ij}}\vartheta_{\mathbf{i}}{}^{4}\vartheta_{\mathbf{j}}{}^{5}
& = 0\, ,
\\
& \nonumber \\
\vartheta_{0}{}^{m}
\left(12\vartheta_{\mathbf{i}}{}^{4}+5\vartheta_{\mathbf{i}}{}^{5} \right)
& = 0\, ,
\\
& \nonumber \\
\vartheta_{\mathbf{j}}{}^{4} \left(\vartheta_{0}^{m} T_{m}
\right)_{\mathbf{i}}{}^{\mathbf{j}}
& = 0\, , 
\end{align}

\noindent
are also satisfied. This set of irreducible quadratic constraints that cannot
be used to solve some deformation parameters in terms of the rest in an
analytic form, and to which the 9-form potentials of the theory may be
associated as explained in Section~\ref{sec-magnetic} is one of our main
results.

%%%%%%%%%%%%%%%%%%%%%%%%%%%%%%%%%%%%%%%%%%%%%%%%%%%%%%%%%%%%%%%%%%%%%%
%%%%%%%%%%%%%%%%%%%%%%%%%%%%%%%%%%%%%%%%%%%%%%%%%%%%%%%%%%%%%%%%%%%%%%
%%%%%%%%%%%%%%%%%%%%%%%%%%%%%%%%%%%%%%%%%%%%%%%%%%%%%%%%%%%%%%%%%%%%%%
%%%%%%%%%%%%%%%%%%%%%%%%%%%%%%%%%%%%%%%%%%%%%%%%%%%%%%%%%%%%%%%%%%%%%%

\section{Summary of results and discussion}
\label{sec-summary}

In the previous section we have constructed order by order in the rank of the
$p$-forms the supersymmetric tensor hierarchy of maximal 9-dimensional
supergravity, up to $p=3$, which covers all the fundamental fields of the
theory.
 
As it usually happens in all maximal supergravity theories, all the
deformation parameters can be expressed in terms of components of the
embedding tensor. Furthermore, we have shown that gauge invariance and local
supersymmetry allow for one triplet, two doublets and one singlet of
independent components of the embedding tensor

\begin{equation}
\label{eq:independent}
\vartheta_{0}{}^{m}\, ,\,\,\, m=1,2,3\, ,\,\,\,
\vartheta_{0}{}^{5}\, ,\,\,\,
\vartheta_{\mathbf{i}}{}^{4}\, ,\,\,\,
\vartheta_{\mathbf{i}}{}^{5}\, ,\,\,\, \mathbf{i}=\mathbf{1},\mathbf{2}\, .
\end{equation}

\noindent 
They can be identified with the deformation parameters studied in
Ref.~\cite{Bergshoeff:2002nv}:

\begin{equation}
  \begin{array}{rclrclrcl}
\vartheta_{0}{}^{m} & = &  m_{m}\, ,\,\,\, (m=1,2,3)\hspace{.5cm}& 
\vartheta_{\mathbf{1}}{}^{4} & = & -m_{11}\, ,\hspace{.5cm}& 
\vartheta_{\mathbf{1}}{}^{5} & = & \tilde{m}_{4}\, ,\,\,\,
\\
& & & & & & & & \\
\vartheta_{0}{}^{5} & = & -\tfrac{16}{3}m_{\rm IIB}\, ,\,\,\,&
\vartheta_{\mathbf{2}}{}^{4} & = & m_{\rm IIA}\, ,\,\,\,& 
\vartheta_{\mathbf{2}}{}^{5} & = & m_{4}\, . 
\end{array}
\end{equation}

\noindent
This proves, on the one hand, that no more deformations are possible and, on
the other hand, that all the deformations of maximal 9-dimensional
supergravity have a higher-dimensional origin, as shown in
Ref.~\cite{Bergshoeff:2002nv}.

Furthermore, we have also shown that it is not possible to give non-zero
values to all the deformation parameters at the same time, since they
must satisfy the quadratic constraints

\begin{align}
\label{eq:irreduciblequadraticconstraints}
\vartheta_{0}{}^{m}
\left(12\vartheta_{\mathbf{i}}{}^{4}+5\vartheta_{\mathbf{i}}{}^{5} \right)
& \equiv \mathcal{Q}^{m}{}_{\mathbf{i}}=0\, ,
\\
& \nonumber \\
\vartheta_{\mathbf{i}}{}^{4}\vartheta_{0}{}^{5}  
& \equiv \mathcal{Q}^{4}{}_{\mathbf{i}}= 0\, ,
\\
& \nonumber \\
\vartheta_{\mathbf{i}}{}^{5}\vartheta_{0}{}^{5}  
& \equiv \mathcal{Q}^{5}{}_{\mathbf{i}} = 0\, ,
\\
& \nonumber \\
\vartheta_{\mathbf{j}}{}^{4} \left(\vartheta_{0}^{m} T_{m}
\right)_{\mathbf{i}}{}^{\mathbf{j}}
& \equiv \mathcal{Q}_{\mathbf{i}} = 0\, , 
\\
& \nonumber \\
\varepsilon^{\mathbf{ij}}\vartheta_{\mathbf{i}}{}^{4}\vartheta_{\mathbf{j}}{}^{5}
& \equiv \mathcal{Q} = 0\, ,
\end{align}

\noindent
all of which are related to gauge invariance. 

Using these results, we can now apply the arguments developed in
Section~\ref{sec-magnetic} to relate the number of symmetries (Noether
currents), deformation parameters, and quadratic constraints to the numbers
(and symmetry properties) of 7-, 8- and 9-forms of the theory. Our results can
be compared with those presented in Ref.~\cite{Bergshoeff:2010xc} (Table~6)
and Ref.~\cite{Bergshoeff:2011zk} (Table~3) and found from $E_{11}$ level
decomposition.

\begin{table}
    \centering
\begin{tabular}{||c||c|c|c|c|c||}
\hline 
\hline 
$\mathbb{R}^{+}$ & 
$j_{1}$ & 
$j_{2}-j_{3}$ & 
$j_{2}+j_{3}$ & 
$j_{4}$ &  $j_{5}$  
\\
\hline 
\hline 
$\alpha$ & 
$0$ & $0$ & $0$ &
$0$ & $0$  
\\
\hline 
$\beta$ &
$0$ & $+3/4$ &  $-3/4$  & 
$0$ & $0$
\\
\hline 
$\gamma$ &
$0$ & $-2$ & $+2$ &
$0$ & $0$ 
\\
\hline 
$\delta$ & 
$0$ & $0$ & $0$  &
$0$ & $0$
\\
\hline
\hline 
\end{tabular}
   \caption{Weights of the Noether currents}
    \label{tab:noetherweights}
\end{table}

Associated to the symmetry group of the equations of motion of the theory,
$SL(2,\mathbb{R})\times \mathbb{R}^{2}$ there are 5 Noether currents $j_{A}$
that fit into one triplet and two singlets of $SL(2,\mathbb{R})$ and are
explicitly given in Appendix~\ref{sec-noether}. Their weights are given in
Table~\ref{tab:noetherweights}. They can be dualized as explained in
Section~\ref{sec-magnetic} into a triplet and two singlets of 7-forms
$\tilde{A}_{(7)}$ whose weights are given in
Table~\ref{tab:789formweights}. In
Refs.~\cite{Bergshoeff:2010xc,Bergshoeff:2011zk} the $\beta$ rescaling has not
been considered. As mentioned before, it corresponds to the so-called trombone
symmetry which may not survive to higher-derivative string corrections. The
associated 7-form singlet $\tilde{A}^{5}_{(7)}$ does not appear in their
analysis. The weights assigned in those references to the fields correspond to
one third of the weight of the $\alpha$ rescaling in our conventions.

\begin{table}[ht]
\centering
\hspace{-1cm}
\begin{tabular}{||c||c|c|c|c|c|c||}
\hline 
\hline 
$\mathbb{R}^{+}$ & 
$\vartheta_{0}{}^{1}$ & 
$\vartheta_{0}{}^{2}-\vartheta_{0}{}^{3}$ & 
$\vartheta_{0}{}^{2}+\vartheta_{0}{}^{3}$ & 
$\vartheta_{\mathbf{1}}{}^{4},\vartheta_{\mathbf{1}}{}^{5}$ &  
$\vartheta_{\mathbf{1}}{}^{4},\vartheta_{\mathbf{2}}{}^{5}$ &
$\vartheta_{0}{}^{5}$ 
\\
\hline 
\hline 
$\alpha$ & 
$-3$ & $-3$ & $-3$ &
$0$ & $0$  & 
$-3$
\\
\hline 
$\beta$ &
$-1/2$ & $-5/4$ &  $1/4$  & 
$3/4$ & $0$ & 
$-1/2$
\\
\hline 
$\gamma$ &
$0$ & $2$ & $-2$ &
$-1$ & $1$ &
$0$ 
\\
\hline 
$\delta$ & 
$0$ & $0$ & $0$  &
$-2$ & $-2$ &
$0$
\\
\hline
\hline 
\end{tabular}
   \caption{Weights of the embedding tensor components}
    \label{tab:embeddingtensorweights}
\end{table}

Associated to each of the $SL(2,\mathbb{R})$ multiplets of independent
embedding-tensor components there is a dual multiplet of 8-forms
$\tilde{A}_{(8)}$ (i.e.~one triplet, two doublets and one singlet) whose
weights are given in Table~\ref{tab:789formweights}. The doublet and singlet
associated to the gauging of the trombone symmetry using the doublet and
singlet of 1-forms are missing in
Refs.~\cite{Bergshoeff:2010xc,Bergshoeff:2011zk}, but the rest of the 8-forms
and their weights are in perfect agreement with those obtained from $E_{11}$.
Given the amount of work that it takes to determine which are the independent
components of the embedding tensor allowed by supersymmetry, this is a quite
non-trivial test of the consistency of the $E_{11}$ and the embedding-tensor
approaches.

\begin{table}[ht]
\centering
\hspace{-1cm}
\begin{tabular}{||c||c|c|c|c|c|c|c|c|c|c|c||}
\hline 
\hline 
$\mathbb{R}^{+}$ & 
$\mathcal{Q}_{\mathbf{1}}{}^{1}$ & 
$\mathcal{Q}_{\mathbf{2}}{}^{1}$ & 
$\mathcal{Q}_{\mathbf{1}}{}^{2-3}$ & 
$\mathcal{Q}_{\mathbf{2}}{}^{2-3}$ & 
$\mathcal{Q}_{\mathbf{1}}{}^{2+3}$ & 
$\mathcal{Q}_{\mathbf{2}}{}^{2+3}$ & 
$\mathcal{Q}_{\mathbf{1}}{}^{4},\mathcal{Q}_{\mathbf{1}}{}^{5}$ &  
$\mathcal{Q}_{\mathbf{2}}{}^{4},\mathcal{Q}_{\mathbf{2}}{}^{5}$ &  
$\mathcal{Q}_{\mathbf{1}}$ &
$\mathcal{Q}_{\mathbf{2}}$ &
$\mathcal{Q}$
\\
\hline 
\hline 
$\alpha$ & 
$-3$ & $-3$ & $-3$ &
$-3$ & $-3$ & $-3$ &
$-3$ & $-3$  & 
$-3$ & $-3$ &
0
\\
\hline 
$\beta$ &
$1/4$ & $-1/2$ & $-1/2$ & $-5/4$ & $1$ & $1/4$ & 
$1/4$ & $-1/2$ & 
$1/4$ & $-1/2$ & 
$3/4$
\\
\hline 
$\gamma$ &
$-1$ & $1$ & $1$ & $3$ & $-3$ & $-1$ &
$-1$ & $1$ &
$-1$ & $1$&
$0$
\\
\hline 
$\delta$ & 
$-2$ & $-2$ & $-2$  &
$-2$ & $-2$ & $-2$  &
$-2$ & $-2$ &
$-2$ & 
$-2$ & 
$-4$
\\
\hline
\hline 
\end{tabular}
   \caption{Weights of quadratic constraints components}
    \label{tab:qqweights}
\end{table}

Finally, associated to each of the quadratic constraints that the components
of the embedding tensor must satisfy
$\mathcal{Q}_{\mathbf{i}}{}^{m},\mathcal{Q}_{\mathbf{i}}{}^{4},
\mathcal{Q}_{\mathbf{i}}{}^{5},\mathcal{Q}_{\mathbf{i}},\mathcal{Q}$ there is
a 9-form potential $\tilde{A}_{(9)}$. The weights of these potentials are
given in Table~\ref{tab:789formweights}. If we set to zero the
embedding-tensor components associated to the trombone symmetry
$\vartheta_{A}{}^{5}$, the only constraints which are not automatically solved
are

\begin{equation}
\mathcal{Q}_{\mathbf{i}}{}^{m} = 12
\vartheta_{0}{}^{m}\vartheta_{\mathbf{i}}{}^{4}=0\, ,
\hspace{1cm}
\mathcal{Q}_{\mathbf{i}}=
\vartheta_{\mathbf{j}}{}^{4} \left(\vartheta_{0}^{m} T_{m}
\right)_{\mathbf{i}}{}^{\mathbf{j}}
=0\, .   
\end{equation}

The first of these constraints can be decomposed into a quadruplet and a
doublet: rewriting $\mathcal{Q}_{\mathbf{i}}{}^{m}$ in the equivalent form

\begin{equation}
\mathcal{Q}_{\mathbf{i}(\mathbf{j}\mathbf{k})}= 
\vartheta_{\mathbf{i}}{}^{4} 
\left(\vartheta_{0}^{m} T_{m}
\right)_{\mathbf{j}}{}^{\mathbf{l}}\varepsilon_{\mathbf{kl}}\, ,  
\end{equation}

\noindent
the quadruplet corresponds to the completely symmetric part
$\mathcal{Q}_{(\mathbf{i}\mathbf{j}\mathbf{k})}$ and the doublet to

\begin{equation}
\varepsilon^{\mathbf{jk}}\mathcal{Q}_{\mathbf{j}(\mathbf{k}\mathbf{i})}= -
\mathcal{Q}_{\mathbf{i}}\, ,
\end{equation}

\noindent
which is precisely the other doublet. Therefore, we get the quadruplet and one
doublet of 9-forms with weight $4$ under $\alpha/3$, while one more doublet is
found in Refs.~\cite{Bergshoeff:2010xc,Bergshoeff:2011zk} .

% \begin{table}
%     \centering
% \begin{tabular}{||c||c|c|c|c|c|c|c|c|c|c|c|c|c|c|c|c||}
% \hline 
% \hline 
% $\mathbb{R}^{+}$ & 
% $\tilde{A}_{(7)}^{1}$ & 
% $\tilde{A}_{(7)}^{2-3}$ & 
% $\tilde{A}_{(7)}^{2+3}$ & 
% $\tilde{A}_{(7)}^{4}$ &  
% $\tilde{A}_{(7)}^{5}$ &
% $\tilde{A}_{(8)}^{1}$ & 
% $\tilde{A}_{(8)}^{2-3}$ & 
% $\tilde{A}_{(8)}^{2+3}$ & 
% $\tilde{A}_{(8)}^{4\, \mathbf{i}}$ &  
% $\tilde{A}_{(8)}^{5\, \mathbf{i}}$ &
% $\tilde{A}_{(8)}^{4}$ &
% $\tilde{A}_{(9)\, m}^{\mathbf{i}}$ &
% $\tilde{A}_{(9)\, 4}^{\mathbf{i}}$ &
% $\tilde{A}_{(9)\, 5}^{\mathbf{i}}$ &
% $\tilde{A}_{(9)}^{\mathbf{i}}$ &
% $\tilde{A}_{(9)}$ 
% \\
% \hline 
% \hline 
% $\alpha$ & 
% $9$ & $9$ & $9$ &
% $9$ & $9$ 
% &
% $12$ & $12$ & $12$ &
% $9$ & $9$  & 
% $12$ 
% & $12$ & $12$ & $12$  & $12$ &  $9$
% \\
% \hline 
% $\beta$ &
% $0$ & $-3/4$ &  $3/4$  & 
% $0$ & $0$
% &
% $1/2$ & $5/4$ &  $-1/4$  & 
% $-3/4$ & $0$ & 
% $1/2$
% & & $-1/4$ & $1/2$ & & $-3/4$  
% \\
% \hline 
% $\gamma$ &
% $0$ & $2$ & $-2$ &
% $0$ & $0$ 
% &
% $0$ & $-2$ & $2$ &
% $1$ & $-1$ &
% $0$  
% & & $1$ & $-1$ & & $0$  
% \\
% \hline 
% $\delta$ & 
% $8$ & $8$ & $8$  &
% $8$ & $8$
% &
% $8$ & $8$ & $8$  &
% $2$ & $2$ &
% $8$
% & $10$ & $10$ & $10$ & $10$ & $12$  
% \\
% \hline
% \hline 
% \end{tabular}
%    \caption{Weights of the 7-, 8- and 9-form fields}
%     \label{tab:789formweights}
% \end{table}

\begin{table}
    \centering
\begin{tabular}{||c||c|c|c|c|c|c|c|c|c|c|c|c||}
\hline 
\hline 
$\mathbb{R}^{+}$ & 
$\tilde{A}_{(7)}^{m}$ & 
$\tilde{A}_{(7)}^{4}$ &  
$\tilde{A}_{(7)}^{5}$ &
$\tilde{A}_{(8)}^{m}$ & 
$\tilde{A}_{(8)}^{4\, \mathbf{i}}$ &  
$\tilde{A}_{(8)}^{5\, \mathbf{i}}$ &
$\tilde{A}_{(8)}^{4}$ &
$\tilde{A}_{(9)\, m}^{\mathbf{i}}$ &
$\tilde{A}_{(9)\, 4}^{\mathbf{i}}$ &
$\tilde{A}_{(9)\, 5}^{\mathbf{i}}$ &
$\tilde{A}_{(9)}^{\mathbf{i}}$ &
$\tilde{A}_{(9)}$ 
\\
\hline 
\hline 
$\alpha$ & 
$9$ & $9$ & $9$  &
$12$ & $9$ & $9$ & $12$ 
& $12$ & $12$ & $12$  & $12$ &  $9$
\\
\hline 
$\delta$ & 
$8$ & $8$ & $8$ & $8$ & $2$ & $2$ & $8$ & 
$10$ & $10$ & $10$ & $10$ & $12$  
\\
\hline
\hline 
\end{tabular}
   \caption{Weights of the 7-, 8- and 9-form fields.}
    \label{tab:789formweights}
\end{table}

This situation is similar to the one encountered in the $N=2$ theories in
$d=4,5,6$ dimensions \cite{Huebscher:2010ib}. In those cases, the
Ka\v{c}-Moody (here $E_{11}$) approach predicts one doublet of $d$-form
potentials more than the embedding-tensor formalism
\cite{Kleinschmidt:2008jj}. However, it can be seen that taking the undeformed
limit of the results obtained in the embedding-tensor formalism, one
additional doublet of $d$-forms arises because some St\"uckelberg shifts
proportional to deformation tensors that could be used to eliminate them, now
vanish. Furthermore, the local supersymmetry algebra closes on them as
independent fields.

By analogy with what happens in the $N=2$ theories in $d=4,5,6$ dimensions,
the same mechanism can make our results compatible with those of the $E_{11}$
approach (up to the trombone symmetry): we expect the existence of two
independent doublets of 9-forms in the undeformed theory but we also expect
new St\"uckelberg transformations in the deformed theory such that one a
combination of them is independent and the supersymmetry algebra closes.

This possibility (and the exclusion of any further 9-forms) can only be proven
by the direct exploration of all the possible candidates to 9-form
supersymmetry transformation rules, to all orders in fermions, something that
lies outside the boundaries of this work.

%%%%%%%%%%%%%%%%%%%%%%%%%%%%%%%%%%%%%%%%%%%%%%%%%%%%%%%%%%%%%%%%%%%%%%
%%%%%%%%%%%%%%%%%%%%%%%%%%%%%%%%%%%%%%%%%%%%%%%%%%%%%%%%%%%%%%%%%%%%%%
%%%%%%%%%%%%%%%%%%%%%%%%%%%%%%%%%%%%%%%%%%%%%%%%%%%%%%%%%%%%%%%%%%%%%%
%%%%%%%%%%%%%%%%%%%%%%%%%%%%%%%%%%%%%%%%%%%%%%%%%%%%%%%%%%%%%%%%%%%%%%

\section{Conclusions}
\label{sec-conclusions}

In this paper we have applied the embedding-tensor formalism to the study of
the most general deformations (\textit{i.e.}~gaugings and massive
deformations) of maximal 9-dimensional supergravity. We have used the complete
global $SL(2,\mathbb{R})\times \mathbb{R}^{2}$ symmetry of its equations of
motion, which includes the so-called \textit{trombone symmetry}. We have found
the constraints that the deformation parameters must satisfy in order to
preserve both gauge and supersymmetry invariance (the latter imposed through
the closure of the local supersymmetry algebra to lowest order in
fermions). We have used most of the constraints to express some components of
the deformation tensors in terms of a few components of the embedding tensor
which we take to be independent and which are given in
Eq.~(\ref{eq:independent}). At that point we have started making contact with
the results of Ref.~\cite{Bergshoeff:2002nv}, since those independent
components are precisely the 8 possible deformations identified there. All of
them have a higher-dimensional origin discussed in detail in
Ref.~\cite{Bergshoeff:2002nv}. The field strengths, gauge transformations and
supersymmetry transformations of the deformed theory, written in terms of the
independent deformation tensors, are collected in Appendix~\ref{sec-final}.

The 8 independent deformation tensors are still subject to quadratic
constraints, given in Eq.~(\ref{eq:irreduciblequadraticconstraints}), but
those constraints cannot be used to express analytically some of them in terms
of the rest, and, therefore, we must keep the 8 deformation parameters and
we must enforce these irreducible quadratic constraints. 

In Section~\ref{sec-summary} we have used our knowledge of the global
symmetries (and corresponding Noether 1-forms), the independent deformation
tensors and the irreducible quadratic constraints of the theory, together with
the general arguments of Section~\ref{sec-magnetic} to determine the possible
7-, 8- and 9-forms of the theory (Table~\ref{tab:789formweights}), which are
dual to the Noether currents, independent deformation tensors and irreducible
quadratic constraints. We have compared this spectrum of higher-rank forms
with the results of Refs.~\cite{Bergshoeff:2010xc,Bergshoeff:2011zk}, based on
$E_{11}$ level decomposition. We have found that, in the sector unrelated to
the trombone symmetry, which was excluded from that analysis, the
embedding-tensor formalism predicts one doublet of 9-forms less than the
$E_{11}$ approach. However, both predictions are not contradictory: the extra
doublet of 9-forms may not survive the deformations on which the
embedding-tensor formalism is built: new 9-form St\"uckelberg shifts
proportional to the deformation parameters may occur that can be used to
eliminate it so only one combination of the two 9-form doubles survives. This
mechanism is present in the $N=2$ $d=4,5,6$ theories \cite{Huebscher:2010ib},
although the physics behind it is a bit mysterious.

We can conclude that we have satisfactorily identified the extended field
content (the tensor hierarchy) of maximal 9-dimensional supergravity and,
furthermore, that all the higher-rank fields have an interpretation in terms
of symmetries and gaugings. This situation is in contrast with our
understanding of the extended field content of the maximal 10-dimensional
supergravities ($N=2A,B$) for which the $E_{11}$ approach can be used to get a
prediction of the higher-rank forms (which turns out to be correct
\cite{Bergshoeff:2005ac,Bergshoeff:2006qw,Bergshoeff:2010mv}) but th
embedding-tensor approach apparently cannot be used\footnote{In the $N=2B$
  case there are no 1-forms to be used as gauge fields and in the $N=2A$ case
  the only 1-form available is not invariant under the only rescaling symmetry
  available.} for this end. This seems to preclude an interpretation for the
9- and 10-form fields in terms of symmetries and gaugings\footnote{The 8-form
  fields are dual to the Noether currents of the global symmetries.}, at least
if we insist in the standard construction of the tensor hierarchy that starts
with the gauging of global symmetries. Perhaps a more general point of view is
necessary.

%%%%%%%%%%%%%%%%%%%%%%%%%%%%%%%%%%%%%%%%%%%%%%%%%%%%%%%%%%%%%%%%%%%%%%
%%%%%%%%%%%%%%%%%%%%%%%%%%%%%%%%%%%%%%%%%%%%%%%%%%%%%%%%%%%%%%%%%%%%%%
%%%%%%%%%%%%%%%%%%%%%%%%%%%%%%%%%%%%%%%%%%%%%%%%%%%%%%%%%%%%%%%%%%%%%%
%%%%%%%%%%%%%%%%%%%%%%%%%%%%%%%%%%%%%%%%%%%%%%%%%%%%%%%%%%%%%%%%%%%%%%

\section*{Acknowledgments}

TO would like to thank E.~Bergshoeff for several useful conversations.  This
work has been supported in part by the Spanish grants FPA2009-07692,
FIS2007-1234 and FPA2008-453, the Comunidad de Madrid grant HEPHACOS
S2009ESP-1473 and the Spanish Consolider-Ingenio 2010 program CPAN
CSD2007-00042 and Comunidad de Murcia-Fundaci\'on Seneca research grants.  The
work of JJFM has been supported by the Spanish Ministry of Education FPU grant
AP2008-00919.  JJFM would like to thank CERN and IFT-UAM/CSIC for their
hospitality dn Prof.~J.~Bernab\'eu for his kind support .  ET wishes to thank
the hospitality of the CERN-TH division. JJFM would like to thank the
hospitality of the CERN-TH Division and the IFT-UAM/CSIC.  TO wishes to thank
M.M.~Fern\'andez for her permanent support.

%%%%%%%%%%%%%%%%%%%%%%%%%%%%%%%%%%%%%%%%%%%%%%%%%%%%%%%%%%%%%%%%%%%%%%
%%%%%%%%%%%%%%%%%%%%%%%%%%%%%%%%%%%%%%%%%%%%%%%%%%%%%%%%%%%%%%%%%%%%%%
%%%%%%%%%%%%%%%%%%%%%%%%%%%%%%%%%%%%%%%%%%%%%%%%%%%%%%%%%%%%%%%%%%%%%%
%%%%%%%%%%%%%%%%%%%%%%%%%%%%%%%%%%%%%%%%%%%%%%%%%%%%%%%%%%%%%%%%%%%%%%
\appendix
%%%%%%%%%%%%%%%%%%%%%%%%%%%%%%%%%%%%%%%%%%%%%%%%%%%%%%%%%%%%%%%%%%%%%%
%%%%%%%%%%%%%%%%%%%%%%%%%%%%%%%%%%%%%%%%%%%%%%%%%%%%%%%%%%%%%%%%%%%%%%
%%%%%%%%%%%%%%%%%%%%%%%%%%%%%%%%%%%%%%%%%%%%%%%%%%%%%%%%%%%%%%%%%%%%%%
%%%%%%%%%%%%%%%%%%%%%%%%%%%%%%%%%%%%%%%%%%%%%%%%%%%%%%%%%%%%%%%%%%%%%%

\section{Conventions}
\label{app-conventions}

We follow the conventions of Ref.~\cite{Bergshoeff:2002nv}.  In particular, we
use mostly plus signature $(-,+,\cdots ,+)$ and the gamma matrices satisfy

\begin{equation}
\gamma^{*}_{a}= -\gamma_{a}\, ,
\hspace{1cm}
\gamma_{a}=\eta_{aa}\gamma_{a}^{\dagger}\, .
\end{equation}

The Dirac conjugate of a spinor $\epsilon$ is defined by

\begin{equation}
\bar{\epsilon} \equiv \epsilon^{\dagger}\gamma_{0}\, .  
\end{equation}

Then, we have

\begin{equation}
 \label{eq:anbn}
\begin{array}{rcl}
(\bar{\epsilon}\gamma^{(n)}\lambda)^{*}
 & = & 
a_{n}\bar{\epsilon}^{*}\gamma^{(n)}\lambda^{*}\, ,
\\
& & \\
(\bar{\epsilon}\gamma^{(n)}\lambda)^{*}
 & = & 
b_{n}\bar{\lambda}\gamma^{(n)}\epsilon \, ,
\\
\end{array}
\end{equation}

\noindent
where the signs $a_{n}$ and $b_{n}$ are  given in Table~\ref{tab:anbn}
  
\begin{table}
    \centering
    \begin{tabular}{||c|c|c|c|c|c|c|c|c|c|c||}
\hline\hline
$n$     & 0   & 1   & 2   & 3   & 4   & 5  & 6   & 7   & 8   & 9   \\
\hline\hline
$a_{n}$ & $-$ & $+$ & $-$ & $+$ & $-$ & $+$ & $-$ & $+$ & $-$ & $+$ \\
\hline        
$b_{n}$ & $+$ & $-$ & $-$ & $+$ & $+$ & $-$ & $-$ & $+$ & $+$ & $-$ \\
\hline\hline 
   \end{tabular}
   \caption{Values of the coefficients $a_{n}$ and $b_{n}$ defined in Eqs.~(\ref{eq:anbn}).}
    \label{tab:anbn}
\end{table}

%%%%%%%%%%%%%%%%%%%%%%%%%%%%%%%%%%%%%%%%%%%%%%%%%%%%%%%%%%%%%%%%%%%%%%
%%%%%%%%%%%%%%%%%%%%%%%%%%%%%%%%%%%%%%%%%%%%%%%%%%%%%%%%%%%%%%%%%%%%%%
%%%%%%%%%%%%%%%%%%%%%%%%%%%%%%%%%%%%%%%%%%%%%%%%%%%%%%%%%%%%%%%%%%%%%%
%%%%%%%%%%%%%%%%%%%%%%%%%%%%%%%%%%%%%%%%%%%%%%%%%%%%%%%%%%%%%%%%%%%%%%

\subsection{Spinor bilinears}
\label{sec-bilinears}

We define the following real bilinears of the supersymmetry parameters
$\epsilon_{1}$ and $\epsilon_{2}$:

\begin{eqnarray}
\bar{\epsilon}_{2}\epsilon_{1}
& \equiv &
a+ib\, ,
\\
& & \nonumber \\   
\bar{\epsilon}_{2}\epsilon^{*}_{1}
& \equiv &
c+id\, ,
\\
& & \nonumber \\   
\bar{\epsilon}_{2}\gamma_{\mu_{1}\cdots \mu_{n}}\epsilon_{1}
& \equiv &
\xi_{\mu_{1}\cdots \mu_{n}}+i\zeta_{\mu_{1}\cdots \mu_{n}}\, ,
\\
& & \nonumber \\   
\bar{\epsilon}_{2}\gamma_{\mu_{1}\cdots \mu_{n}}\epsilon_{1}^{*}
& \equiv &
\sigma_{\mu_{1}\cdots \mu_{n}}+i\rho_{\mu_{1}\cdots \mu_{n}}\, ,
\end{eqnarray}

%%%%%%%%%%%%%%%%%%%%%%%%%%%%%%%%%%%%%%%%%%%%%%%%%%%%%%%%%%%%%%%%%%%%%%
%%%%%%%%%%%%%%%%%%%%%%%%%%%%%%%%%%%%%%%%%%%%%%%%%%%%%%%%%%%%%%%%%%%%%%
%%%%%%%%%%%%%%%%%%%%%%%%%%%%%%%%%%%%%%%%%%%%%%%%%%%%%%%%%%%%%%%%%%%%%%
%%%%%%%%%%%%%%%%%%%%%%%%%%%%%%%%%%%%%%%%%%%%%%%%%%%%%%%%%%%%%%%%%%%%%%

\subsection{Relation with other conventions}
\label{sec-relationwithotherconventions}

The electric fields used in this paper are related to those used in
Ref.~\cite{Meessen:1998qm} (which uses a mostly minus signature) as
follows:

\begin{eqnarray}
K 
& = & 
e^{\frac{\sqrt{7}}{3}\varphi}\, ,
\\
& & \nonumber \\
\lambda \equiv C^{(0)} +ie^{-\varphi} 
& = & 
\tau \equiv \chi +ie^{-\phi}\, ,
\\
& & \nonumber \\ 
A_{(1)} 
& = &
A^{0}\, ,
\\
& & \nonumber \\
\mathbf{A}_{(1)} 
& = &
A^{\mathbf{i}}\, ,
\\
& & \nonumber \\
\mathbf{A}_{(2)} 
& = &
B^{i} +\tfrac{1}{2}A^{0\mathbf{i}}\, ,
\\
& & \nonumber \\
A_{(3)} 
& = &
-C +\tfrac{1}{2}\varepsilon_{\mathbf{i}j}A^{\mathbf{i}}\wedge B^{j} 
-\tfrac{1}{12} \varepsilon_{\mathbf{ij}} A^{0\mathbf{ij}}\, ,
\\
& & \nonumber \\
A_{(4)} 
& = &
-\tilde{C} +C\wedge A^{0} -\tfrac{1}{4}\varepsilon_{i\mathbf{j}}B^{i}\wedge
A^{0\mathbf{j}}\, .
\end{eqnarray}

\noindent
The field strengths are related by

\begin{eqnarray}
F_{(2)} 
& = &
F^{0}\, ,
\\
& & \nonumber \\
\mathbf{F}_{(2)} 
& = &
F^{\mathbf{i}}\, ,
\\
& & \nonumber \\
\mathbf{F}_{(3)} 
& = &
H^{i}\, ,
\\
& & \nonumber \\
F_{(4)} 
& = &
-G\, ,
\\
& & \nonumber \\
F_{(5)} 
& = &
-\tilde{G}\, .
\end{eqnarray}

The relation with the fields used in Ref.~\cite{Bergshoeff:2002nv}
(which also uses mostly plus signature) is given by (our fields are in
the r.h.s.~of these equations)

\begin{eqnarray}
B^{i}
& = &
-(B^{i} +\tfrac{1}{2}A^{0\mathbf{i}})\, ,   
\\
& & \nonumber \\
C
& = & 
-(C -\tfrac{1}{6}\varepsilon_{\mathbf{ij}}A^{0\mathbf{ij}})\, ,
\end{eqnarray}

\noindent
while the field strengths are related by 

\begin{eqnarray}
H^{i} & = & -H^{i}\, ,\\
& & \nonumber \\
G & = & -G\, .
\end{eqnarray}

\noindent
The rest of the fields are identical.

\section{Noether currents}
\label{sec-noether}

The Noether 1-form currents of the undeformed theory $j_{A}$ are given by

\begin{align}
\star j_{m}
& =
\star d\mathcal{M}_{ij}\left(\mathcal{M}^{-1}\right)_{jk}T_{mi}{}^k
+e^{\frac{4}{\sqrt7}\varphi}(
\mathcal{M}^{-1}_{\mathbf{ij}})T_{m\mathbf{k}}{}^\mathbf{i} 
A^\mathbf{k}\wedge \star F^\mathbf{j}
\nonumber \\
& \nonumber \\
&
+T_{mk}{}^i\left[ 
	e^{-\frac{1}{\sqrt7}\varphi}\mathcal{M}^{-1}_{ij}\left(
		B^k-\tfrac{1}{2}A^{0k}
	\right)\wedge\star H^j
+\tfrac{1}{2}\varepsilon_{ij}\left(
	-2e^{\frac{2}{\sqrt7}\varphi}A^\mathbf{j}\wedge B^k\wedge\star G
\right.\right.
\nonumber \\
& \nonumber \\
&
\left.\left.
\phantom{e^{\frac{2}{\sqrt7}\varphi}}
	+\left(
		B^j-A^{0j}
	\right)\wedge B^k\wedge G
	+\varepsilon_{ln}A^l\wedge B^{jk}\wedge\left(
		H^n-\tfrac{1}{2}A^n\wedge F^0
	\right)
\right.\right.
\nonumber \\
& \nonumber \\
&
\left.\left.
\phantom{e^{\frac{2}{\sqrt7}\varphi}}
	+\tfrac{1}{4}\varepsilon_{ln}A^{0ln}\wedge B^k\wedge H^j
\right)\right]\, ,
\end{align}

\begin{align}
\star j_4
&=
\tfrac{6}{\sqrt7}\star d\varphi
+3\left[
	e^{\frac{4}{\sqrt7}\varphi}A^0\wedge\star F^0+e^{-\frac{1}{\sqrt7}\varphi}\mathcal{M}^{-1}_{ij}\left(
		B^i+\tfrac{1}{2}A^{0i}
	\right)\wedge\star H^j
	+e^{\frac{2}{\sqrt7}\varphi}\left(
		C-\tfrac{1}{6}\varepsilon_\mathbf{ij}A^{0\mathbf{ij}}
	\right)\wedge\star G
\right.
\nonumber \\
& \nonumber \\
&
\left.
\phantom{e^{\frac{2}{\sqrt7}\varphi}}
	+A^0\wedge\left(
		C+\varepsilon_{\mathbf{i}j}A^{\mathbf{i}}\wedge B^j
	\right)\wedge G
\right]
+\tfrac{3}{2}\varepsilon_{ij}\left[
	\left(
		-C
		+\varepsilon_{kl}A^k\wedge B^l
		-\tfrac{7}{12}\varepsilon_{kl}A^{0kl}
	\right)\wedge B^i\wedge H^j
\right.
\nonumber \\
& \nonumber \\
&
\left.
	-\tfrac{3}{2}A^{0i}\wedge C\wedge H^j
	+\left(
		A^i\wedge B^j
		-\tfrac{1}{2} A^{0ij}
	\right)\wedge F^0\wedge C
\right]\, ,
\end{align}

\begin{align}
\star j_5
& =
\tfrac{\sqrt7}{4}\star d\varphi
-\tfrac{3}{8}\star\frac{\tau d\bar\tau+\textrm{c.c.}}{(\Im\textrm{m}\tau)^2}
+e^{\frac{4}{\sqrt7}\varphi}T_{50}{}^0 A^0\wedge\star F^0
+e^{\frac{3}{\sqrt7}\varphi}T_{5\mathbf{k}}{}^\mathbf{i}\mathcal{M}^{-1}_{ij}A^\mathbf{k}\wedge\star F^\mathbf{j}
\nonumber \\
& \nonumber \\
&
+e^{-\frac{1}{\sqrt7}\varphi}\mathcal{M}^{-1}_{ij}\left[
		T_{5k}{}^i\left(
			B^k-\tfrac{1}{2}A^{0k}
		\right)
+\tfrac{1}{4}A^{0i}
\right]\wedge\star H^j
\nonumber \\
& \nonumber \\
&
+e^{\frac{2}{\sqrt7}\varphi}\left(
	T_5  C
	-\tfrac{1}{12}\varepsilon_{ij}A^{0ij}
	-T_{5k}{}^i\varepsilon_{ij}\left(
		A^k\wedge B^j
		-\tfrac{1}{6}A^{0kj}
	\right)
\right)\wedge\star G
\nonumber \\
& \nonumber \\
&
+\tfrac{1}{4}\varepsilon_{ij}\left[
	T_{5k}{}^i\left(
		-2B^{jk}
		+3A^{0j}\wedge B^k
		-5A^{0k}\wedge B^j
	\right)
	-\tfrac{1}{2}A^{0i}\wedge B^j
	\right]\wedge G
\nonumber \\
& \nonumber \\
&
+\tfrac{1}{4}\varepsilon_{ij}\left[
	T_{5k}{}^i\left(
		+2\varepsilon_{ln}A^l\wedge B^{nk}
		-\varepsilon_{ln}A^{0ln}\wedge B^k
	\right)
	-T_5\left(6A^{0i}+B^i\right)\wedge C
	-\tfrac{1}{12}\varepsilon_{kl}A^{0kl}\wedge B^i
\right]\wedge H^j
\nonumber \\
& \nonumber \\
&
+\varepsilon_{ij}\varepsilon_{ln}T_{5k}{}^i\left[
	\tfrac{5}{6}A^{0jk}\wedge B^l
	-A^{0lj}\wedge B^k
	+\tfrac{1}{2}A^k\wedge B^{jl}
\right]\wedge H^n
\nonumber \\
& \nonumber \\
&
+T_5\left[
	A^0\wedge C\wedge G
	+\tfrac{1}{2}\varepsilon_{ij}
		\left(B^j+\tfrac{1}{2}A^{0j}\right)\wedge A^i\wedge F^0\wedge C		
\right]
\end{align}

%%%%%%%%%%%%%%%%%%%%%%%%%%%%%%%%%%%%%%%%%%%%%%%%%%%%%%%%%%%%%%%%%%%%%%
%%%%%%%%%%%%%%%%%%%%%%%%%%%%%%%%%%%%%%%%%%%%%%%%%%%%%%%%%%%%%%%%%%%%%%
%%%%%%%%%%%%%%%%%%%%%%%%%%%%%%%%%%%%%%%%%%%%%%%%%%%%%%%%%%%%%%%%%%%%%%
%%%%%%%%%%%%%%%%%%%%%%%%%%%%%%%%%%%%%%%%%%%%%%%%%%%%%%%%%%%%%%%%%%%%%%

\section{Final results}
\label{sec-final}

In this Appendix we give the final form of the deformed covariant field
strengths, covariant derivatives, gauge and supersymmetry transformations in
terms of the independent deformation parameters given in
Eq.~\ref{eq:independent}. We must bear in mind that they are assumed to
satisfy the irreducible quadratic constraints given in
Eq.~(\ref{eq:irreduciblequadraticconstraints}) and only then the field
strengths etc.~have the right transformation properties.

The covariant derivatives of the scalar fields are given by

\begin{align}
\mathfrak{D} \varphi
& =
-\tfrac{137}{24\sqrt{7}}\vartheta_{0}{}^{5} A^{0}
+\left(
-\tfrac{\sqrt{7}}{4}\vartheta_{\mathbf{i}}{}^{4}
+\tfrac{6}{\sqrt{7}}\vartheta_{\mathbf{i}}{}^{5}
\right)
A^{\mathbf{i}}\, ,
\\
& \nonumber \\
\mathfrak{D} \tau
& =
\vartheta_{0}{}^{m} k_{m}{}^{\tau} A^{\mathbf{1}}
-\tfrac{3}{4}\vartheta_{0}{}^{5}\tau A^{0}
+\tfrac{3}{4}\left(
\vartheta_{\mathbf{1}}{}^{5}\tau
+\vartheta_{\mathbf{2}}{}^{5}
\right)
\left(A^{\mathbf{1}}-\tau A^{\mathbf{2}}
\right)\, ,
\end{align}

\noindent
and their gauge transformations are explictly given by 

\begin{align}
\delta_{\Lambda} \varphi
& =
-\tfrac{137}{24\sqrt{7}}\vartheta_{0}{}^{5}\Lambda^{0}
+\left(
-\tfrac{\sqrt{7}}{4}\vartheta_{\mathbf{i}}{}^{4}
+\tfrac{6}{\sqrt{7}}\vartheta_{\mathbf{i}}{}^{5}
\right)
\Lambda^{\mathbf{i}}\, ,
\\
& \nonumber \\
\delta_{\Lambda} \tau
& =
\vartheta_{0}{}^{m} k_{m}{}^{\tau} \Lambda^{0}
-\tfrac{3}{4}\vartheta_{0}{}^{5}\tau\Lambda^{0}
+\tfrac{3}{4}
\left(
\vartheta_{\mathbf{1}}{}^{5}\tau
+\vartheta_{\mathbf{2}}{}^{5}
\right)
\left(
\Lambda^{\mathbf{1}}
-\tau\Lambda^{\mathbf{2}}
\right)\, .
\end{align}

The deformed $p$-form field strengths are given by 

\begin{eqnarray}
F^{0}
& = &
dA^{0}
-\tfrac{1}{2}\left(
3\vartheta_{\mathbf{i}}{}^{4}
+\tfrac{1}{2}\vartheta_{\mathbf{i}}{}^{5}
\right)A^{0\mathbf{i}}
+\left(
3\vartheta_{\mathbf{i}}{}^{4}
+\tfrac{1}{2}\vartheta_{\mathbf{i}}{}^{5}
\right)B^{\mathbf{i}}\, ,
\\
& & \nonumber \\
F^{\mathbf{i}}
& = &
dA^{\mathbf{i}}
+\tfrac{1}{2}\left(
\vartheta_{0}{}^{m}(T_{m}^{(3)})_{\mathbf{j}}{}^{\mathbf{i}}A^{0\mathbf{j}}
-\tfrac{3}{4}\delta_{\mathbf{1}}{}^\mathbf{i}\vartheta_{0}{}^{5} A^{0\mathbf{1}}
+\tfrac{3}{2}\varepsilon^{\mathbf{ij}}\vartheta_{\mathbf{j}}{}^{5} A^{\mathbf{12}}
\right)
\nonumber \\
& & \nonumber \\
& &
+\vartheta_{0}{}^{m}(T_{m}^{(3)})_{\mathbf{j}}{}^{\mathbf{i}} B^{j}
-\tfrac{3}{4}\delta_{1}{}^{\mathbf{i}}\vartheta_{0}{}^{5}B^{1}\, ,
\end{eqnarray}

\begin{eqnarray}
H^{i}
& = & 
\mathfrak{D}B^{i}
+\tfrac{1}{2}
\left(
A^{0}\wedge dA^{\mathbf{i}}+A^{\mathbf{i}}\wedge dA^{0}
\right)
+\tfrac{1}{6}
\varepsilon^{i\mathbf{j}}
\left(
3\vartheta_\mathbf{j}{}^{4}
+\tfrac{1}{2}\vartheta_{\mathbf{j}}{}^{5}
\right)A^{0\mathbf{12}}
\nonumber \\
& & \nonumber \\
& & 
+\varepsilon^{i\mathbf{j}} \left(
3\vartheta_{\mathbf{j}}{}^{4}
-\tfrac{1}{4}\vartheta_{\mathbf{j}}{}^{5}
\right)C\, ,
\\
& & \nonumber \\
G
& = &
\mathfrak{D}C
-\varepsilon_{\mathbf{i}j} \left[ F^{\mathbf{i}}\wedge B^{j} 
-\tfrac{1}{2} \delta^{j}{}_{\mathbf{j}}
\left(A^{\mathbf{i}}\wedge dA^{\mathbf{j} }
-\tfrac{1}{3}d(A^{0\mathbf{ij}}) \right) \right]
\nonumber \\
& & \nonumber \\
& &
+\tfrac{1}{2}\left(
\varepsilon_{ij}\vartheta_0{}^m(T_m^{(2)})_k{}^i B^{jk}
-\tfrac{3}{4}\vartheta_0{}^5 B^{12}
\right)
+Z\tilde C\, ,
\end{eqnarray}

\noindent
where the covariant derivatives acting on the different fields are given by 

\begin{eqnarray}
\mathfrak{D}B^{i}
& = &
dB^{i}
+\vartheta_{0}{}^{m}(T_{m}^{(2)})_{j}{}^{i} A^{0}\wedge B^{j}
-\tfrac{3}{4}\delta_{1}{}^{i}\vartheta_{0}{}^{5} A^{0}\wedge B^{1}
\nonumber \\
& & \nonumber \\
& &
+\left(
3\vartheta_{\mathbf{k}}{}^{4}-\tfrac{1}{4}\vartheta_{\mathbf{k}}{}^{5}
\right)
A^{\mathbf{k}}\wedge B^{i}
+\tfrac{3}{4}\delta_{\mathbf{j}}{}^{i}\vartheta_{k}{}^{5} A^{\mathbf{j}} \wedge B^{k}\, ,
\\
& & \nonumber \\
\mathfrak{D}C
& = & 
dC
-\tfrac{3}{4}\vartheta_{0}{}^{5} A^{0}\wedge C
+\left(
3\vartheta_{\mathbf{i}}{}^{4}
-\tfrac{1}{4}\vartheta_{\mathbf{i}}{}^{5}
\right)A^{\mathbf{i}}\wedge C\, .
\end{eqnarray}

The field strengths transform covariantly under the gauge transformations

\begin{eqnarray}
\delta_{\Lambda} A^{0}
& = & 
-\mathfrak{D}\Lambda^{0}
+\left(
3\vartheta_{\mathbf{i}}{}^{4}
+\tfrac{1}{2}\vartheta_{\mathbf{i}}{}^{5}
\right)\Lambda^{i}\, ,
\\
& & \nonumber \\
\delta_{\Lambda} A^{\mathbf{i}}
& = & 
-\mathfrak{D}\Lambda^{\mathbf{i}}
+\vartheta_{0}{}^{m}(T_{m}^{(3)})_{\mathbf{j}}{}^{\mathbf{i}}\Lambda^{j}
-\tfrac{3}{4}\delta_{1}{}^{\mathbf{i}}\vartheta_{0}{}^{5}\Lambda^{1}\, ,
\\
& & \nonumber \\
\delta_{\Lambda} B^{i}
& = & 
-\mathfrak{D}\Lambda^{i}
+F^{0}\wedge \Lambda^{\mathbf{i}}
+F^{\mathbf{i}} \Lambda^{0}
+\tfrac{1}{2}\left(
A^{0}\wedge \delta_{\Lambda} A^{\mathbf{i}}
+A^{\mathbf{i}}\wedge \delta_{\Lambda} A^{0}
\right)
\nonumber \\
& & \nonumber \\
& & 
+\varepsilon^{i\mathbf{j}}\left(
3\vartheta_{\mathbf{j}}{}^{4}
-\tfrac{1}{4}\vartheta_{\mathbf{j}}{}^{5}
\right)\Lambda\, ,
\\
& & \nonumber \\
\delta_{\Lambda} \left( C -\tfrac{1}{6}\varepsilon_{\mathbf{ij}}A^{0\mathbf{ij}}\right)
& = &
-\mathfrak{D}\Lambda
-\varepsilon_{\mathbf{i}j}\left(
 \Lambda^{\mathbf{i}}H^{j}
+F^\mathbf{i}\wedge \Lambda^{j}
-\delta_{\Lambda} A^{\mathbf{i}}\wedge B^{j}
\right)
\nonumber \\
& & \nonumber \\
& &
-\tfrac{1}{2}\varepsilon_{\mathbf{ij}}A^{0\mathbf{i}}\delta_{\Lambda}A^{\mathbf{j}}
+Z \tilde{\Lambda}\, ,
\end{eqnarray}

\noindent
where the covariant derivatives of the different gauge parameters are given by 

\begin{eqnarray}
\mathfrak{D}\Lambda^{0}
& = & 
d\Lambda^{0}
+\left(
3\vartheta_{\mathbf{i}}{}^{4}
+\tfrac{1}{2}\vartheta_{\mathbf{i}}{}^{5}
\right)A^{\mathbf{i}}\Lambda^{0}\, ,
\\
& & \nonumber \\
\mathfrak{D}\Lambda^{\mathbf{i}}
& = &
d\Lambda^{\mathbf{i}}
+\vartheta_{0}{}^{m}(T_{m}^{(3)})_{\mathbf{j}}{}^{\mathbf{i}}A^{0}\Lambda^{\mathbf{j}}
-\tfrac{3}{4}\delta_{1}{}^\mathbf{i}\vartheta_{0}{}^{5} A^{0}\Lambda^{\mathbf{1}}
+\tfrac{3}{4}\varepsilon^{\mathbf{ij}}\varepsilon_{\mathbf{kl}}\vartheta_{\mathbf{j}}{}^{5}
A^{\mathbf{k}}\Lambda^{\mathbf{l}}\, ,
\\
& & \nonumber \\
\mathfrak{D}\Lambda^{i}
& = &
d\Lambda^{i}
+\vartheta_{0}{}^{m}(T_{m}^{(2)})_{j}{}^{i} A^{0} \wedge \Lambda^{j}
+\left(
3\vartheta_{\mathbf{k}}{}^{4}
-\tfrac{1}{4}\vartheta_{\mathbf{k}}{}^{5}
\right)
A^{\mathbf{k}}\wedge \Lambda^{i}
\nonumber \\
& & \nonumber \\
& &
+\tfrac{3}{4}\delta_{\mathbf{j}}{}^{i}\vartheta_{\mathbf{k}}{}^{5}
A^{\mathbf{j}}\wedge \Lambda^{k}\, ,
\\
& & \nonumber \\
\mathfrak{D}\Lambda
& = &
d\Lambda
-\tfrac{3}{4}\vartheta_{0}{}^{5} A^{0}\wedge\Lambda
+\left(
3\vartheta_{\mathbf{i}}{}^{4}
-\tfrac{1}{4}\vartheta_{\mathbf{i}}{}^{5}
\right) A^{\mathbf{i}}\wedge\Lambda\, .
\end{eqnarray}

The supersymmetry transformation rules of the fermion fields are given by 

\begin{eqnarray}
\delta_{\epsilon}\psi_{\mu}
& = & 
\mathfrak{D}_{\mu}\epsilon
+f\gamma_{\mu} \epsilon
+k\gamma_{\mu} \epsilon^{*}
+\tfrac{i}{8\cdot 2!}e^{-\frac{2}{\sqrt{7}}\varphi}
\left(\tfrac{5}{7}\gamma_{\mu}\gamma^{(2)} 
-\gamma^{(2)}\gamma_{\mu} \right)F^{0}\epsilon  
\nonumber \\
& & \nonumber \\
& & 
-\tfrac{1}{8\cdot 2!}e^{\frac{3}{2\sqrt{7}}\varphi+\frac{1}{2}\phi}
\left(\tfrac{5}{7}\gamma_{\mu}\gamma^{(2)} 
-\gamma^{(2)}\gamma_{\mu} \right)(F^{1}-\tau F^{2})\epsilon^{*}  
\nonumber \\
& & \nonumber \\
& & 
-\tfrac{i}{8\cdot 3!}e^{-\frac{1}{2\sqrt{7}}\varphi}
\left(\tfrac{3}{7}\gamma_{\mu}\gamma^{(3)} 
+\gamma^{(3)}\gamma_{\mu} \right)(H^{1}-\tau H^{2})\epsilon^{*}  
\nonumber \\
& & \nonumber \\
& & 
-\tfrac{1}{8\cdot 4!}e^{\frac{1}{\sqrt{7}}\varphi}
\left(\tfrac{1}{7}\gamma_{\mu}\gamma^{(4)} 
-\gamma^{(4)}\gamma_{\mu} \right)G \epsilon\, ,
\end{eqnarray}

\begin{eqnarray}
\delta_{\epsilon}\tilde{\lambda}
& = & 
i\not\!\!\mathfrak{D} \varphi\epsilon^{*} +\tilde{g}\epsilon  +\tilde{h}\epsilon^{*}
-\tfrac{1}{\sqrt{7}}e^{-\frac{2}{\sqrt{7}}\varphi}\not\! F^{0}\epsilon^{*}
-\tfrac{3i}{2\cdot 2!\sqrt{7}}e^{\frac{3}{2\sqrt{7}}\varphi +\frac{1}{2}\phi}
(\not\! F^{1}-\tau^{*}\not\! F^{2})\epsilon
\nonumber \\
& & \nonumber \\
& & 
-\tfrac{1}{2\cdot 3!\sqrt{7}}e^{-\frac{1}{2\sqrt{7}}\varphi +\frac{1}{2}\phi}
(\not\!\! H^{1}-\tau^{*}\not\!\! H^{2})\epsilon
-\tfrac{i}{4!\sqrt{7}}e^{\frac{1}{\sqrt{7}}\varphi}\not\! G\epsilon^{*}\, ,
\end{eqnarray}

\begin{eqnarray}
\delta_{\epsilon}\lambda
& = & 
-e^{\phi}\not\!\!\mathfrak{D} \tau\epsilon^{*} +g\epsilon  +h\epsilon^{*}
-\tfrac{i}{2\cdot 2!}e^{\frac{3}{2\sqrt{7}}\varphi +\frac{1}{2}\phi}
(\not\! F^{1}-\tau\not\! F^{2})\epsilon
\nonumber \\
& & \nonumber \\
& & 
+\tfrac{1}{2\cdot 3!}e^{-\frac{1}{2\sqrt{7}}\varphi +\frac{1}{2}\phi}
(\not\!\! H^{1}-\tau\not\!\! H^{2})\epsilon\, ,
\end{eqnarray}

\noindent
where

\begin{eqnarray}
\mathfrak{D}_{\mu}\epsilon  
 & = & 
\left\{ 
\nabla_{\mu}
+\tfrac{i}{2}
\left[
\tfrac{1}{2}e^{\phi}
\mathfrak{D}^{5}_{\mu}\chi
+A^{I}{}_{\mu}\vartheta_{I}{}^{m}\mathcal{P}_{m}
\right]
+\tfrac{9}{14}\gamma_{\mu}\not\!\!A^{I}\vartheta_{I}{}^{4}
\right\}\epsilon\, ,
\\
& & \nonumber \\
\mathfrak{D}^{5}_{\mu}\chi
& = & 
\partial_{\mu}\chi
-\tfrac{3}{4}A^{I}{}_{\mu}\vartheta_{I}{}^{5} \chi\, ,
\end{eqnarray}

\noindent
and where the fermion shifts are given by 

\begin{eqnarray}
f
& = &
\tfrac{1}{14}e^{\frac{2}{\sqrt{7}}\varphi}
\left(
\vartheta_{0}{}^{m}\mathcal{P}_{m}
+\tfrac{3i}{2}\vartheta_{0}{}^{5}
\right)\, ,
\\
& & \nonumber \\
k
& = &
-\tfrac{9i}{14}e^{-\frac{3\varphi}{2\sqrt{7}}+\frac{\phi}{2}}
\left( \vartheta_{\mathbf{1}}{}^{4}\tau +\vartheta_{\mathbf{2}}{}^{4}
\right)\, ,
\\
& & \nonumber \\
\tilde{g}
& = &
e^{-\frac{3\varphi}{2\sqrt{7}}+\frac{\phi}{2}}
\left[
\tfrac{6}{\sqrt{7}}
\left( 
\vartheta_{\mathbf{1}}{}^{4}\tau^{*} +\vartheta_{\mathbf{2}}{}^{4}
\right)
+\tfrac{\sqrt{7}}{4}
\left(
\vartheta_{\mathbf{1}}{}^{5}\tau^{*} +\vartheta_{\mathbf{2}}{}^{5}
\right)
\right]\, ,
\\
& & \nonumber \\
\tilde{h}
& = &
\tfrac{4}{\sqrt{7}}e^{\frac{2}{\sqrt{7}}\varphi}
\left(
\tfrac{3}{16}\vartheta_{0}{}^{5}+\vartheta_{0}{}^{m}\mathcal{P}_{m}
\right)\, ,
\\
& & \nonumber \\
g
& = &
\tfrac{3}{4}e^{-\frac{3\varphi}{2\sqrt{7}}+\frac{\phi}{2}}
\left(
\vartheta_{\mathbf{1}}{}^{5} \tau +\vartheta_{\mathbf{2}}{}^{5}
\right)\, ,
\\
& & \nonumber \\
h
& = &
i e^{\frac{2\varphi}{\sqrt{7}}+\phi}
\left(
\vartheta_{0}{}^{m} k_{m}{}^{\tau}
-\tfrac{3}{4}\vartheta_{0}{}^{5} \tau
\right)\, .
\end{eqnarray}

The supersymmetry transformations of the bosonic fields are

\begin{eqnarray}
\delta_{\epsilon}\varphi 
& = & 
-\tfrac{i}{4}\bar{\epsilon}\tilde{\lambda}^{*}+\mathrm{h.c.}\, ,
\\
& & \nonumber \\  
\delta_{\epsilon}\tau 
& = & 
-\tfrac{1}{2}e^{-\phi}\bar{\epsilon}^{*}\lambda\, ,
\end{eqnarray}

\begin{eqnarray}
\delta_{\epsilon}A^{0}{}_{\mu}
& = & 
\tfrac{i}{2}e^{\frac{2}{\sqrt{7}}\varphi}\bar{\epsilon}
\left(\psi_{\mu} -\tfrac{i}{\sqrt{7}}\gamma_{\mu}\tilde{\lambda}^{*}\right) 
+\mathrm{h.c.}
\, ,
\\ 
& & \nonumber \\ 
\delta_{\epsilon}A^{\mathbf{1}}{}_{\mu}
& = & 
\tfrac{i}{2}\tau^{*}e^{-\frac{3}{2\sqrt{7}}\varphi+\frac{1}{2}\phi}
\left(
\bar{\epsilon}^{*}\psi_{\mu} 
-\tfrac{i}{4}\bar{\epsilon}\gamma_{\mu}\lambda
+\tfrac{3i}{4\sqrt{7}}\bar{\epsilon}^{*}\gamma_{\mu}\tilde{\lambda}^{*}
\right)
+\mathrm{h.c.}
\, ,
\\ 
& & \nonumber \\ 
\delta_{\epsilon}A^{\mathbf{2}}{}_{\mu}
& = & 
\tfrac{i}{2}e^{-\frac{3}{2\sqrt{7}}\varphi+\frac{1}{2}\phi}
\left(
\bar{\epsilon}^{*}\psi_{\mu} 
-\tfrac{i}{4}\bar{\epsilon}\gamma_{\mu}\lambda
+\tfrac{3i}{4\sqrt{7}}\bar{\epsilon}^{*}\gamma_{\mu}\tilde{\lambda}^{*}
\right)
+\mathrm{h.c.}
\end{eqnarray}

\begin{eqnarray}
\delta_{\epsilon}B^{1}
& = &
\tau^{*} e^{\frac{1}{2\sqrt{7}}\varphi +\frac{1}{2}\phi} 
\left[ 
\bar{\epsilon}^{*}\gamma_{[\mu}\psi_{\nu]} 
-\tfrac{i}{8}\bar{\epsilon}\gamma_{\mu\nu}\lambda 
-\tfrac{i}{8\sqrt{7}}\bar{\epsilon}^{*}\gamma_{\mu\nu}\tilde{\lambda}^{*}
\right]   
+\mathrm{h.c.}
\nonumber \\
& & \nonumber \\
& & 
-\delta^{1}{}_{\mathbf{i}}\left(A^{0}{}_{[\mu|}\delta_{\epsilon}A^{\mathbf{i}}{}_{|\nu]}
+A^{\mathbf{i}}{}_{[\mu|}\delta_{\epsilon}A^{0}{}_{|\nu]}\right)\, ,
\\
& & \nonumber \\  
\delta_{\epsilon}B^{2}
& = &
e^{\frac{1}{2\sqrt{7}}\varphi +\frac{1}{2}\phi} 
\left[ 
\bar{\epsilon}^{*}\gamma_{[\mu}\psi_{\nu]} 
-\tfrac{i}{8}\bar{\epsilon}\gamma_{\mu\nu}\lambda 
-\tfrac{i}{8\sqrt{7}}\bar{\epsilon}^{*}\gamma_{\mu\nu}\tilde{\lambda}^{*}
\right]   
+\mathrm{h.c.}
\nonumber \\
& & \nonumber \\
& & 
-\delta^{2}{}_{\mathbf{i}}
\left(A^{0}{}_{[\mu|}\delta_{\epsilon}A^{\mathbf{i}}{}_{|\nu]}
+A^{\mathbf{i}}{}_{[\mu|}\delta_{\epsilon}A^{0}{}_{|\nu]}\right)\, ,
\end{eqnarray}

\begin{eqnarray}
\delta_{\epsilon}C_{\mu\nu\rho} 
& = & 
-\tfrac{3}{2} e^{-\frac{1}{\sqrt{7}}\varphi} \bar{\epsilon} \gamma_{[\mu\nu}
\left(\psi_{\rho]} +\tfrac{i}{6\sqrt{7}}\tilde{\lambda}^{*}
\right)   
+\mathrm{h.c.}
\nonumber \\
& & \nonumber \\  
& & 
+ 3\delta_{\epsilon}A^{I}{}_{[\mu|}
\left(
g_{Ii}B^{i}{}_{|\nu\rho]}
+\tfrac{2}{3} h_{IJ}{}^{i} g_{Ki} A^{JK}{}_{|\nu\rho]}
\right)\, .
\end{eqnarray}

%%%%%%%%%%%%%%%%%%%%%%%%%%%%%%%%%%%%%%%%%%%%%%%%%%%%%%%%%%%%%%%%%%%%%%
%%%%%%%%%%%%%%%%%%%%%%%%%%%%%%%%%%%%%%%%%%%%%%%%%%%%%%%%%%%%%%%%%%%%%%
%%%%%%%%%%%%%%%%%%%%%%%%%%%%%%%%%%%%%%%%%%%%%%%%%%%%%%%%%%%%%%%%%%%%%%
%%%%%%%%%%%%%%%%%%%%%%%%%%%%%%%%%%%%%%%%%%%%%%%%%%%%%%%%%%%%%%%%%%%%%%

\end{document}